\documentclass[journal]{IEEEtran}

\usepackage{etex}
\usepackage[utf8]{inputenc} 
\usepackage[T1]{fontenc}    
\usepackage{hyperref}       
\usepackage{url}            
\usepackage{amsfonts}       
\usepackage{microtype}      
\usepackage{hyperref}
\usepackage{pgfplots}
\usepackage{amsmath, amssymb, amsthm, algorithm, algorithmic,epstopdf}
\usepackage{mathtools}
\usepackage{bm}
\usepackage{epsfig,color,epstopdf,graphicx,multirow,array,bbm}
\usepackage{subcaption,booktabs}

\usepackage{tabu}
\usepackage{wrapfig}

\usepackage{tikz}
\usepackage{pgfplots}
\usepgfplotslibrary{groupplots}

\pgfplotsset{
        my stylecompare/.style={
			width=.4\textwidth,
            height=4.5cm,
            label style={font=\Large},
            title style={font=\Large},
            x tick label style={font =\small, /pgf/number format/1000 sep=},
	    axis lines=left,
	    major x tick style = transparent,
	    major y tick style = transparent,
	    every y tick label/.style={
 		   xshift=-.7cm, yshift=-2pt,anchor=south west,inner sep=0pt,font=\small,
 		   scaled y ticks=false,
	    },
	    every x tick label/.style={font = \small},
            scaled x ticks=false,
        },
        my legend style compare/.style={
            legend entries={
            		GRASTA ($0.6$),
            		ORPCA ($7.6$),
            		s-ReProCS ($1.3$),	
            		NORST ($1.0$),	
            		Offline-NORST ($1.7$),
            		Alt Proj ($70.8$),
            		RPCA-GD ($92.2$),
            },
            legend style={
                at={(0.1,1.5)},
                anchor=north west,
            },
            legend columns=2,
	    legend style={font=\tiny},
        },
        cycle multi list={
        {blue, line width=0.6pt, mark=o,mark size=3pt},
        {black, line width=0.6pt, mark=square,mark size=2.5pt},
        {teal, line width=0.6pt, mark=oplus,mark size=2.5pt},
        {red, line width=0.6pt, mark=triangle,mark size=2.5pt},
        {red, solid, line width=0.5pt, mark=oplus,mark size=2.8pt},
        {olive, line width=0.6pt, mark=10-pointed star,mark size=2.5pt},
        {cyan, line width=0.6pt, mark=Mercedes star,mark size=3pt},
        {teal, line width=0.6pt, mark=oplus,mark size=2.5pt},
		},
}

\usetikzlibrary{decorations.markings}
\usetikzlibrary{decorations.pathreplacing}
\def\MarkLt{4pt}
\def\MarkSep{2pt}

\tikzset{
  TwoMarks/.style={
    postaction={decorate,
      decoration={
        markings,
        mark=at position #1 with
          {
              \begin{scope}[xslant=0.2]
              \draw[line width=\MarkSep,white,-] (0pt,-\MarkLt) -- (0pt,\MarkLt) ;
              \draw[-] (-0.5*\MarkSep,-\MarkLt) -- (-0.5*\MarkSep,\MarkLt) ;
              \draw[-] (0.5*\MarkSep,-\MarkLt) -- (0.5*\MarkSep,\MarkLt) ;
              \end{scope}
          }
       }
    }
  },
  TwoMarks/.default={0.5},
}

\usetikzlibrary{positioning}

\usetikzlibrary{calc}

\tikzstyle{block}  = [rectangle, draw, rounded corners, text width=5cm, text centered, minimum height=1em]
\tikzstyle{smallblock}  = [rectangle, draw, rounded corners,text width=1.8cm, text centered, minimum height=1em]
\tikzstyle{input}  = [rectangle, draw, text width=1.2cm, text centered, minimum height=1em]
\tikzstyle{output}  = [rectangle, draw, text width=1.2cm, text centered, minimum height=1em]
\tikzstyle{block1}  = [rectangle, draw, rounded corners,text width=5cm, text centered, minimum height=1em]
\tikzstyle{blockl1}  = [rectangle, draw, rounded corners,text width=4cm, text centered, minimum height=1em]

\newcommand{\norm}[1]{\left\|#1\right\|}

\newtheorem{theorem}{Theorem}

\newtheorem{definition}[theorem]{Definition}
\newtheorem{remark}[theorem]{Remark}

\newcommand{\bi}{\begin{itemize}}
\newcommand{\ei}{\end{itemize}}
\newcommand{\ben}{\begin{enumerate}}
\newcommand{\een}{\end{enumerate}}

\newcommand{\bean}{\begin{eqnarray*} }
\newcommand{\eean}{\end{eqnarray*} }
\newcommand{\bea}{\begin{eqnarray} }
\newcommand{\eea}{\end{eqnarray} }
\newcommand{\ba}{\begin{align*} }
\newcommand{\ea}{\end{align*} }

\newcommand{\nn}{\nonumber}

\newcommand{\dif}{{\text{dif}}}

\newcommand{\bl}{\begin{frame}}
\newcommand{\el} {\end{frame}}

\newcommand{\cred}{\color{red}} 

\newcommand{\svdeq}{\overset{\mathrm{SVD}}=}

\newcommand{\tmax}{d} 

\newcommand{\mt}{\bm{m}_t}
\newcommand{\xt}{\bm{s}_t}
\newcommand{\st}{\bm{s}_t}
\newcommand{\s}{\bm{s}}

\newcommand{\x}{\bm{s}}
\newcommand{\xhat}{\hat{\x}}
\newcommand{\xhatt}{\xhat_t}
\renewcommand{\l}{\bm{\ell}}
\newcommand{\lt}{\bm{\ell}_t}
\newcommand{\lhat}{\hat{\bm{\ell}}}

\newcommand{\lhatt}{\hat{\l}_t}   
\newcommand{\y}{\bm{m}}
\newcommand{\yt}{\y_t}
\newcommand{\tty}{{\tilde{\y}}}

\renewcommand{\v}{\bm{w}}
\newcommand{\vt}{\v_t}
\newcommand{\V}{\bm{W}}

\renewcommand{\a}{\bm{a}}

\newcommand{\et}{\bm{e}_t}

\newcommand{\new}{\mathrm{new}}

\newcommand{\at}{\bm{a}_t}

\newcommand{\I}{\bm{I}}

\newcommand{\Lam}{\bm{\Lambda}}

\newcommand{\T}{\mathcal{T}}

\newcommand{\pP}{\bm{\mathcal{P}}}

\newcommand{\A}{\bm{A}}

\newcommand{\Shat}{\hat{\bm{S}}}
\newcommand{\shat}{\hat{\bm{s}}}
\newcommand{\Lhat}{\hat{\bm{L}}}

\renewcommand{\P}{\bm{P}}
\newcommand{\LU}{\bm{U}}
\newcommand{\LV}{\bm{V}}

\newcommand{\G}{{\bm{G}}}

\newcommand{\tU}{\tilde{\bm{U}}}
\newcommand{\tV}{\tilde{\bm{V}}}
\newcommand{\tUhat}{\hat{\tU}}
\newcommand{\tVhat}{\hat{\tV}}

\renewcommand{\b}{\bm{b}}
\newcommand{\Phat}{\hat{\bm{P}}}

\newcommand{\Span}{\operatorname{span}} 

\newcommand{\E}{\mathbb{E}}

\newcommand{\train}{\mathrm{train}}
\newcommand{\That}{\hat{\mathcal{T}}}

\newcommand{\SE}{\mathrm{SE}}

\newcommand{\that}{{\hat{t}}}

\newcommand{\M}{\bm{M}}

\renewcommand{\L}{\bm{L}}
\renewcommand{\S}{\bm{S}}
\newcommand{\X}{\bm{S}}
\newcommand{\Y}{\bm{M}}
\newcommand{\W}{\bm{W}}

\newcommand{\tL}{\tilde{\bm{L}}}
\newcommand{\tS}{\tilde{\bm{S}}}

\newcommand{\outfracrow}{\small\text{max-outlier-frac-row}}
\newcommand{\outfraccol}{\small\text{max-outlier-frac-col}}

\renewcommand{\Re}{\mathbb{R}}

\newcommand{\rmat}{r_L} 

\newcommand{\xmint}{\rho_t}
\newcommand{\init}{\mathrm{init}}

\renewcommand{\xmint}{s_{\min}}
\newcommand{\zz}{\epsilon}
\renewcommand{\dif}{\Delta}
\newcommand{\bpsi}{\bm\Psi}
\IEEEoverridecommandlockouts

\begin{document}

\title{Robust Subspace Learning: Robust PCA, Robust Subspace Tracking, and Robust Subspace Recovery}
\author{Namrata Vaswani, Thierry Bouwmans, Sajid Javed, Praneeth Narayanamurthy
\thanks{N. Vaswani and P. Narayanamurthy are with the Dept. of Electrical and Computer Engineering, Iowa State University, Ames, USA. T. Bouwmans is with Maitre de Conf{\'e}rences, Laboratoire MIA, Universit{\'e} de La Rochelle, La Rochelle, France. Sajid Javed is with the Tissue Analytics Lab, Dept. of Computer Science, University of Warwick, UK. N. Vaswani is the corresponding author. Email: namrata@iastate.edu}}

\maketitle

\begin{abstract}
PCA is one of the most widely used dimension reduction techniques. A related easier problem is ``subspace learning'' or ``subspace estimation''. Given relatively clean data, both are easily solved via singular value decomposition (SVD). The problem of subspace learning or PCA in the presence of outliers is called robust subspace learning or robust PCA (RPCA). For long data sequences, if one tries to use a single lower dimensional subspace to represent the data, the required subspace dimension may end up being quite large. For such data, a better model is to assume that it lies in a low-dimensional subspace that can change over time, albeit gradually. The problem of tracking such data (and the subspaces) while being robust to outliers is called robust subspace tracking (RST). This article provides a magazine-style overview of the entire field of robust subspace learning and tracking. In particular solutions for three problems are discussed in detail: RPCA via sparse+low-rank matrix decomposition (S+LR), RST via S+LR, and ``robust subspace recovery (RSR)''. RSR assumes that an entire data vector is either an outlier or an inlier. The S+LR formulation instead assumes that outliers occur on only a few data vector indices and hence are well modeled as sparse corruptions.
\end{abstract}


\begin{IEEEkeywords}
Robust PCA, Robust Subspace Tracking, Robust Subspace Recovery, Robust Subspace Learning
\end{IEEEkeywords}


\section{Introduction} 

Principal Components Analysis (PCA) finds application in a variety of scientific and data analytics' problems ranging from exploratory data analysis and classification, e.g., image retrieval or  face recognition, to a variety of modern applications such as video analytics, recommendation system design, and understanding social networks dynamics. PCA finds a small number of orthogonal basis vectors, called principal components, along which most of the variability of the dataset lies.
In most applications, only the subspace spanned by the principal components (principal subspace) is of interest. For example, this is all that is needed for dimension reduction. This easier problem is typically called ``subspace learning'' or ``subspace estimation''.  
Sometimes the two terms are used interchangeably though.
%
%
%
Given a matrix of  clean data, PCA is easily accomplished via singular value decomposition (SVD) on the data matrix. The same problem becomes much harder if the data is corrupted by even a few outliers. The reason is that SVD is sensitive to outliers. We show an example of this in Fig. \ref{PCA_outliers}. In today's big data age, since data is often acquired using a large number of inexpensive sensors, outliers are becoming even more common. They occur due to various reasons such as node or sensor failures, foreground occlusion of video sequences, or abnormalities or other anomalous behavior on certain nodes of a network. 
The harder problem of PCA or subspace learning for outlier corrupted data is called {\em robust PCA or robust subspace learning}. 


Since the term ``outlier'' does not have a precise mathematical meaning, the robust PCA problem was, until recently, not well defined. Even so, many classical heuristics existed for solving it, e.g., see \cite{rpca_neu,Torre03aframework}, and references therein. 
In recent years, there have been multiple attempts to qualify this term. Most popular among these is the idea of treating an outlier as an additive sparse corruption which was popularized in the work of Wright and Ma \cite{error_correction_PCP_l1}. This is a valid definition because it models the fact that outliers occur infrequently and allows them to have any magnitude. In particular their magnitude can be much larger than that of the true data points. Using this definition, a nice recent work by Candes, Wright, Li, and Ma \cite{rpca} defined robust PCA as the problem of decomposing a given data matrix, $\Y$, into the sum of a low rank matrix, $\L$, whose column subspace gives the principal components, and a sparse matrix (outliers' matrix), $\X$. This definition, which is often referred to as the sparse+low-rank (S+LR) formulation, has lead to a large amount of interesting new work on robust PCA solutions, many of which are provably correct under simple assumptions, e.g., \cite{rpca,rpca2,rpca_zhang,rrpcp_perf,robpca_nonconvex,rrpcp_aistats,rpca_gd}.
A key motivating application is video analytics: decomposing a video into a slowly changing background video and a sparse foreground video \cite{Torre03aframework,rpca}. We show an example in Fig. \ref{fig_video}. The background changes slowly and the changes are usually dense (not sparse). It is thus well modeled as a dense vector that lies in low dimensional subspace of the original space \cite{rpca}. The foreground usually consists of one or more moving objects, and is thus correctly modeled as the sparse outlier.

Often, for long data sequences, e.g., long surveillance videos, or long dynamic social network connectivity data sequences, if one tries to use a single lower dimensional subspace to represent the data, the required subspace dimension may end up being quite large. For such data, a better model is to assume that it lies in a low-dimensional subspace that can change over time, albeit gradually. 
The problem of tracking a (slowly) changing subspace over time is often referred to as ``subspace tracking'' or ``dynamic PCA''. The problem of tracking it in the presence of additive sparse outliers can thus be called either {\em ``robust subspace tracking''} or {\em ``dynamic robust PCA''} \cite{rrpcp_allerton,rrpcp_tsp,rrpcp_perf}.

\pgfplotstableread[col sep = comma]{small_noise_twod.dat}\sninp
\pgfplotstableread[col sep = comma]{large_noise_twod.dat}\lninp

\pgfplotsset{every axis title/.style={text width=.5\textwidth, align=center, at={(.5, -0.2)}, below}}
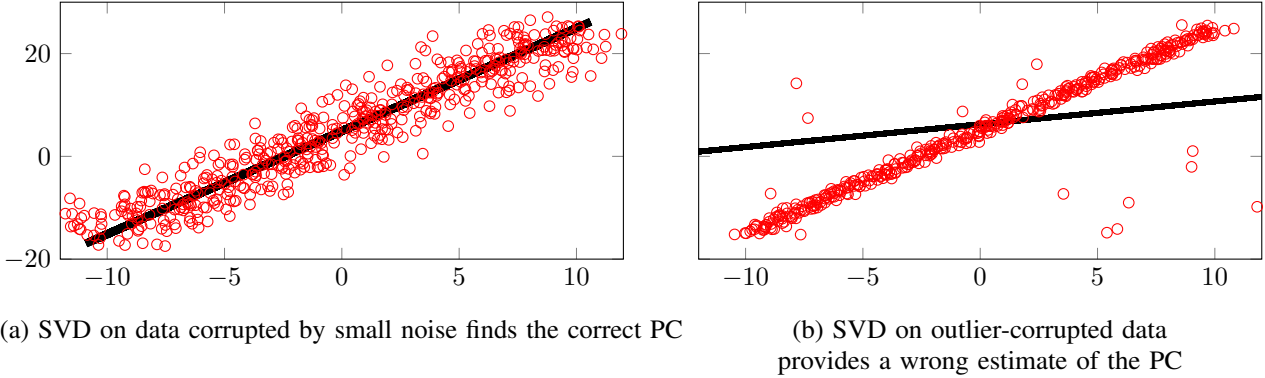
\begin{figure*}[t!]
\begin{center}
\begin{tikzpicture}
    \begin{groupplot}[
        group style={
            group size=2 by 1,
            x descriptions at=edge bottom,
            y descriptions at=edge left,
        },
        height=5cm,
        width=.5\textwidth,
        %
        xmin=-12,xmax=12,
        ymin=-20, ymax=30,
    ]
       \nextgroupplot[
       axis x line*=box,
		axis y line*=box,
            title={(a) SVD on data corrupted by small noise finds the correct PC}
        ]
                \node [text width=1em,anchor=north west] at (rel axis cs: 0.45,-1.2)
                {\subcaption{\label{fig:small_noise}}};

	        \addplot[%
	        scatter=false,
	        only marks,
	        mark=o,
	        color=red,
	        on layer=background] table[x index = {0}, y index = {1}]{\sninp};
	
	            	        \addplot[black,
	        line width=3pt,
	        on layer=foreground] table[x index = {2}, y index = {3}]{\sninp};

        \nextgroupplot[
               axis x line*=box,
		axis y line*=box,
            title={(b) SVD on outlier-corrupted data provides a wrong estimate of the PC}
        ]
        \node [text width=1em,anchor=north west] at (rel axis cs: 0.45,-1.2)
                {\subcaption{\label{fig:large_noise}}};

        \addplot[%
	        scatter=false,
	        only marks,
	        mark=o,
	        color=red,
	        on layer=main] 	
	        table[x index = {0}, y index = {1}]{\lninp};
	        \addplot[black,
	        line width=2pt,
	        on layer=main]
	        table[x index = {2}, y index = {3}]{\lninp};
	
	    \end{groupplot}
\end{tikzpicture}
\end{center}
\caption{\small{(a) PCA in small noise: the SVD solution works. The black line is the estimated principal component computed using the observed data. 
(b) PCA in outliers: the SVD solution fails to correctly find the direction of largest variance of the true data. Instead its estimate is quite far from the true principal component.
}}
\label{PCA_outliers}
\end{figure*}

Another way to interpret the word ``outlier'' is to assume that either an entire data vector is an outlier or it is an inlier. This a more appropriate model for outliers due to malicious users in recommendation system design or due to malicious participants in survey data analysis who enter all wrong answers in a survey or in rating movies.
%
For this problem setting, it is clearly impossible to recover the entire low rank matrix $\L$, it is only possible to correctly estimate its column subspace (and not the individual principal components). In modern literature, this problem is referred to as ``robust subspace recovery'' \cite{novel_m_estimator}.
%

Henceforth, the terms ``robust PCA (RPCA)'' and ``robust subspace tracking (RST)'' only refer to the S+LR definitions while robust subspace recovery (RSR) is the above problem.
This article provides a magazine-style overview of solutions for all these problems. For a detailed review, please see \cite{rrpcp_proc} for RPCA and RST via S+LR, and to \cite{lerman_review} for RSR.%

Two important extensions of RPCA are (a) robust matrix completion (or robust PCA with missing data or low rank matrix completion with sparse outliers)  \cite{rmc,rmc_gd} 
and (b) compressive or under-sampled robust PCA which involves robust PCA from undersampled projected linear measurements \cite{rrpcp_allerton11,SpaRCS,compressivePCP,rpca_Giannakis,candes_mri}. This is useful in dynamic or functional MRI applications.%

%


\begin{figure*}[t!]
\begin{subfigure}[t]{0.5\linewidth}
\centering
\begin{tabular}{@{}c@{}c@{}c@{}}
\\    \newline
\includegraphics[scale=1.02]{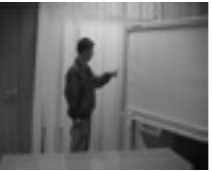} &
\includegraphics[scale=1.08]{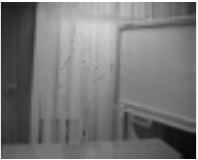} &
\includegraphics[scale=1.02]{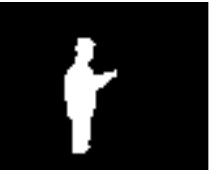} \\ \newline
\includegraphics[scale=1.01]{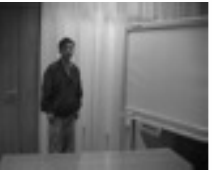} &
\includegraphics[scale=1.06]{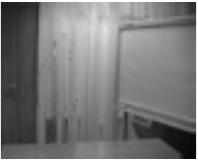} &
\includegraphics[scale=1.08]{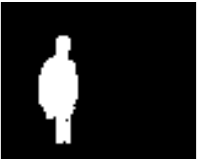} \\ \newline
\includegraphics[scale=1.01]{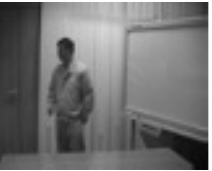} &
\includegraphics[scale=1.02]{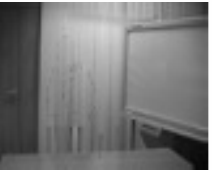} &
\includegraphics[scale=1.1]{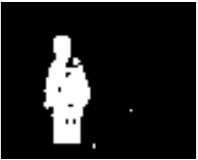}
\\ \newline
\small{Original} & \small{Background} & \small{Foreground}
\end{tabular}
\caption{Video Analytics}
\label{fig_video}
\end{subfigure}%
\begin{subfigure}[t]{0.5\linewidth}
\centering
\begin{tabular}{@{}c@{}c@{}c@{}}
\\    \newline
\includegraphics[scale=0.63, trim={2.3cm, 2cm, 2.3cm, 1cm}, clip]{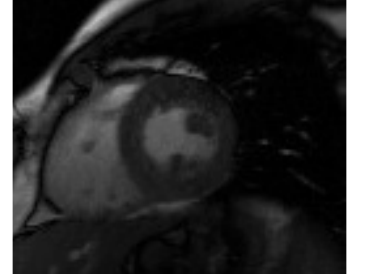} &
\includegraphics[scale=0.63, trim={2.3cm, 2cm, 2.3cm, 1cm}, clip]{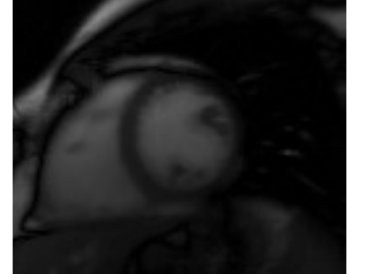} &
\includegraphics[scale=0.63, trim={2.3cm, 2cm, 2.3cm, 1cm}, clip]{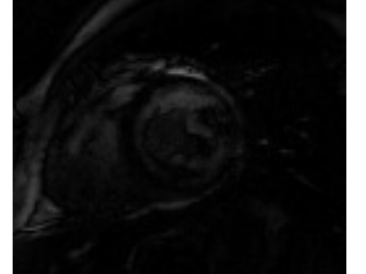} \\ \newline
\includegraphics[scale=0.63, trim={2.3cm, 2cm, 2.3cm, 1cm}, clip]{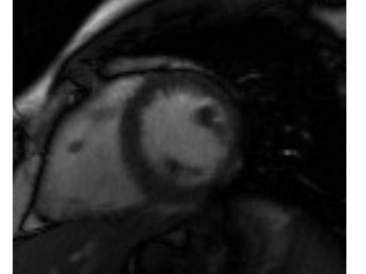} &
\includegraphics[scale=0.63, trim={2.3cm, 2cm, 2.3cm, 1cm}, clip]{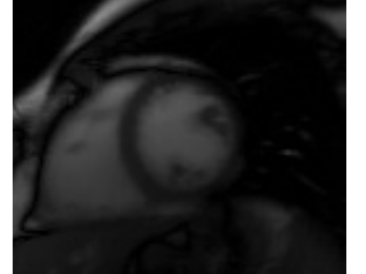} &
\includegraphics[scale=0.63, trim={2.3cm, 2cm, 2.3cm, 1cm}, clip]{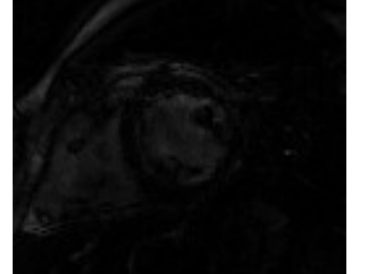} \\ \newline
\includegraphics[scale=0.63, trim={2.3cm, 2cm, 2.3cm, 1cm}, clip]{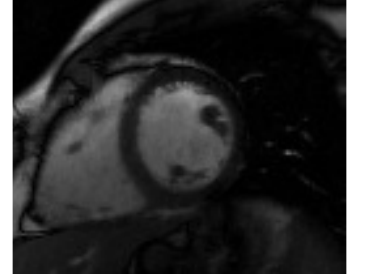} &
\includegraphics[scale=0.63, trim={2.3cm, 2cm, 2.3cm, 1cm}, clip]{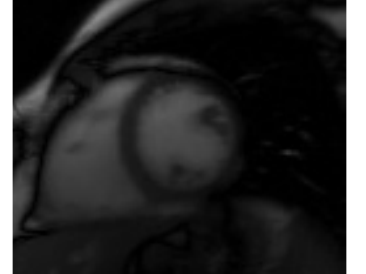} &
\includegraphics[scale=0.63, trim={2.3cm, 2cm, 2.3cm, 1cm}, clip]{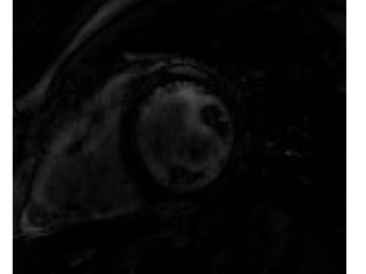} \\ \newline
\small{Original} & \small{Background} & \small{Sparse ROI}
\end{tabular}
\caption{Dynamic MRI}
\label{fig_mri}
\end{subfigure}
\caption{\footnotesize{(a) A video analytics application: Video layering (foreground-background separation) in videos can be posed as a Robust PCA problem. This is often the first step to simplify many computer vision and video analytics' tasks. For one such example, see Fig. \ref{fig_denoise}. We show three frames of a video in the first column. The background images for these frames are shown in the second column. Notice that they all look very similar and hence are well modeled as forming a low rank matrix. The foreground support is shown in the third column. This clearly indicates that the foreground is sparse and changes faster than the background. Result taken from \cite{rrpcp_tsp}, code at \url{http://www.ece.iastate.edu/~hanguo/PracReProCS.html}.
(b) Low-rank and sparse matrix decomposition for accelerated dynamic MRI \cite{candes_mri}. 
The first column shows three frames of cardiac cine data. The second column shows the slow changing background part of this sequence, while the third column shows the fast changing sparse region of interest (ROI). This is also called the ``dynamic component''. These are the reconstructed columns obtained from 8-fold undersampled data. They were reconstructed using under-sampled stable PCP \cite{candes_mri}.%
}}
\end{figure*}

\subsubsection{Article Organization} We describe the various applications of RPCA next followed by a brief discussion of desirable algorithm properties. In the following section (Section II), we provide a concise overview of all the three problems  - RPCA, RST, and RSR - and solutions for them along with one representative theoretical guarantee for each. This section is aimed at the reader who is already familiar with the problems and would like to see a summary of the solution approaches and a flavor of the guarantees. In the two sections after this (Sections III and IV) we provide detailed explanation of solution approaches for RPCA and RST via S+LR, and for RSR respectively. A discussion of how to set parameters in practice and quantitative experimental comparisons on real videos are given in Section V. We conclude in Section VI with a detailed discussion of open questions.

\subsection{Applications}
We describe some of the modern applications where the robust PCA problem occurs.

{\em Computer vision and video analytics. }
A large class of videos, e.g., surveillance videos, consist of a sparse foreground layer which contains one or more moving objects or people and a slow changing background scene. We show an example in Fig. \ref{fig_video}.
Assume that all images are arranged as 1D vectors. Then, the $t$-th image forms the $t$-th column, $\yt$, of the data matrix, $\M$.
If the image size is denoted by $n$ and the total number of images in the video by $\tmax$, then $\M$ is an $n \times \tmax$ matrix.
Assuming that the background scene changes slowly over time, the $t$-th background image forms the $t$-th column, $\l_t$, of the low rank matrix $\L$. Let $\rmat$ denote its rank.
If the background never changes, then $\L$ will be rank one. The low rank assumption implies that the background changes depend on a much small number, $\rmat$, of factors than either the number of images $\tmax$ or the image size $n$.
Let $\T_t$ denote the support (set of indices of the nonzero pixels) of the foreground frame $t$.
To get the $\M = \L + \X$ formulation, we define $\s_t$ as a sparse vector with support $\T_t$ and with nonzero entries equal to the difference between foreground and background intensities on this support.

\begin{figure*}[t!]
				\begin{subfigure}{0.49\textwidth}
				\centering
					\begin{tabular}{cc}
					\includegraphics[height=15mm]{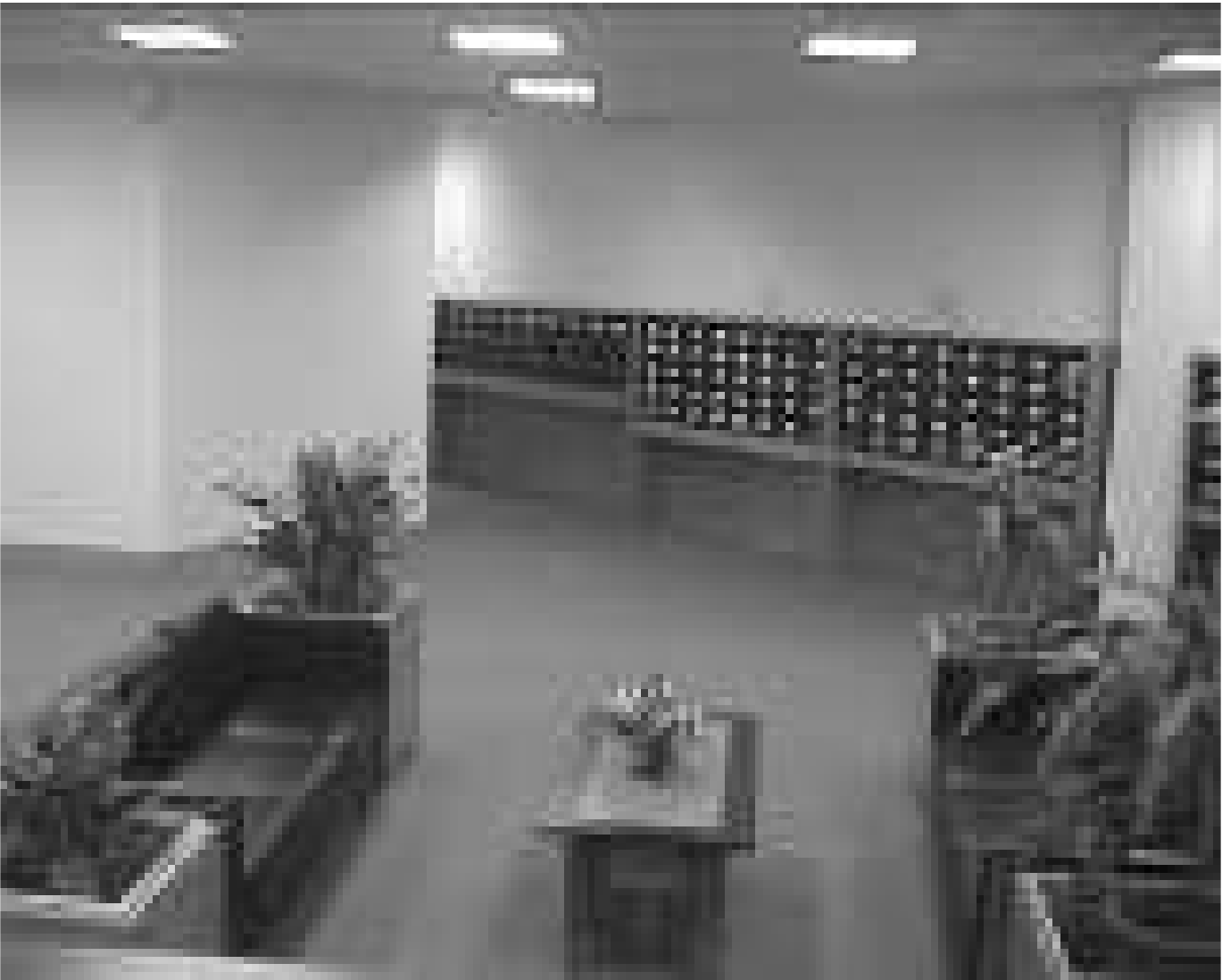}
					\includegraphics[height=15mm]{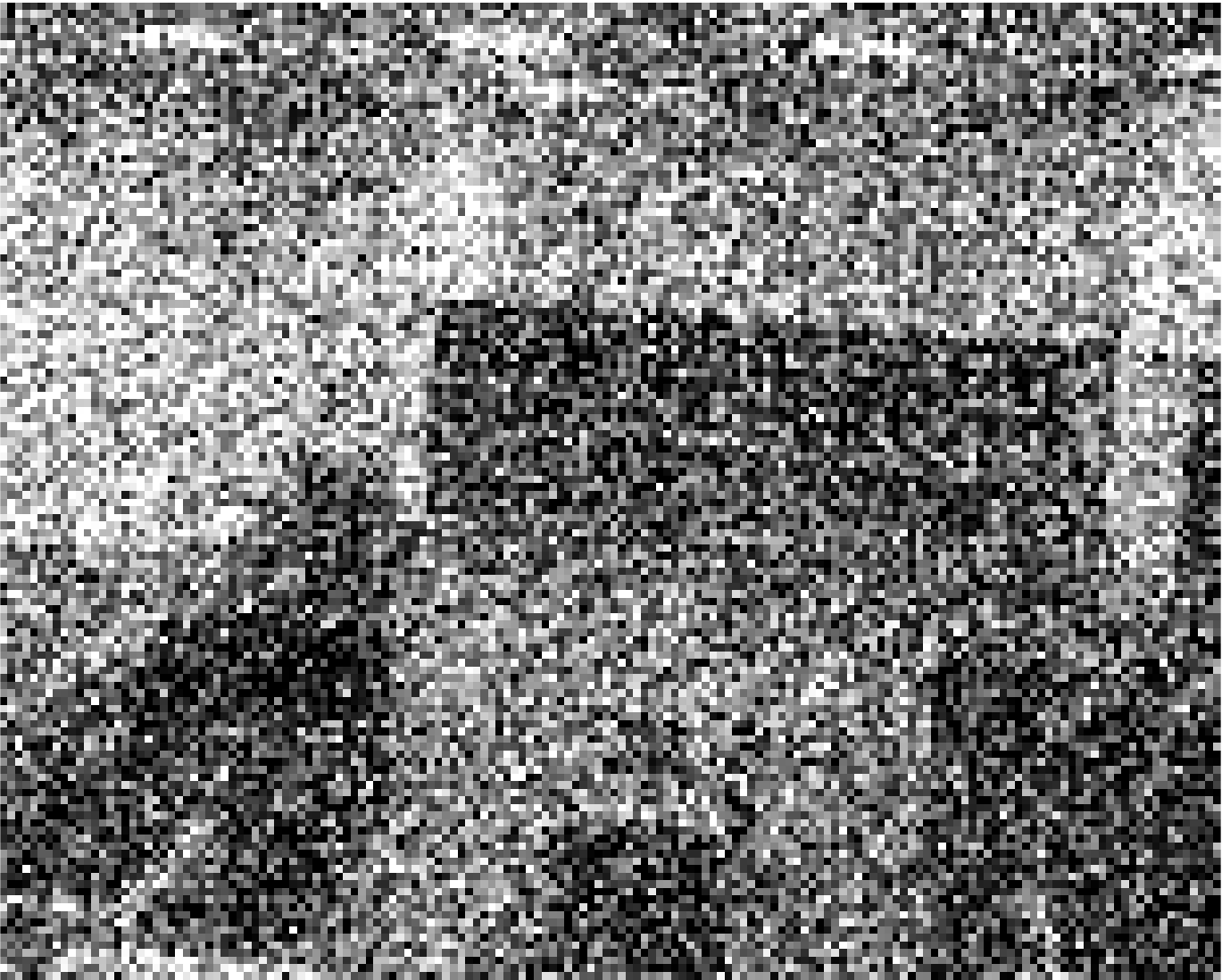}	
					\includegraphics[height=15mm]{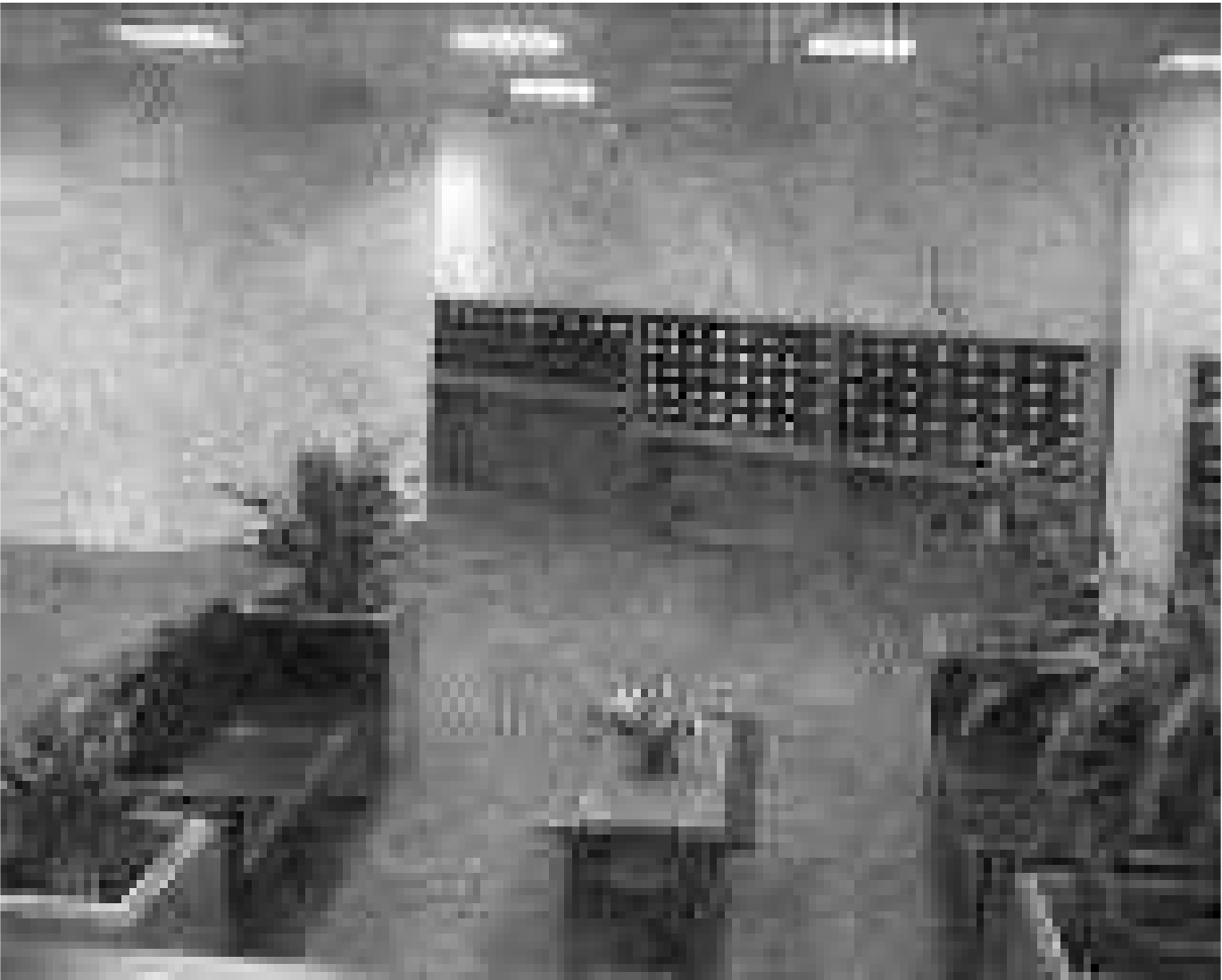}
					\includegraphics[height=15mm]{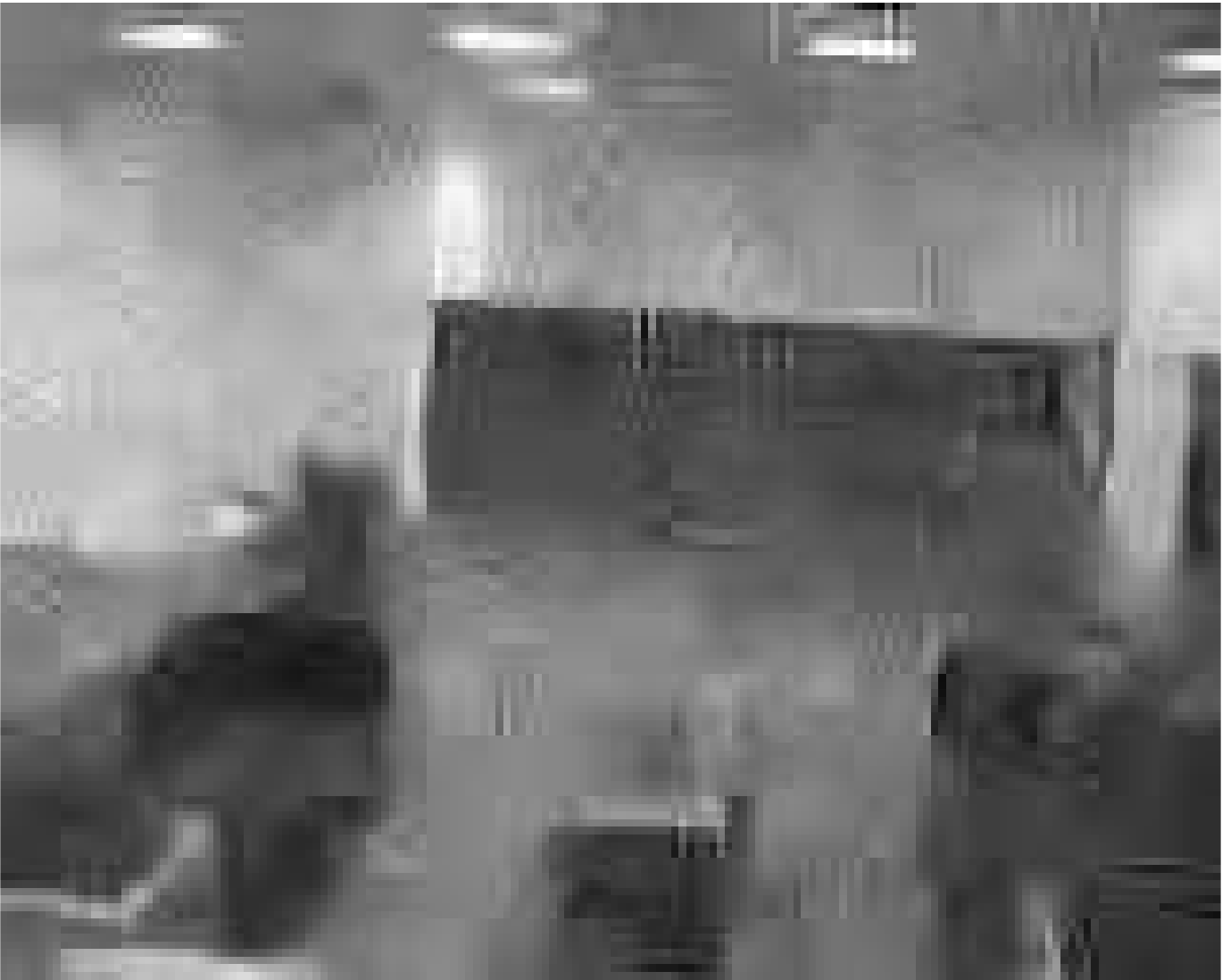}\\
					\includegraphics[height=15mm]{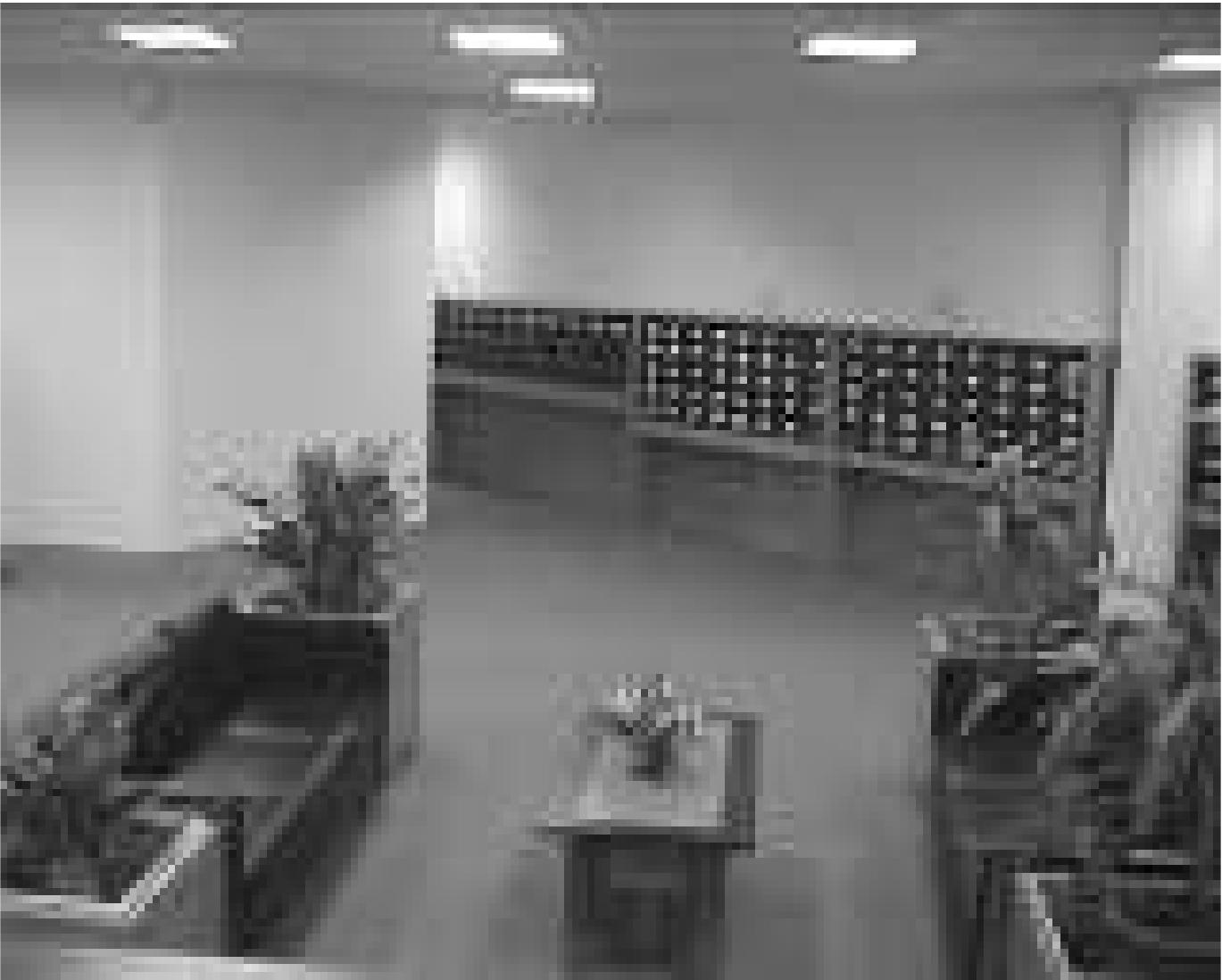}
					\includegraphics[height=15mm]{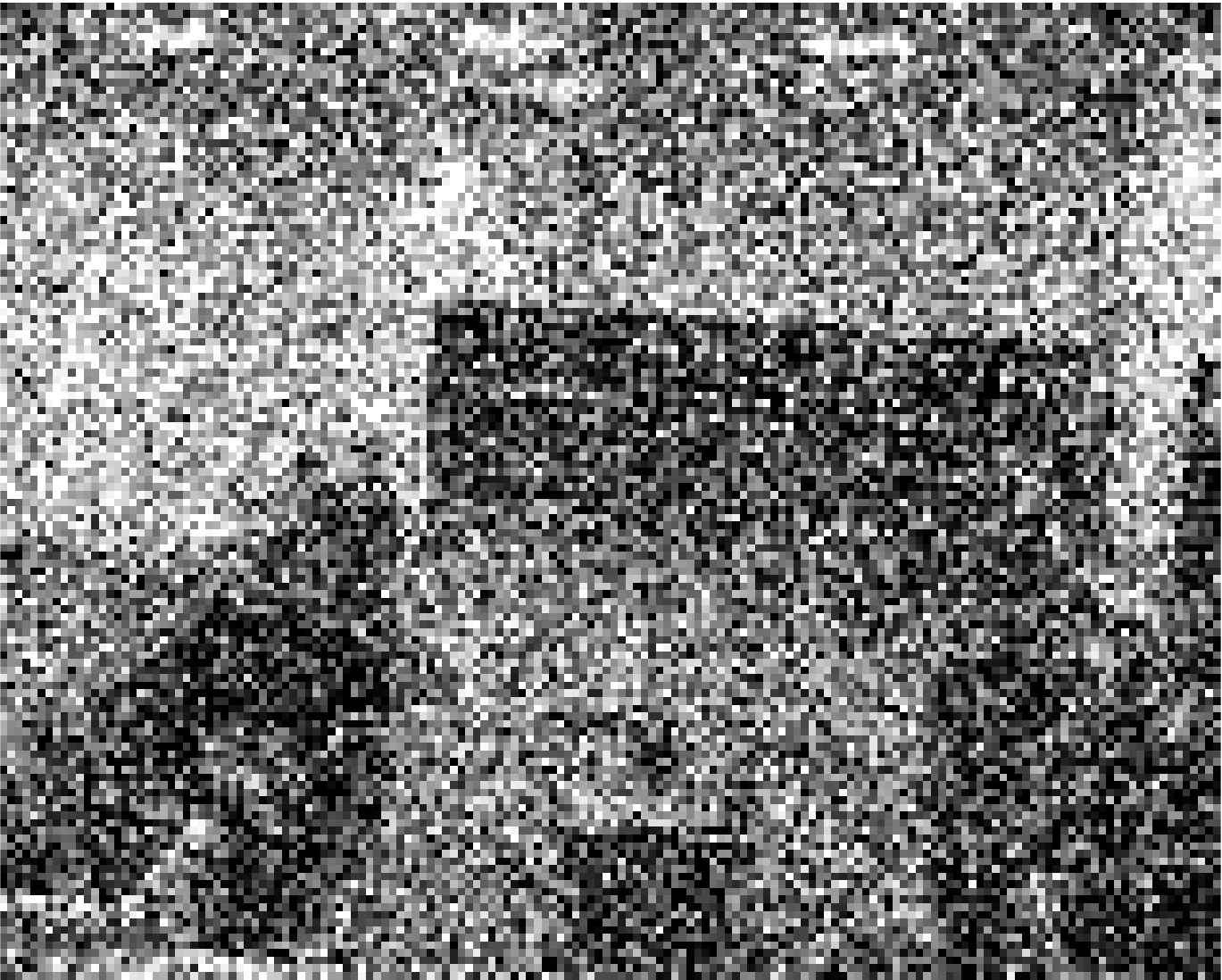}	
					\includegraphics[height=15mm]{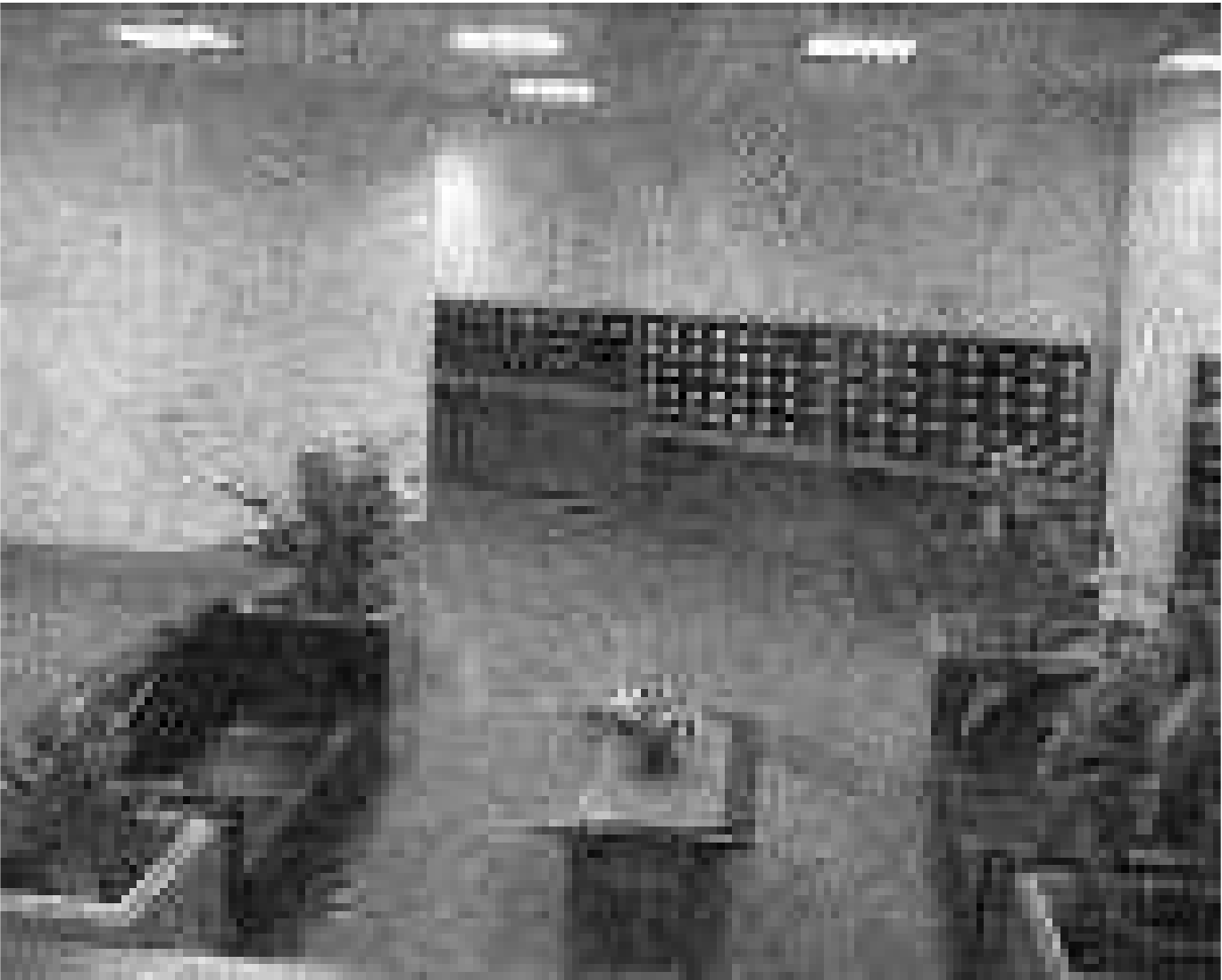}
					\includegraphics[height=15mm]{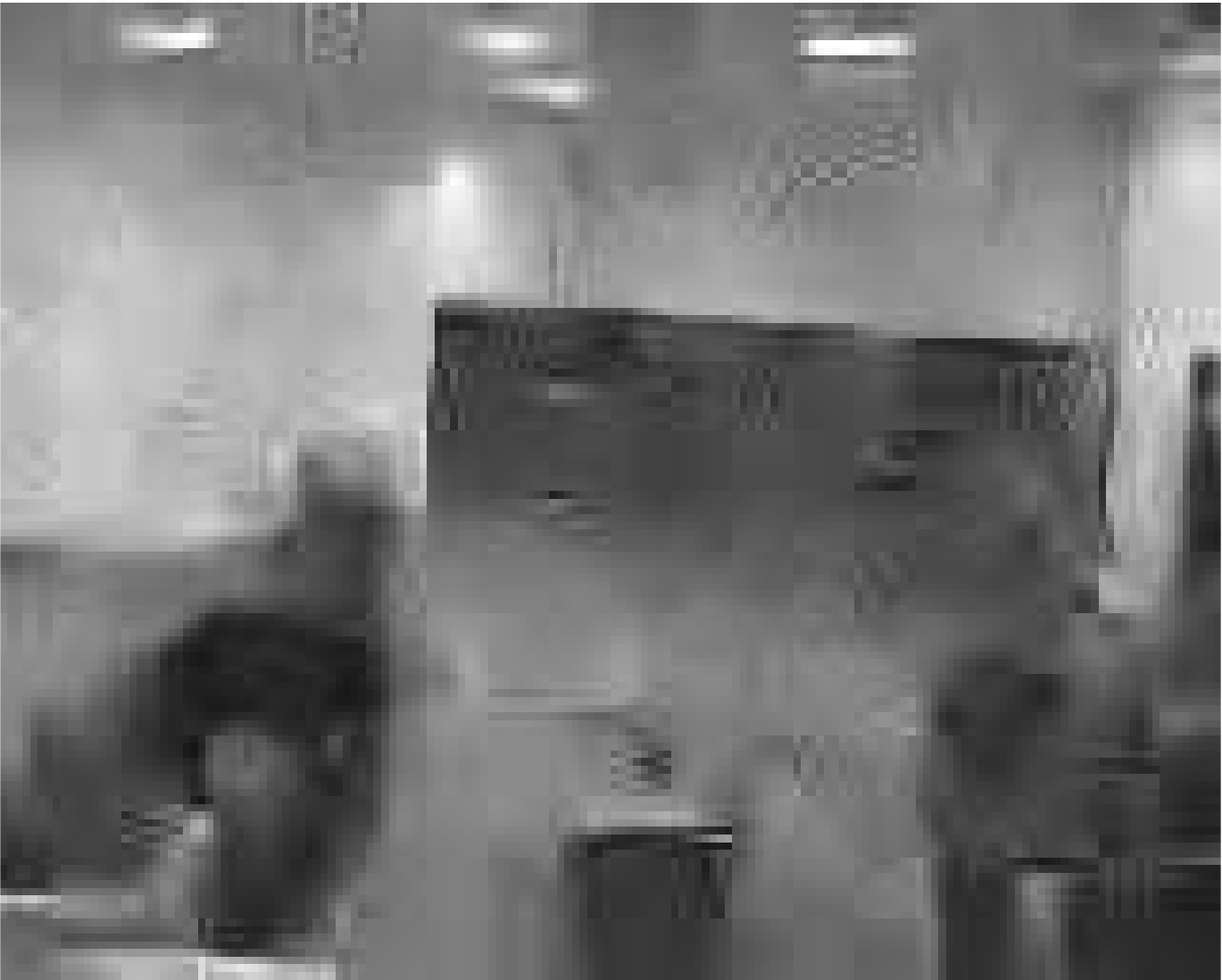}\\
					\includegraphics[height=15mm]{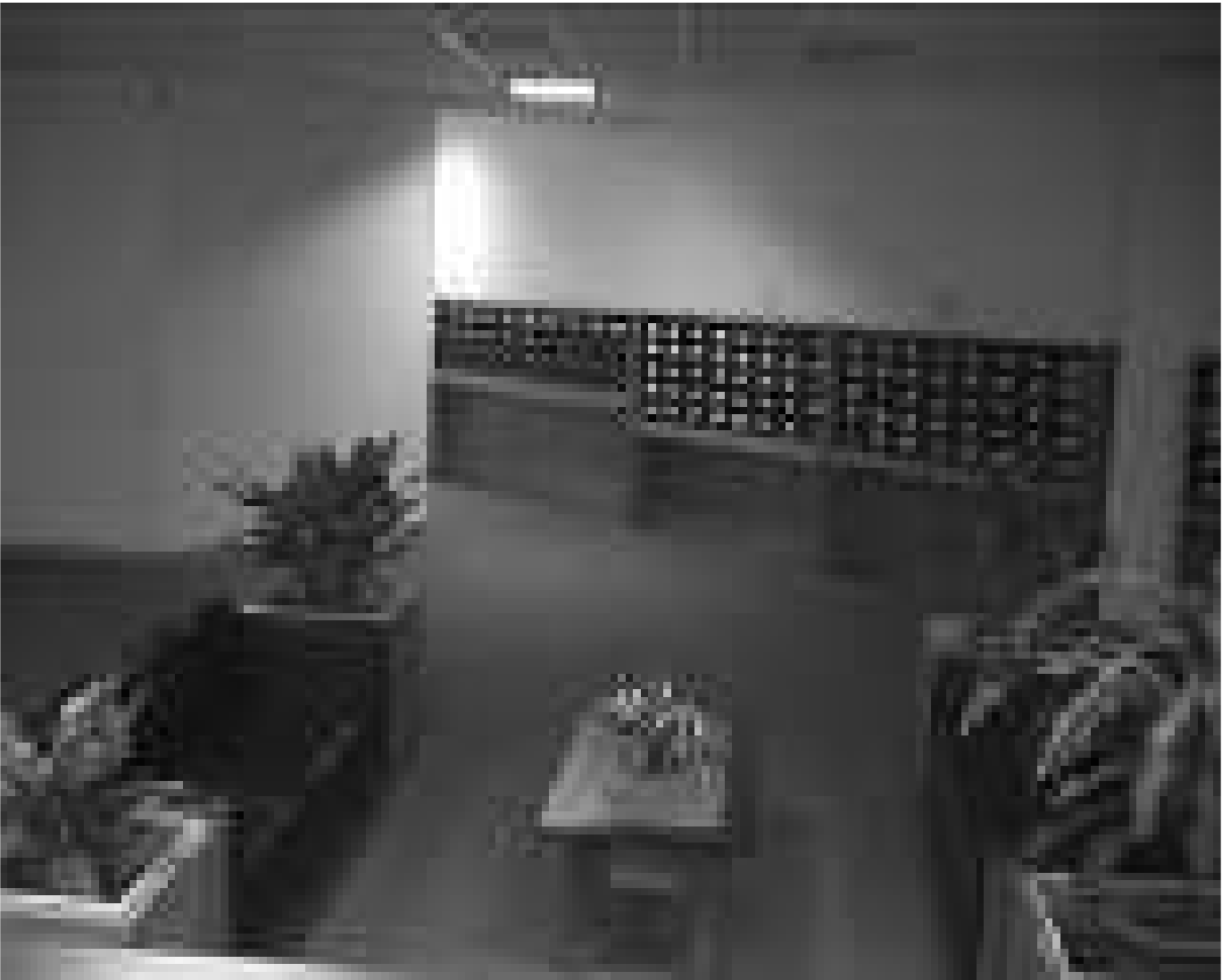}
					\includegraphics[height=15mm]{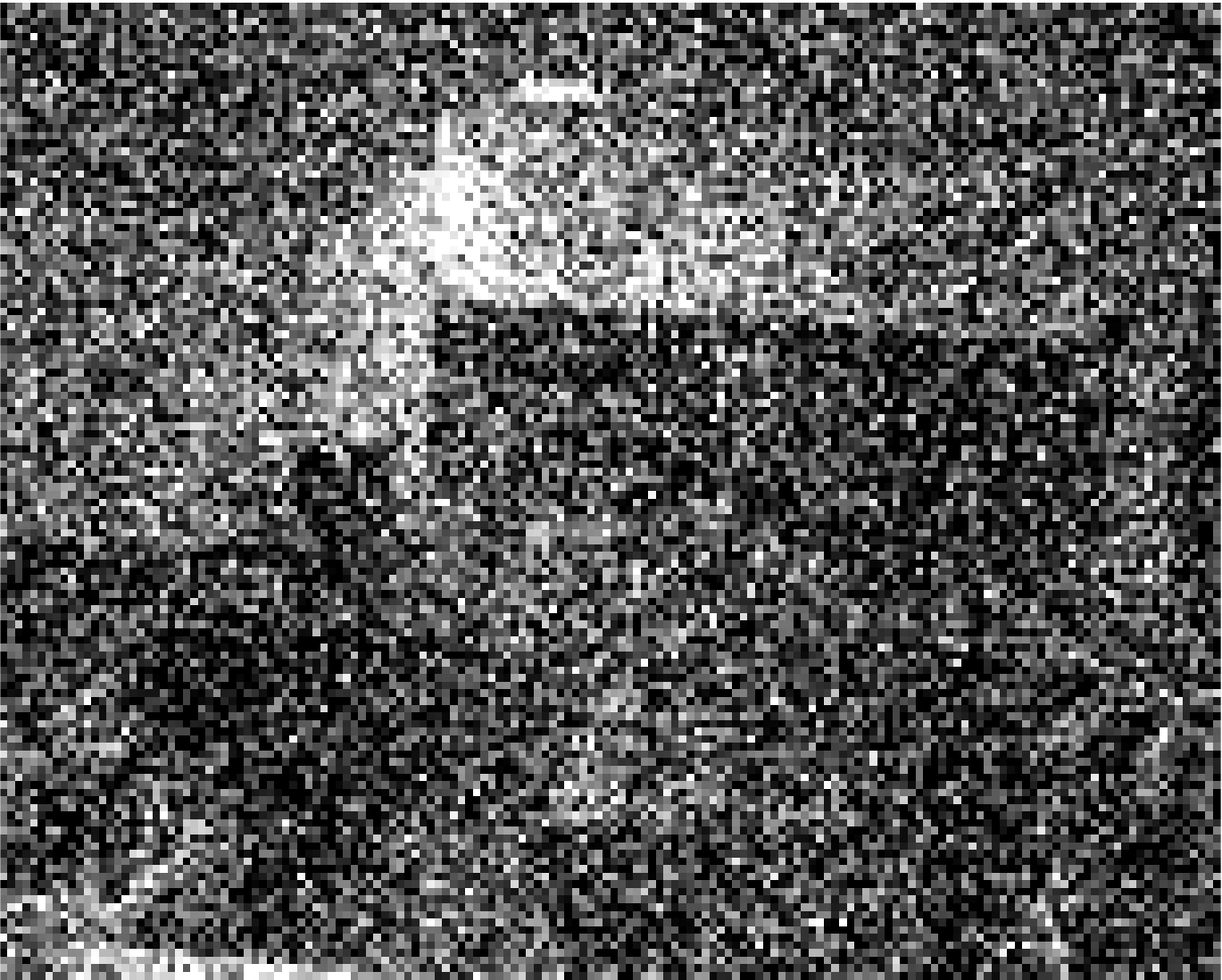}	
					\includegraphics[height=15mm]{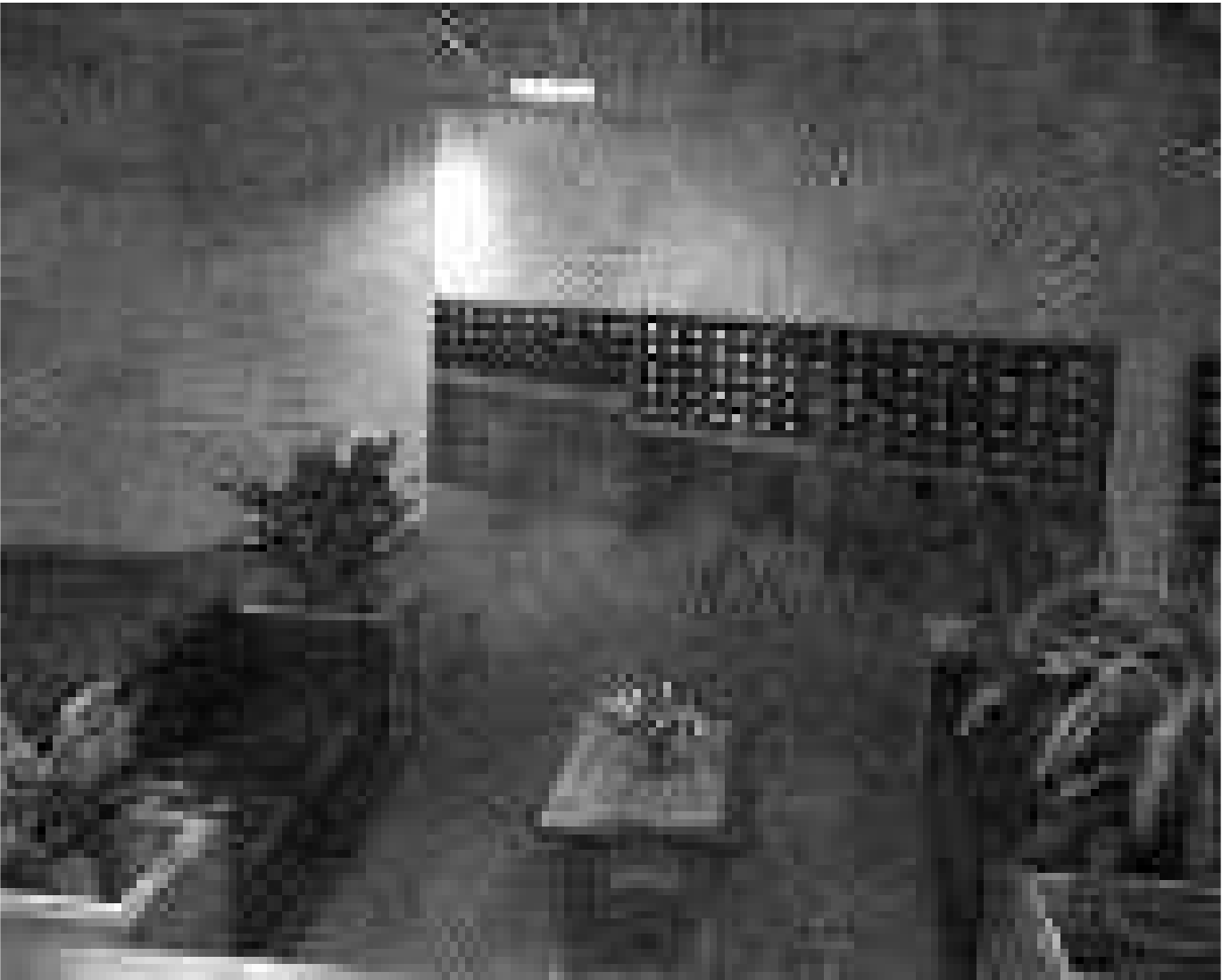}
					\includegraphics[height=15mm]{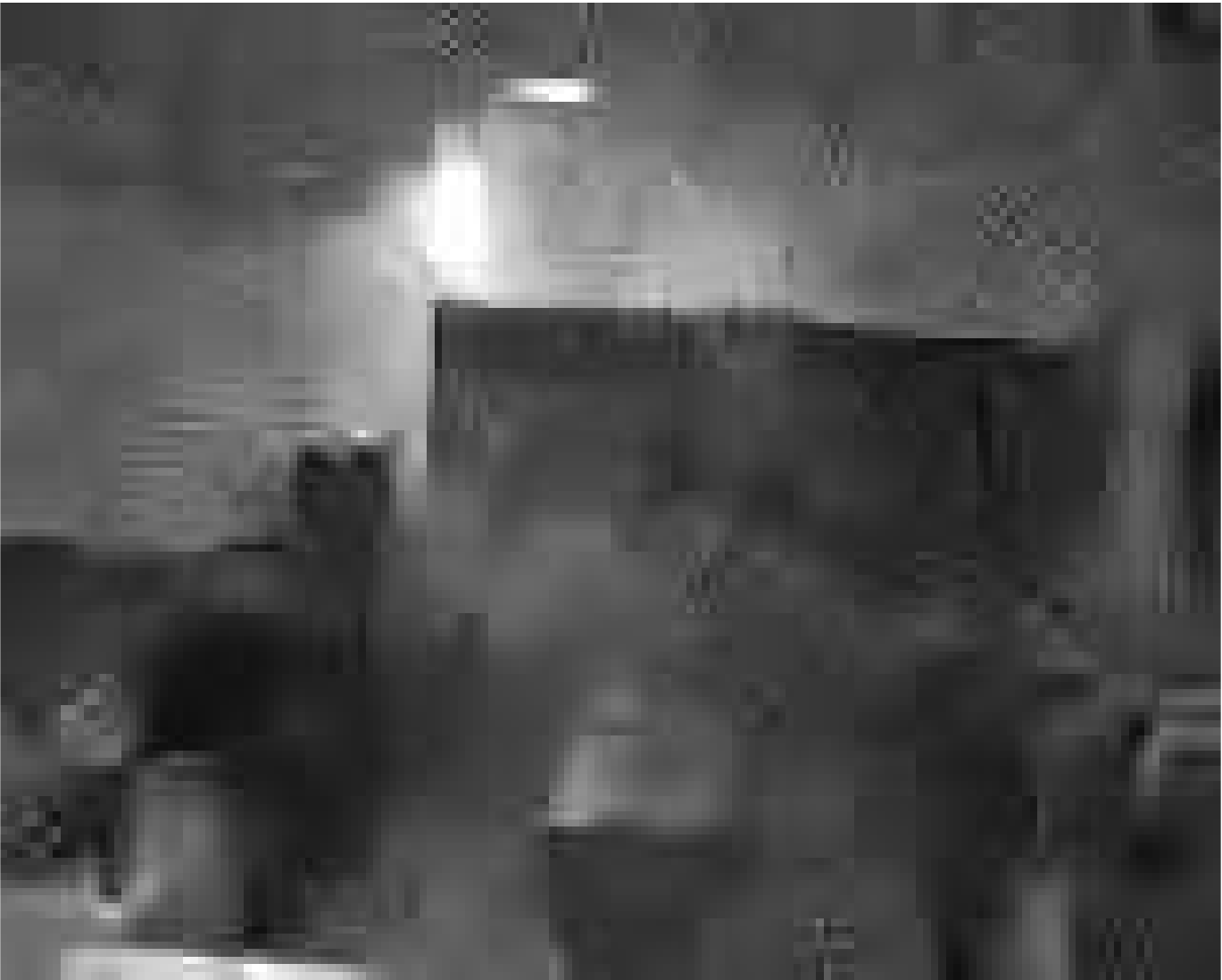}\\
					\hspace*{\fill}\makebox[0pt]{\footnotesize original}\hspace*{\fill}
        			\hspace*{\fill}\makebox[0pt]{\footnotesize noisy}\hspace*{\fill}
        			\hspace*{\fill}\makebox[0pt]{\footnotesize RPCA-VBM3D}\hspace*{\fill}
        			\hspace*{\fill}\makebox[0pt]{\footnotesize VBM3D}\hspace*{\fill}
        \\
					\hspace*{\fill}\makebox[0pt]{ \ }\hspace*{\fill}
        			\hspace*{\fill}\makebox[0pt]{\ }\hspace*{\fill} 
        			\hspace*{\fill}\makebox[0pt]{\footnotesize (PSNR=$30$dB)}\hspace*{\fill}
        			\hspace*{\fill}\makebox[0pt]{\footnotesize (PSNR=$25$dB)}\hspace*{\fill}
						\end{tabular}
					\caption{\small{Denoising a very noisy video. PSNR shown in parenthesis}}\label{VisualLobby}
				\end{subfigure}
\begin{subfigure}{0.49\textwidth}
\centering
\begin{tabular}{cc}
		\includegraphics[height=15mm]{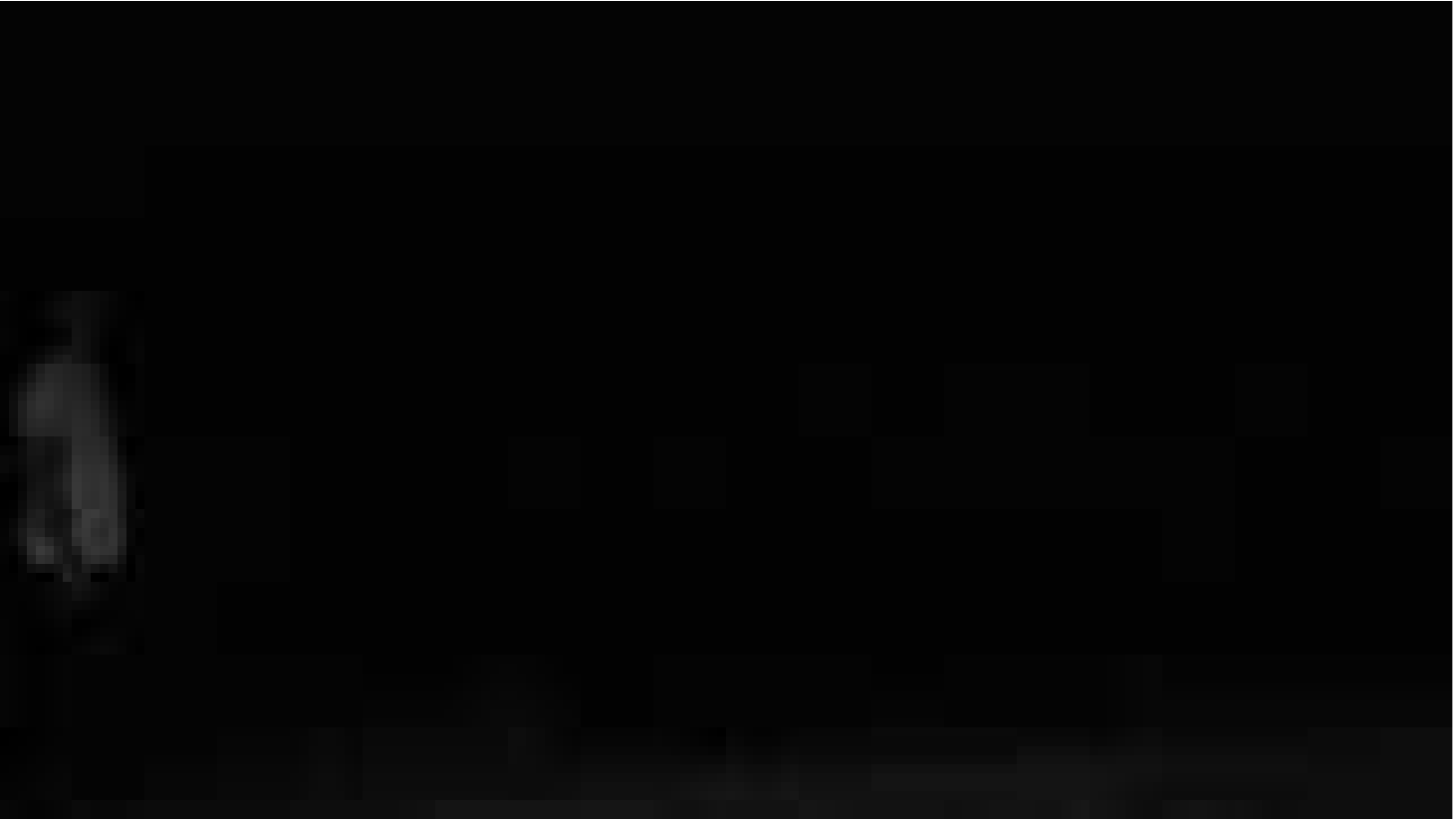}
		\includegraphics[height=15mm]{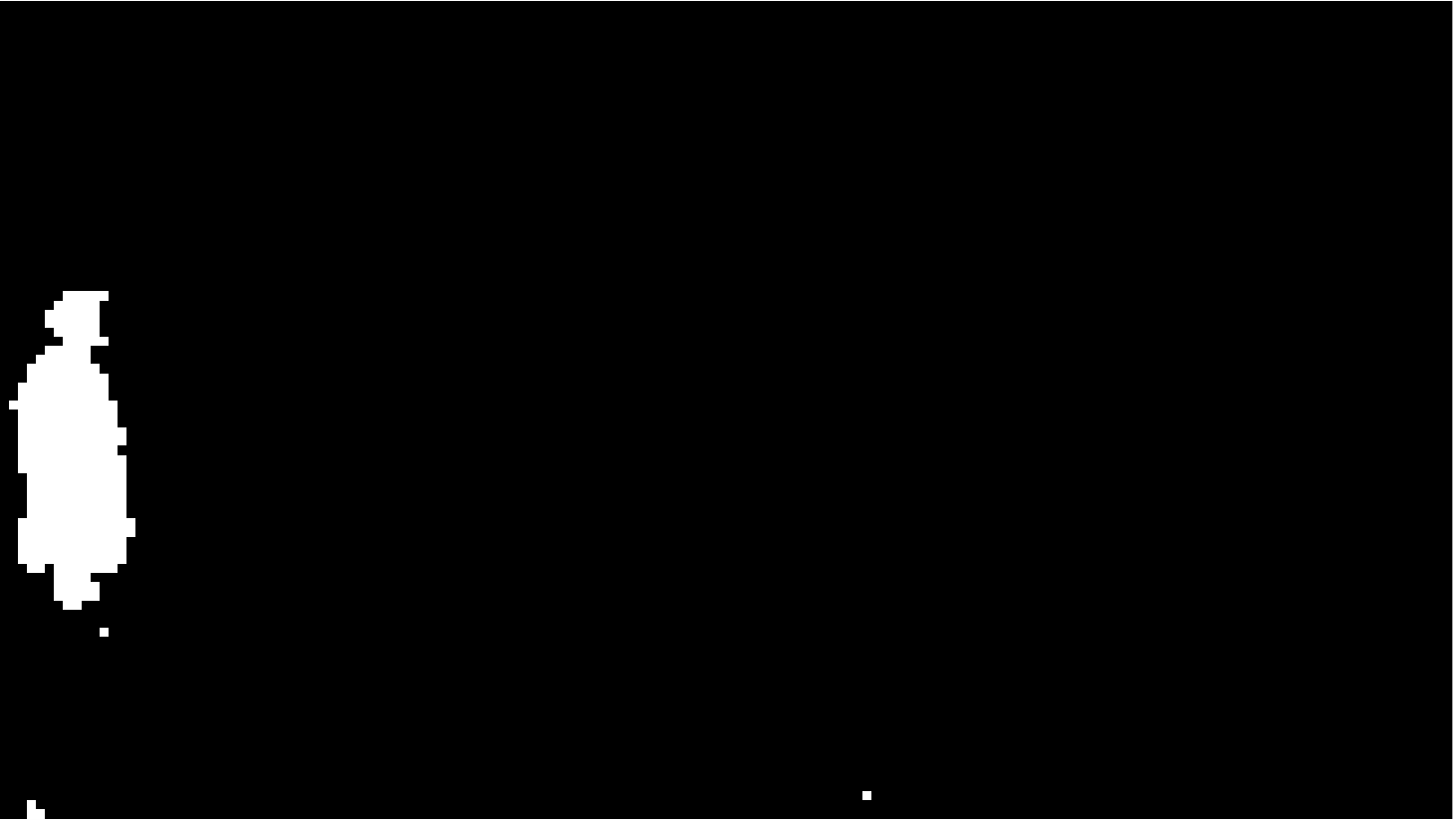}
		\includegraphics[height=15mm]{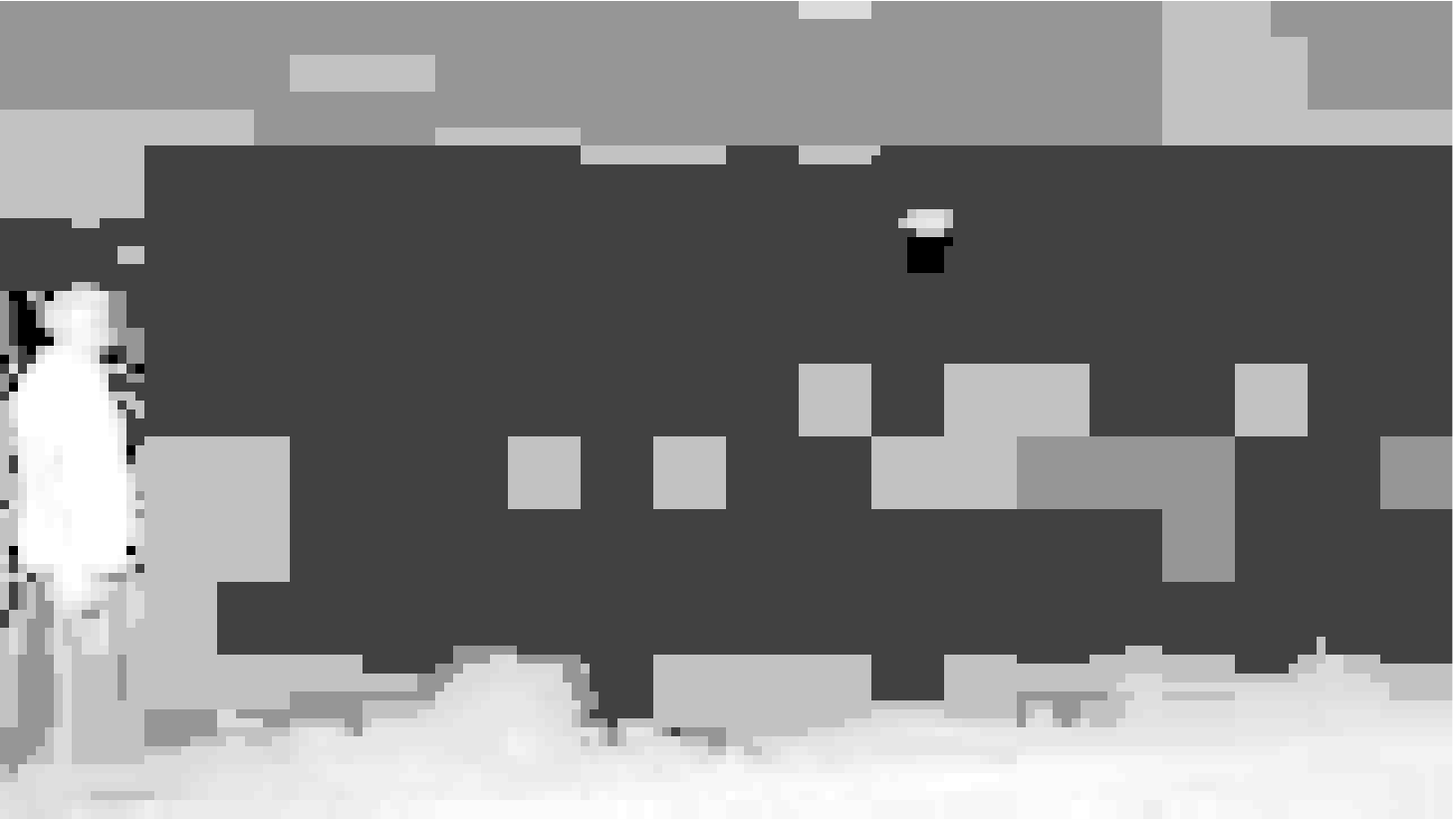}\\

		\includegraphics[height=15mm]{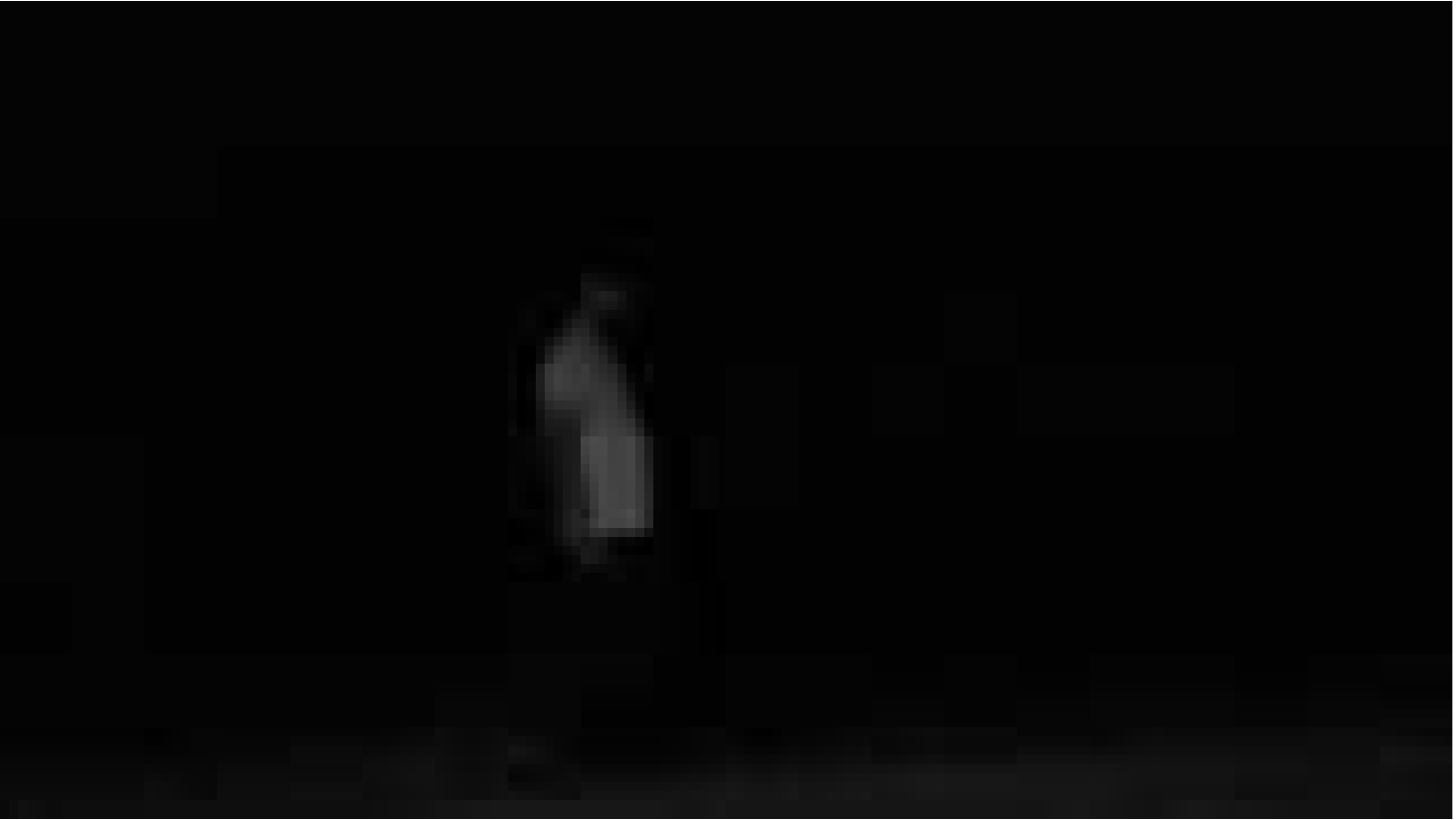}
		\includegraphics[height=15mm]{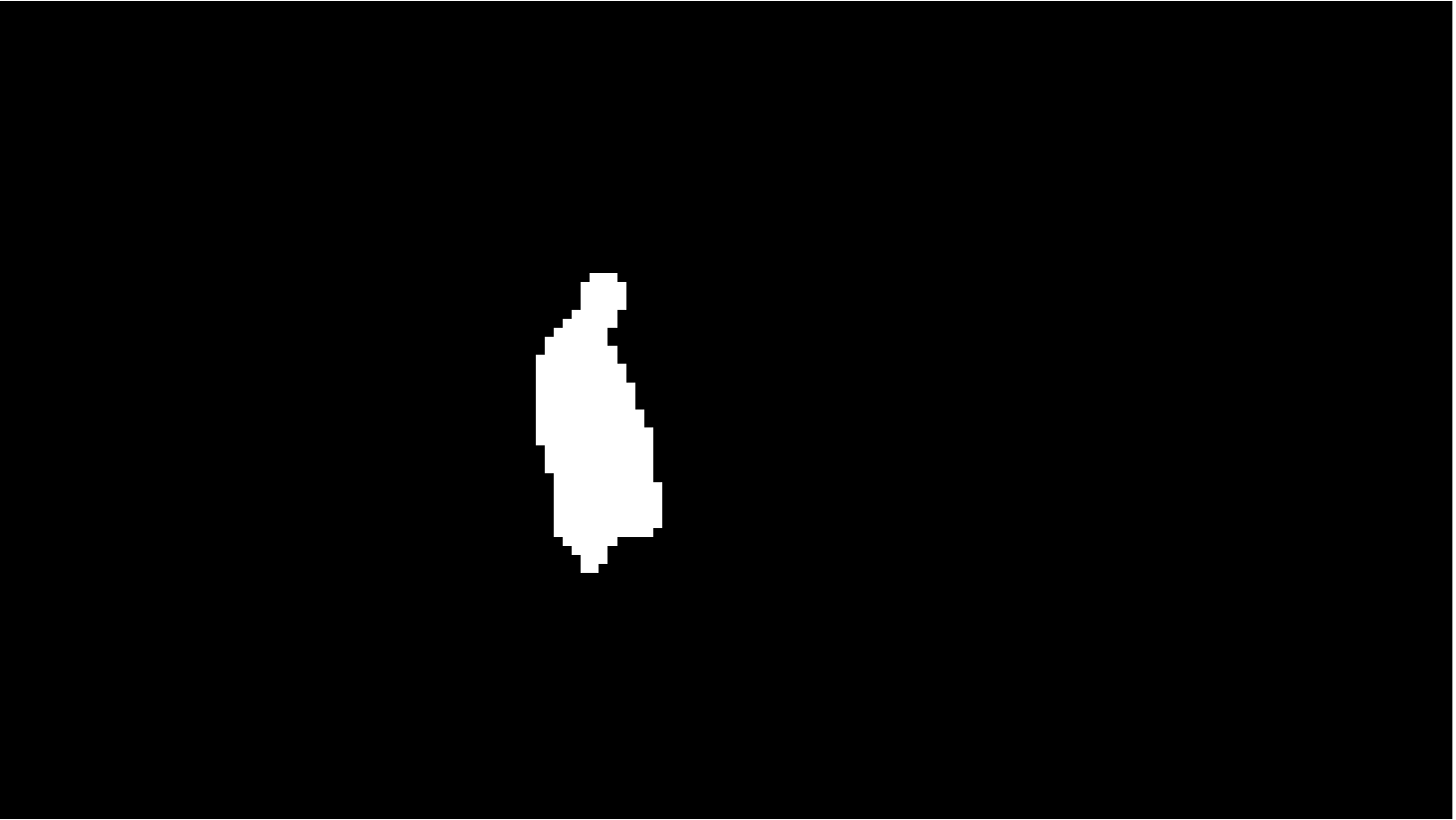}
		\includegraphics[height=15mm]{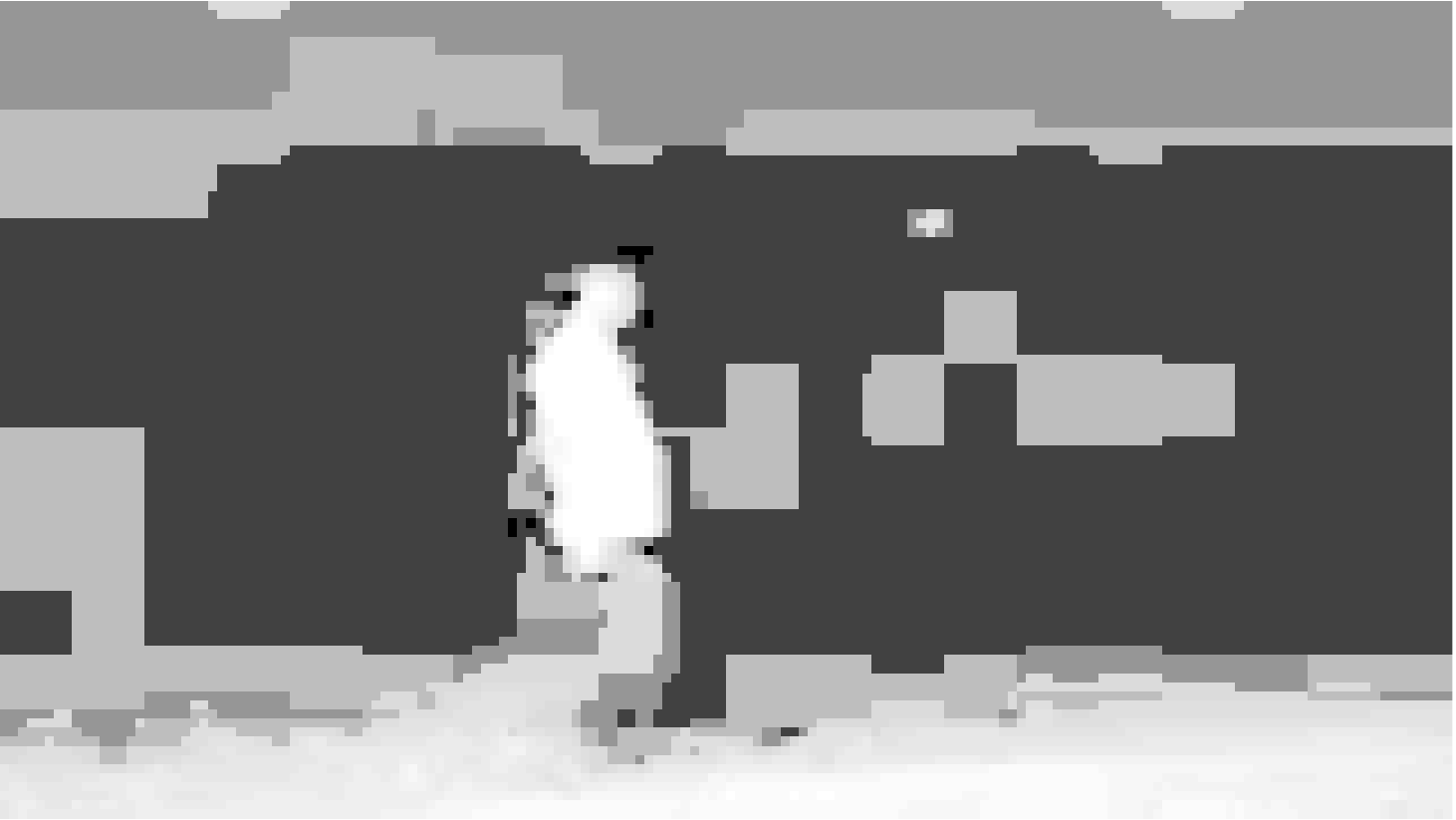}\\
		
		\includegraphics[height=15mm]{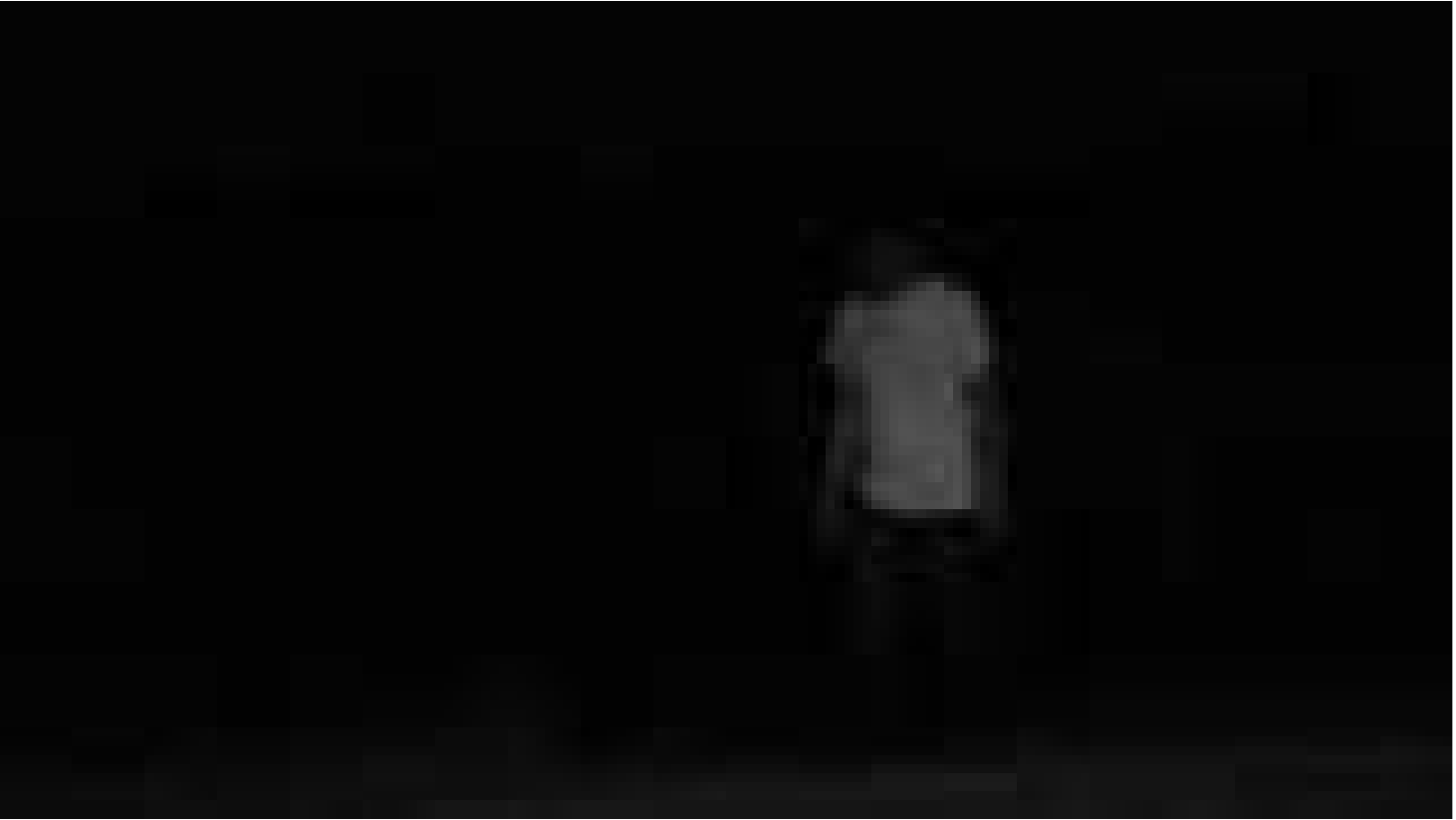}
		\includegraphics[height=15mm]{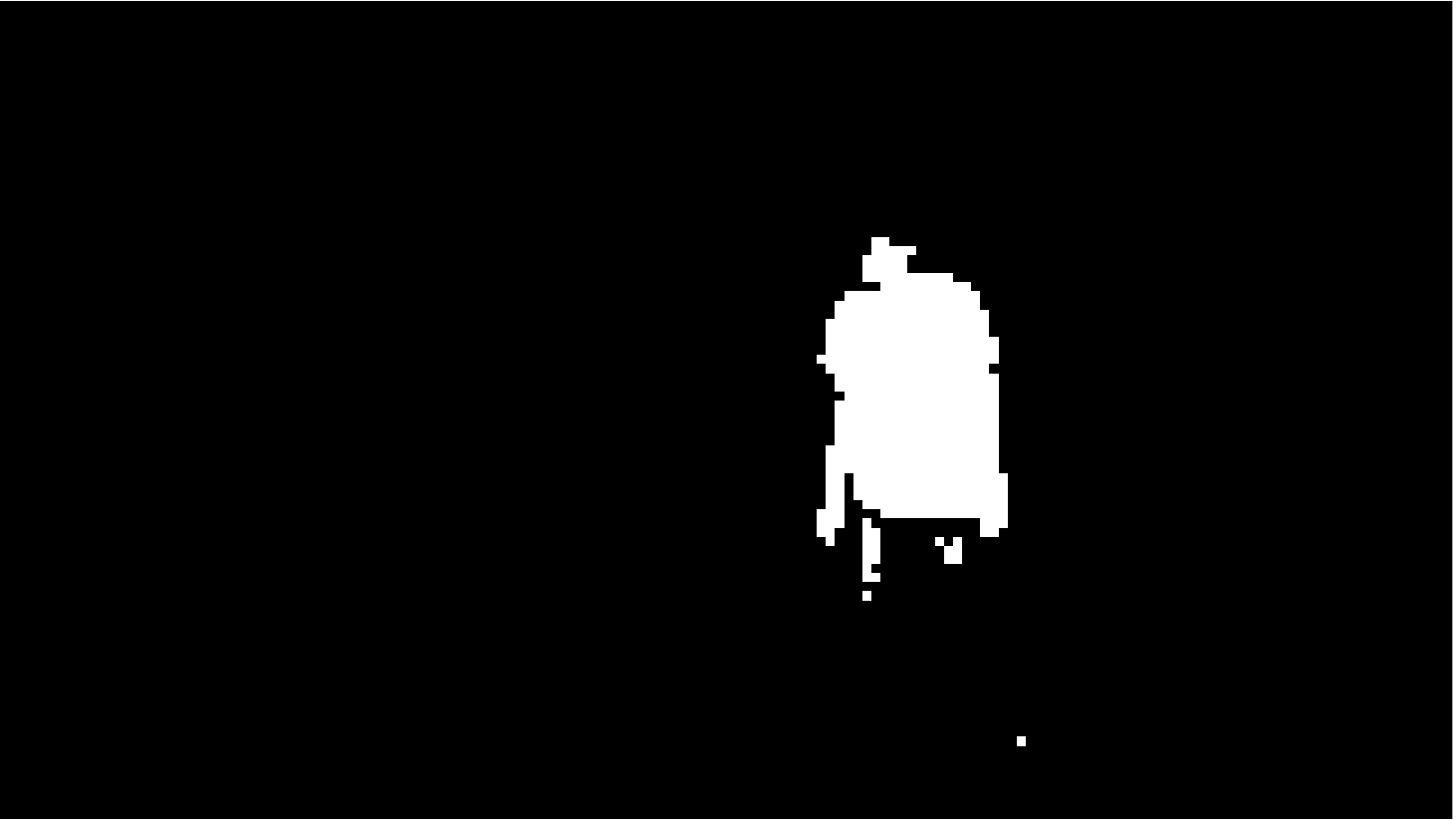}
		\includegraphics[height=15mm]{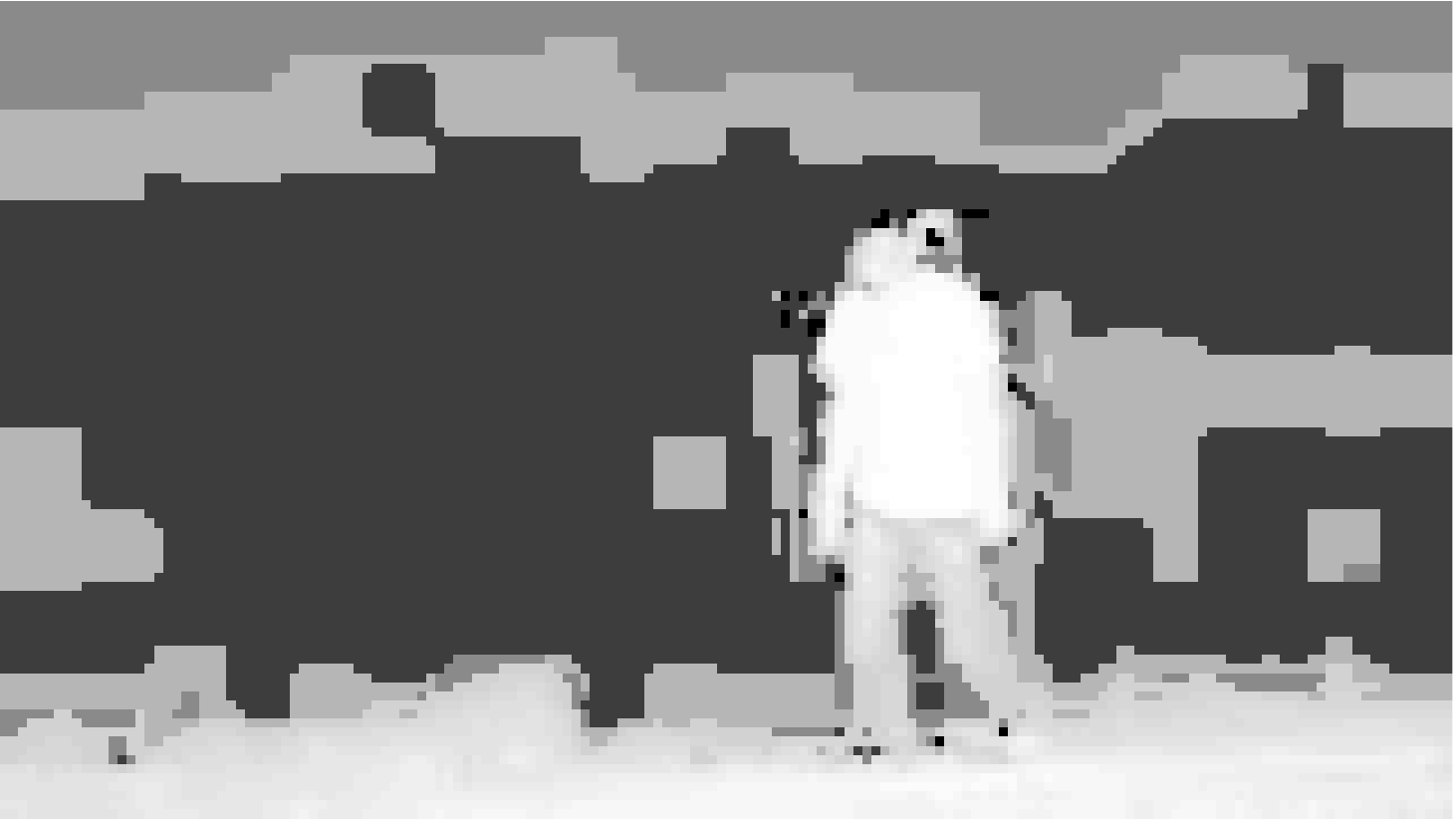}\\
	    \hspace*{\fill}\makebox[0pt]{original }\hspace*{\fill}
		\hspace*{\fill}\makebox[0pt]{RPCA}\hspace*{\fill}
		\hspace*{\fill}\makebox[0pt]{Hist-Eq}\hspace*{\fill}
   \\
					\hspace*{\fill}\makebox[0pt]{\footnotesize  \ }\hspace*{\fill}
        			\hspace*{\fill}\makebox[0pt]{\footnotesize  \ }\hspace*{\fill}
        			\hspace*{\fill}\makebox[0pt]{\footnotesize  \ }\hspace*{\fill}
	\end{tabular}
	\caption{Video enhancement: ``seeing" in the dark}\label{VisualLobby}
\end{subfigure}
\caption{\small{RPCA to enable video enhancement and denoising \cite{rrpcp_denoise,rrpcp_tsp}:
(a) Denoising a very noisy video (video with Gaussian noise of standard deviation $\sigma=70$ and hence PSNR 11dB): we show original and noisy videos in first two columns. The noisy video had PSNR 11dB. RPCA-VM3D uses an RPCA algorithm (ReProCS) to first separate it into sparse and low-rank layers, and then applies the best denoiser from literature, VBM3D, on each layer. Last column directly applies VBM3D to the video. Notice that this gives a much more blurred denoised image. Peak Signal to Noise Ratio (PSNR) is noted below each figure too.
(b) Video enhancement via RPCA: an application where an RPCA algorithm (ReProCS) helps to ``see" in the dark. A standard approach for such low light data is to do contrast enhancement using histogram equalization. As can be seen from the third column (Hist-Eq), this fails.
The code for this experiment is downloadable from \url{http://www.ece.iastate.edu/~hanguo/ReLD_Denoising.zip}.
}}
\label{fig_denoise}				
\end{figure*}

Being able to correctly  solve the video layering problem (separate a video into foreground and background layers) enables better solutions to many other video analytics and computer vision applications. For example, the foreground layer directly provides a video surveillance or object tracking solution, while the background layer and its subspace estimate are useful in video editing or animation applications. Also, an easy low bandwidth video conferencing solution would be to transmit only the layer of interest (usually the foreground). As another example, automatic video retrieval to look for videos of moving waters or other natural scenes will become significantly easier if the retrieval algorithm is applied to only the background layer where as if the goal is to find a certain videos of a certain dog or cat breed, the algorithm can be applied to only the foreground layer. Finally, denoising and enhancement of very noisy videos becomes easier if the denoiser is applied to each layer separately (or to only the layer of interest). We show an example of this in Fig. \ref{fig_denoise}.


{\em Dynamic and functional MRI  \cite{candes_mri}. }
The sparse+low-rank model is a good one for many dynamic MRI sequences. The changing region-of-interest (ROI) forms the sparse outlier for this problem while everything else that is slowly changing forms the low-rank component \cite{candes_mri}. We show a cardiac sequence in Fig. \ref{fig_mri}. The beating heart valves form the ROI in this case. This model can be used to both accurately recover a dynamic MRI sequence from undersampled data (solve the ``compressive" MRI problem)  and to correctly separate out the ROI. Similarly, in fMRI based brain activity imaging, only a sparse brain region is activated in response to given stimulus. This is the changing ROI (sparse outlier) for this problem. There is always some background brain activity in all the brain voxels. This is well modeled as being slowly changing and being influenced by only small number of factors, $\rmat$.

{\em Detecting anomalies in computer and social networks. }
Another application is in detecting anomalous connectivity patterns in social networks or in computer networks \cite{mateos_anomaly,selin_reprocs}. This application is best solved using tensors rather than matrices, however to explain the idea, we will just use the simpler matrix formulation. In this case, $\lt$ is the vector (or tensor) of network link ``strengths'' at time $t$ when no anomalous behavior is present while $\st$ is the vector (or tensor) of outlier entries \cite{selin_reprocs}. Outliers occur due to occasional anomalous behavior or due to node failures on a few edges. Hence they are well modeled as being sparse.

{\em Recommendation system design. }
This  actually requires solving the robust matrix completion problem. To understand the problem, consider a specific example, say the Netflix problem which involves designing a movie recommendation system. Suppose there are $n$ movies and $\tmax$ users. We use $\lt$ to denote the vector of true movie preferences of user $t$. The vector $\st$ would contain the outlier entries (if any). The matrix $\L$ is well modeled as being low rank under the assumption that user preferences are governed by only a few factors, $\rmat$.
The outliers $\st$ occur because some users enter some incorrect ratings due to laziness, malicious intent, or just typographical errors \cite{rpca}. These are clearly sparse. The data matrix $\M:=\L+\X$ is also incomplete since any given user does not rate all movies.
The goal in this case is to complete the true movie preferences' matrix $\L$ while being robust to the outliers. This is then used to recommend movies to the users.

{\em Survey data analysis. }
A related application is in survey data analysis. Consider a survey with $n$ questions given to $\tmax$ participants. If all participants answer questions truthfully and carefully (no malicious or typographical errors), then the the survey answers of the $t$-th participant form $\lt$. This is a valid model under the assumption that the responses to all survey questions are governed by much fewer factors than $n$ or $\tmax$, and hence the resulting matrix formed by the true survey responses is low rank. The goal of the survey is in fact to find these few factors (principal components of this matrix). The problem becomes one of robust PCA because some of the participants' entries will be wrong due to mistakes and typos. These are modeled as the sparse corruptions $\st$. If the entries are wrong due to malicious participants then the column-sparse model is a more appropriate one.


\subsection{Desirable algorithm properties}

In terms of algorithm design, the following are important questions to ask.
\bi
\item {\em Provably correct and under what assumptions? }
It is useful to know if an algorithm is guaranteed to be correct under simple and practically valid assumptions. We say that a guarantee is a {\em complete correctness result} or {\em complete guarantee} if it only makes assumptions on the algorithm input (the input data, parameters, and the initialization, if any) {\em and} proves that the algorithm output is within a small error of the true value of the quantity of interest.
In all other situations, we say that the guarantee is a {\em partial guarantee}. This could mean many things: it could mean that the result makes assumptions on intermediate algorithm estimates; or that the guarantee only proves that the algorithm converges to a local minimum (or just to some stationary point) of a cost function. A partial guarantee by itself is usually not very useful. However sometimes the proof technique used for obtaining it may be useful and may help obtain a complete correctness result in later work.%


\item {\em Practical accuracy. } Performance guarantees are often just sufficient conditions. Also the assumptions required can be hard to verify in practice. Hence practical accuracy is an equally important measure of algorithm success. Commonly used metrics for RPCA include normalized recovery error of $\L$ or the error in recovering its column span. If the outlier matrix $\X$ is the ``quantity'' of interest, e.g., in case of foreground object tracking, then the error in recovering its support is often a more useful metric.%

\item {\em Time complexity. } Both order-wise computational complexity and actual experimental comparisons of time taken are important. The former is useful because it can take too long to do experimental comparisons for very large sized problems. The latter is useful because order-wise complexity ignores constants that can sometimes be very large in practice. We discuss both types of time comparisons in this article.

\item {\em Memory complexity and/or number of passes. } In today's big data age, this is often the most important concern. It can be the bottleneck even for applications for which real time output is not needed and hence slow algorithms can be tolerated. 
    We say that an algorithm is {\em memory-optimal} if it needs the same order of memory as the algorithm output. So for robust PCA, an algorithm would be memory optimal if it needed memory of order $O(nr)$. It is nearly memory-optimal if its memory complexity is within logarithmic factors of the optimal.
 If the dataset needs to be loaded from a storage device while running the algorithm, another related concern is the number of passes required. This number then governs the number of times the data needs to be re-loaded. For processing very large datasets, one needs algorithms that are either nearly memory-optimal or need few passes through the data, ideally both.

\item {\em Online (or recursive or causal) versus Batch. } Online algorithms are algorithms for settings where inputs or data are arriving one at a time, and we need to make decisions on the fly or with small delays, without knowing what will happen in the future. In signal processing or controls literature, such an algorithm would be called ``recursive''. The notion of online algorithm also requires that the algorithm output quality improves as more data comes in. For robust PCA, this means that the the subspace estimate quality improves over time. A batch method, on the other hand, needs to wait for all data to come in first.

\ei

\section{An Overview}
This section provides a short overview of the problem setting and key solutions for all the three problems discussed in this article along with three sample theoretical guarantees. It is a quick summary for the reader who is already somewhat familiar with the topic. In the two later sections, we explain the solution approaches in detail. Before we begin, we should mention that the code for all the robust PCA and robust subspace tracking solutions is downloadable from the github library of Andrew Sobral \cite{lrslibrary2015}. The link is \url{https://github.com/andrewssobral/lrslibrary}.

In our guarantees and elsewhere, we use the letters $C,c$ to denote different numerical constants in each use. To keep our statements simple, condition numbers are treated as numerical constants (and combined with $C$ or $c$).
%

\subsection{Robust PCA and Robust Subspace Tracking via S+LR: problem setting}
\subsubsection{Static RPCA \cite{rpca}}
Given an $n \times \tmax$ observed data matrix, 
\[
\M := \L + \S + \V,
\]
where $\L$ is a low rank matrix (true data), $\S$ is a sparse matrix (outlier matrix), and $\V$ is small unstructured noise/modeling error, the goal is to estimate $\L$, and hence its column space. We use $\rmat$ to denote the rank of $\L$. The maximum fraction of nonzeros in any row (column) of the outlier matrix $\S$ is denoted by $\outfracrow$ ($\outfraccol$). 
The column space of $\Lhat$ then gives the PCA solution.


\subsubsection{Robust Subspace Tracking (RST) or dynamic RPCA \cite{rrpcp_perf}}
At each time $t$, we get a data vector $\yt \in \Re^n$ that satisfies%
\bea
\mt := \lt + \s_t + \v_t, \text{ for } t = 1, 2, \dots, \tmax. \nn
\label{orpca_eq}
\eea
where  $\v_t$ is small unstructured noise, $\xt$ is the sparse outlier vector, and $\lt$ is the true data vector that lies in a fixed or slowly changing low-dimensional subspace of $\Re^n$, i.e., $\lt = \P_{(t)} \a_t$ where $\P_{(t)}$ is an $n \times r$ {\em basis matrix}\footnote{matrix with mutually orthonormal columns} with $r \ll n$ and with $\|(\I - \P_{(t-1)}\P_{(t-1)}{}')\P_{(t)}\|$ small compared to $\|\P_{(t)}\|=1$.
We use $\T_t$ to denote the support set of $\xt$.
%
%
Given an initial subspace estimate, $\Phat_{0}$, the goal is to track $\Span(\P_{(t)})$ and $\lt$ either immediately or within a short delay. A by-product is that $\lt$, $\x_t$, and $\T_t$ can also be tracked on-the-fly.
The initial subspace estimate, $\Phat_0$, can be computed by using only a few iterations of any of the static RPCA solutions (described below), e.g., PCP \cite{rpca} or AltProj \cite{robpca_nonconvex}, applied to the first $t_\train$ data samples $\Y_{[1,t_\train]}$. Typically,  $t_\train=Cr$ suffices.
Here and below $[a, b]$ refers to all integers between $a$ and $b$, inclusive, $[a,b): = [a,b-1]$, and $\M_\T$ denotes a sub-matrix of $\M$ formed by its columns indexed by entries in the set $\T$.
Alternatively, in some applications, e.g., video surveillance, it is valid to assume that outlier-free data is available. In these situations, simple SVD can be used too.

Technically, {\em Dynamic RPCA} is the offline version of the above problem. Define matrices $\L,\S,\V,\Y$ with $\L = [\l_1,\l_2, \dots \l_{\tmax}]$ and $\Y,\S,\V$ similarly defined.  The goal is to recover the matrix $\L$ and its column space with $\epsilon$ error.

We use $\SE(\Phat,\P):=\|(\I - \Phat \Phat')\P\|$ to measure the Subspace Error (distance) between the subspaces spanned by the columns of $\P$ and $\Phat$. This measures the sine of the maximum principal angle between the two subspaces. It is symmetric when the two subspaces have equal dimension. Here and elsewhere $\|.\|$ denotes the induced $l_2$ norm and $'$ denotes matrix transpose.

\subsubsection{Identifiability and other assumptions}
The above problem definitions do not ensure identifiability since either of $\L$ or $\S$ can be both low-rank and sparse. For the unfamiliar reader, we explain this point in detail in Sec. \ref{identif}.
One way to ensure that $\L$ is not sparse is by requiring that its left and right singular vectors are dense or ``incoherent'' w.r.t. a sparse vector \cite{rpca}: 
\begin{definition}[$\mu$-Incoherence/Denseness]
We say that an $n \times r$ basis matrix  (matrix with mutually orthonormal columns) $\P$ is $\mu$-incoherent if
\bea
\max_{i=1,2,\dots, n} \|\P^{(i)}\|^2 \le {\mu \frac{\rmat}{n}} \nn
\eea
where $\P^{(i)}$ denotes the $i$-th row of $\P$ and $\mu \ge 1$ is the (in)coherence parameter. It quantifies the non-denseness  of $\P$.%
\end{definition}

One way to ensure that $\S$ is not low-rank is by imposing upper bounds on $\outfracrow$ and $\outfraccol$.

Consider the Robust Subspace Tracking (RST) problem. If the $r$-dimensional subspaces change at each time, there are too many unknown parameters making the subspace tracking problem unidentifiable. 
One way to ensure identifiability is to assume that they are piecewise constant \cite{rrpcp_perf,rrpcp_medrop}, i.e., that
\[
\P_{(t)} = \P_{j} \text{ for all } t \in [t_j, t_{j+1}),  \  j=1,2,\dots, J,
\]
with $t_0=1$ and $t_{J+1}=\tmax$, and to lower bound $t_{j+1}-t_j$ by a number more than $r$. 
With this model,  $\rmat = r J$ in general (except if subspace directions are repeated more than once).%

Left and right incoherence of $\L$ and outlier fraction bounds on $\S$ are again needed.
The union of the column spans of all the $\P_j$'s is equal to the span of the left singular vectors of $\L$. Thus, left incoherence is equivalent to assuming that the $\P_j$'s are $\mu$-incoherent. One can replace right incoherence by an independent identically distributed assumption on the $\at$'s, along with bounded-ness of each element of $\at$. As explained in \cite{rrpcp_medrop}, the two assumptions are similar. The latter is more suited for RST since it involves online estimation. 
Since RST algorithms are online and often operate on mini-batches of data, we also need to re-define $\outfracrow$ as the maximum nonzero fraction per row in any $\alpha$-consecutive-column sub-matrix of $\S$. We refer to it as $\outfracrow^\alpha$. Here $\alpha$ is the mini-batch size used by the RST algorithm.


Lastly, as we will see, if one also assumes a lower bound on outlier magnitudes (uses the model that outliers are large magnitude corruptions), it can help significantly relax the required upper bound on $\outfracrow^\alpha$, while also allowing for a fast, online, and memory-efficient algorithm.

\subsection{Static Robust PCA (RPCA): a summary}
The first provably correct and polynomial complexity solution to RPCA was introduced in \cite{rpca}.  It involves solving a convex program which they called Principal Component Pursuit (PCP). 
The same program was introduced and studied in parallel \cite{rpca2} and later work \cite{rpca_zhang} as an S+LR solution.
While polynomial complexity is better than exponential, it is still too slow for today's big datasets. Moreover, the number of iterations needed for a convex program solver to get to within an $\epsilon$ ball of the true solution of the convex program is $O(1/\epsilon)$ and thus the typical complexity for a PCP solver is $O(n d^2/\epsilon)$ \cite{robpca_nonconvex}.
To address this issue, a provably correct alternating minimization (alt-min) solution called AltProj was introduced in \cite{robpca_nonconvex}. This has much lower time complexity and still works under the same assumptions as PCP. The following holds \cite{robpca_nonconvex}.
\begin{theorem}[AltProj guarantee for RPCA]
Let $\L \svdeq \LU \bm\Sigma \LV'$. If
(1) $\LU$, $\LV$ are $\mu$-incoherent,
(2)  $\max(\outfracrow,\outfraccol) \le \frac{c}{ \mu \rmat }$,
(3) $\|\V\|_F^2 \le C \epsilon^2$, and
(4) algorithm parameters are appropriately set,
then AltProj returns  $\Lhat,\Shat$ with $\|\Lhat-\L\|_F \le \epsilon$, and $\|\Shat-\S\|_{\max} \le \epsilon/\sqrt{md}$ in time $O( n d \rmat^2 \log(1/\epsilon))$.
\end{theorem} 

An even faster solution that relied on projected gradient descent (GD), called RPCA-GD, was proposed in \cite{rpca_gd}. If condition numbers are treated as constants, its complexity is only $O(n d \rmat \log(1/\epsilon))$. This is comparable to that of vanilla $r$-SVD for simple PCA; however this approach needs outlier fraction bounds to be $\sqrt{\rmat}$ times tighter. Another algorithm that relies on the recursive projected compressive sensing (ReProCS) approach and has the same time complexity as RPCA-GD is ReProCS-NORST \cite{rrpcp_merop,rrpcp_medrop}. This is also online (after initialization), has much lower, and nearly-optimal, memory complexity of $O(n \rmat \log n \log(1/\epsilon))$), {\em and} has outlier tolerance  that is better than even AltProj. But it needs two extra assumptions: lower bound on most outlier magnitudes, and fixed or slow subspace change.
For a similar setting, a recent Robust Matrix Completion (RMC) solution, called NO-RMC, is also applicable \cite{rmc_gd}. It is the fastest RPCA solution with a time complexity of just $O(n \rmat^2 \log^2 n \log^2(1/\epsilon))$, but it needs $\tmax \approx n$. As we explain later, this is a strong assumption for video data. It achieves this speed-up by deliberately under-sampling $\M$ and solving RMC.
%
We compare the theoretical guarantees for these solutions in Table \ref{compare_assu} and experimental performance (for video analytics) in Table \ref{table1}. As can be seen, among the provable solutions, ReProCS has the best experimental performance {\em and} is the fastest. AltProj is about three time slower. GD and NO-RMC are, in fact, slower than AltProj.

\subsection{Robust Subspace Tracking (RST): a summary}
In \cite{rrpcp_perf,rrpcp_tsp,rrpcp_aistats,rrpcp_dynrpca,rrpcp_medrop}, a novel online solution framework called Recursive Projected Compressive Sensing (ReProCS) was introduced to solve the RST or the dynamic RPCA problem.
%
%
We give below the best ReProCS guarantee. This is for a ReProCS-based algorithm that we call Nearly Optimal RST via ReProCS (ReProCS-NORST) because its tracking delay is nearly optimal \cite{rrpcp_merop,rrpcp_medrop}. At each time, it outputs $\Phat_{(t)}$, $\lhat_t$, $\shat_t$, $\That_t$. It also outputs $\that_j$ as the time at which it detects the $j$-th subspace change.

\begin{theorem}[ReProCS-NORST guarantee for RST]
Let $\alpha := C  r \log n$,
$\Lam:= \E[\a_1 \a_1{}']$, $\lambda^+:=\lambda_{\max}(\Lam)$, $\lambda^-:=\lambda_{\min}(\Lam)$, 
and let $\xmint:=\min_t \min_{i \in \T_t} (\xt)_i$ denote the minimum outlier magnitude.
Pick an $\zz \leq \min(0.01,0.4 \min_j \SE(\P_{j-1}, \P_j)^2/f)$.
If
\ben
\item $\P_j$'s are $\mu$-incoherent; and $\at$'s are zero mean, mutually independent over time $t$, have identical covariance matrices, i.e. $\E[\at \at{}'] = \Lam$, are
element-wise uncorrelated ($\Lam$ is diagonal), are element-wise bounded (for a numerical constant $\eta$, $(\at)_i^2 \le \eta \lambda_i(\Lam))$, and are independent of all outlier supports $\T_t$,

\item  $\|\vt\|^2 \le c r \|\E[\vt \vt{}']\|$, $\|\E[\vt \vt{}']\| \le c \zz^2 \lambda^-$, zero mean, mutually independent, independent of $\xt,\lt$;

\item $\outfraccol \le  {c_1}/{\mu r}$, $\outfracrow^\alpha \le c_2$, 

\item subspace change: let $\dif:=\max_j \SE(\P_{j-1}, \P_j)$,
\ben
\item $t_{j+1}-t_j > C r \log n \log(1/\zz)$, and
\item $\dif \le 0.8$ and $C_1 \sqrt{r \lambda^+} (\dif + 2\zz) \le \xmint$;
\een

\item init\footnote{This can be satisfied by applying $C \log r$ iterations of AltProj \cite{robpca_nonconvex} on the first $C r$ data samples and assuming that these have outlier fractions in any row or column bounded by $c/r$.}:
{$\SE(\Phat_0,\P_0) \le 0.25$, $C_1 \sqrt{r \lambda^+}  \SE(\Phat_0,\P_0) \le \xmint$;}%

\een
and (6) algorithm parameters are appropriately set;
then, with probability (w.p.) $\ge 1 - 10 \tmax n^{-10} $, 
$\SE(\Phat_{(t)}, \P_{(t)}) \le $
\[
 \left\{
\begin{array}{ll}
(\zz + \dif) & \text{ if }  t \in [t_j, \that_j+\alpha), \\
 (0.3)^{k-1} (\zz + \dif) & \text{ if }  t \in [\that_j+(k-1)\alpha, \that_j+ k\alpha), \\
\zz   & \text{ if }  t \in [\that_j+ K\alpha+\alpha, t_{j+1}),
\end{array}
\right.
\]
where $K := C \log (1/\zz)$.
Memory complexity is $O(n r \log n \log (1/\zz) )$ and time complexity is $O(n \tmax r \log (1/\zz) )$.%
\end{theorem}
In the above theorem statement, the condition number, $f:=\lambda^+/\lambda^-$ is treated as a numerical constant and ignored.
Under the assumptions of the above theorem, the following also hold:
\ben
\item {$\|\xhat_t-\xt\| = \|\lhat_t-\lt\| \le 1.2 (\SE(\Phat_{(t)}, \P_{(t)}) + \zz) \|\lt\|$,} 

\item  at all times, $t$, $\That_t = \T_t$,

\item $t_j \le \that_j \le t_j+2 \alpha$,

\item Offline-NORST: $\SE(\Phat_{(t)}^{off}, \P_{(t)})  \le \zz$, $\|\xhat_t^{off}-\xt\| = \|\lhat_t^{off}-\lt\| \le \zz \|\lt\|$ at all $t$.
\een

Observe that ReProCS-NORST can automatically detect and track subspace changes with delay that is only slightly more than $r$ (near-optimal). Also, its memory complexity is within log factors of $nr$ (memory needed to output the subspace estimate).
%
%
The second key point is that,  after the first $Cr$ data samples (used for initialization), for $\alpha$-sample mini-batches of data, it tolerates a constant maximum fraction of outliers per row without any assumption on outlier support. This is possible because it makes two other extra assumptions: fixed or slow subspace change and a lower bound on $\xmint$ (see last two assumptions of the Theorem given above). As long as the initial subspace estimate is accurate enough and the subspace changes slowly enough so that both $\Delta$ and $\SE(\Phat_0,\P_0)$ are $O(1/\sqrt{r})$, the lower bound just requires that $\xmint$ be of the order of $\sqrt{\lambda^+}$ or larger which is reasonable. Moreover, as explained later in the ``Relax outlier magnitudes lower bond assumption'' remark, this can be relaxed further.
The requirement on $\Delta$ and $\SE(\Phat_0,\P_0)$ is not restrictive either because $\SE(.)$ is only measuring the largest principal angle. It still allows the chordal distance between the two subspaces ($l_2$ norm of the vector containing the sine of all principal angles) to be $O(1)$.
%

Slow subspace change is a natural assumption for static camera videos (with no sudden scene changes). The outlier lower bound is a mild requirement because, by definition,  an ``outlier'' is a large magnitude corruption. The small magnitude ones get classified as small noise $\vt$. Moreover, as we explain later (see the ``Relax outlier magnitudes lower bond assumption'' remark), the requirement can be relaxed further. The looser requirement on outlier fractions per row means that ReProCS can tolerate slow moving or occasionally static objects better than other approaches (does not confuse these for the background).

Another approach that can solve RST is Modified-PCP \cite{zhan_pcp_jp}. This was designed as a solution to the problem of RPCA with partial subspace knowledge \cite{zhan_pcp_jp}. It solves RST by using the previous subspace estimate as the ``partial subspace knowledge".


\subsection{Dynamic versus Static RPCA}
While robust subspace tracking or dynamic RPCA is a different problem than static RPCA (it assumes a time-varying subspace and slow subspace change), from the point of view of applications, especially those for which a natural time sequence exists, e.g., video analytics for videos from static camera(s), or dynamic MRI based region-of-interest tracking \cite{candes_mri}, both models are equally applicable. For long videos from static cameras, dynamic is generally a better model to use. The same is true for videos involving slow moving or occasionally static foreground objects since these result in a larger fraction of outliers per row of the data matrix. 
On the other hand, for videos involving sudden scene changes, the original RPCA problem is a better fit, since in this case the slow subspace change assumption absolutely does not hold. 

For recommendation system design or survey data analysis, if the initial dataset has a certain number of users, but as time goes on, more users get added to the system, the following strategy is best. Use a static RPCA solution on the initial dataset. As more users get added, use an RST solution to both update the solution and to detect when subspace changes occur. 
In both applications, subspace change would occur when the survey reaches a different previously unrepresented demographic.

\subsection{Robust Subspace Recovery (RSR): problem and summary}
There is a long body of old and recent work that studies the following problem: given a set of $\tmax$ points in $\Re^n$, suppose that the {\em inlier} points lie in an unknown low-dimensional subspace, while {\em outlier} points are all points that do not. Thus, an entire data vector is either an inlier or an outlier.
Also suppose that the fraction of outliers is less than 50\% of all data points. Then, how does one find the low-dimensional subspace spanned by the inliers? In recent literature, this body of work has been referred to as ``mean absolute deviation rounding (MDR)'', \cite{mccoy_tropp11} ``outlier-robust PCA'' \cite{xu2013outlier}, ``PCA with contaminated data''  \cite{xu_nips2013_2} or more recently as ``robust subspace recovery'' \cite{novel_m_estimator}.
%
In historical work, it was called just ROBPCA \cite{ROBPCA_Hubert}.
%
This is a harder problem to solve especially when the fraction of outliers is large since, in this case, entire data vectors get classified as outliers and thrown away.

While many algorithms have been proposed in recent literature, complete guarantees under simple assumptions are hard to obtain. An exception is the outlier pursuit solution  \cite{xu_nips2013_2}. It solves this problem by using a nice extension of the PCP idea: it models the outliers as additive ``column-sparse corruptions''. 
Xu et al \cite{xu_nips2013_2} prove the following.
\begin{theorem}[Outlier pursuit guarantee for RSR]
Let $\gamma$ denote the fraction of corrupted columns (outlier fraction).
Outlier pursuit correctly recovers the column space of $L$ and correctly identifies the outlier support if 
 \ben
 \item  $\gamma \in O(1/r)$; and
 \item the matrix $\L$ is column-incoherent, i.e., the row norm of the matrix of its right singular vectors is bounded by $ \mu r / ((1-\gamma) n)$.
 \een
 \label{out_pursuit_thm}
\end{theorem}
This is a nice and simple correctness guarantee but it requires a tight bound on outlier fractions (just like the S+LR literature). 
Most other guarantees (described later) only prove asymptotic convergence of the iterates of the algorithm to either a stationary point or a local minimum of the chosen cost function; and/or lower bound some performance metric for the solution, e.g., explained variance or mean absolute deviation.

\section{Robust PCA and Subspace Tracking via Sparse + Low-Rank Matrix Decomposition (S+LR)}
Before we begin, we should mention that the code for all the methods described in this section is downloadable from the github library of Andrew Sobral \cite{lrslibrary2015}. The link is
\url{https://github.com/andrewssobral/lrslibrary}.


\subsection{Understanding Identifiability Issues}\label{identif}
The S+LR formulation by itself does not ensure identifiability. For example, it is impossible to correctly separate $\M$ into $\L+\X$ if $\L$ is also sufficiently sparse in addition to being low-rank or if $\S$ is also sufficiently low rank in addition to being sparse. In particular, there is a problem if $\S$ has rank that is equal to or lower than that of $\L$ or if $\L$ has support size that is smaller than or equal to that of $\S$.

For example, if $\S$ is sparse with support that never changes across columns, then $\S$ will be a matrix with rank at most $s$ (where $s$ is the column support size). If, in addition, the nonzero entries of $\st$ are the same for all $t$ (all columns), then, in fact, $\S$ will be a rank one matrix. In this case, without extra assumptions, there is no way to correctly identify $\S$ and $\L$. This situation occurs in case of a video with a foreground object that never moves and whose pixel intensities are such that $\s_t$ never changes.
As a less extreme example, if the object remains static for $b_0 \tmax$ frames at a time before moving, and its pixel intensities are such that $\s_t$ itself is also constant for these many frames, then the rank of $\S$ will be $\tmax/(b_0 \tmax) = 1/b_0$. If this number is less than $\rmat$, then again, the problem is not identifiable without extra assumptions. Observe that, in this example, $\outfracrow$ equals $b_0$. Thus $\outfracrow$ needs to be less than $1/\rmat$ for identifiability. By a similar argument, $\outfraccol$ needs the same bound.

The opposite problem occurs if the left or right singular vectors of $\L$ are sparse. If the left singular vectors are sparse, $\L$ will be a row sparse matrix (some rows will be zero). If the right singular vectors are sparse, then it will be a column sparse matrix. If both left and singular vectors are sparse, then $\L$ will be a general sparse and low rank matrix. In all these situations, if the singular vectors are sparse enough, it can happen that $\L$ is sparser than $\S$ and then it becomes unidentifiable.

It is possible, though, to impose simple assumptions to ensure that neither of the above situations occur. We can ensure that $\S$ is not lower rank than $\L$ by requiring that the fraction of outliers in {\em each} row and {\em each} column be at most $c/\rmat$ with $c<1$ \cite{rpca2,rpca_zhang,robpca_nonconvex}. Alternatively, one can assume that the outlier support is generated from a uniform distribution; with this assumption, just a bound on its size suffices \cite{rpca}.

We can ensure that $\L$ is not sparse by requiring that its left and right singular vectors be dense. This is called the ``denseness'', or, more generally the ``incoherence'' assumption. The term ``incoherence'' is used to imply that this condition helps ensure that the matrix $\L$ is ``different" enough from the sparse matrix $\S$ (in the sense that the normalized inner product between the two matrices is not too large). 
%
%
Let $\L \svdeq \LU \bm\Sigma \LV'$ be the reduced (rank-$\rmat$) SVD of $\L$. Since the left and right singular vectors, $\LU$  and $\LV$, have unit Euclidean norm columns, the simplest way to ensure that the columns are dense (non-sparse) is to assume that the magnitude of each entry of $\LU$ and $\LV$ is small enough. In the best (most dense) case, all their entries would be equal and would just differ in signs. Thus, all entries of $\LU$ will have magnitude $1/\sqrt{n}$ and all those of $\LV$ will have magnitude $1/\sqrt{d}$.
A slightly weaker assumption than this, but one that suffices, is to assume an upper bound on the Euclidean norms of each row of $\LU$ and of $\LV$. Consider $\LU$ which is $n \times \rmat$. In the best (most dense) case, each of its rows will have norm $\sqrt{\rmat/n}$. Incoherence or denseness means that each of its rows has a norm that is within a constant fraction of this number. Using the definition of the incoherence parameter $\mu$ given above, this is equivalent to saying that $\LU$ is $\mu$-incoherent with $\mu$ being a numerical constant.
%
All RPCA guarantees require that both $\LU$ and $\LV$ are $\mu$-incoherent. One of the first two (parallel) guarantees for PCP \cite{rpca} 
also made the following stronger assumption which we will refer to as {\em ``strong incoherence''}.
\bea
\max_{i=1,2,\dots, n, j=1,2,\dots,\tmax} |(\LU \LV')_{i,j}| \le \sqrt{\mu \frac{\rmat}{n \tmax}} 
\label{strong_incoh}
\eea
This says that the inner product between a row of $\LU$ and a row of $\LV$ is upper bounded.
Observe that the required bound is $1/\sqrt{\rmat}$ times what left and right incoherence would imply (by using Cauchy-Schwartz inequality). 

\subsection{Older RPCA solutions: Robust Subspace Learning (RSL) \cite{Torre03aframework} and variants}
The S+LR definition for RPCA was introduced in \cite{rpca}. However, even before this, many good heuristics existed for solving RPCA. The ones that were motivated by video applications did implicitly model outliers as sparse corruptions in the sense that a data vector (image) was assumed to contain pixel-wise outliers (occluded pixels). The most well known among these is the Robust Subspace Learning (RSL) \cite{Torre03aframework} approach. In later work, other authors also tried to develop incremental versions of this approach \cite{Li03anintegrated,ipca_weightedand}.  
The main idea of all these algorithms is to  detect the outlier data entries and either replace their values using nearby values \cite{ipca_weightedand} or weight each data point in proportion to its reliability (thus soft-detecting and down-weighting the detected outliers) \cite{Torre03aframework,Li03anintegrated}. Outlier detection is done by projecting the data vectors orthogonal to the current subspace estimate and thresholding out large entries of the resulting vector. As seen from the  exhaustive experimental comparisons shown in \cite{rrpcp_tsp}, the online heuristics fail in all experiments -- both for simulated and real data.
On the other hand, the original RSL method \cite{Torre03aframework} remains a good practical heuristic. 


\subsection{Convex optimization based solution: Principal Component Pursuit (PCP)}
The first solution to S+LR was introduced in parallel works by Candes et al. \cite{rpca} (where they called it a solution to robust PCA) and by Chandrasekharan et al. \cite{rpca2}. Both proposed to solve the following convex program which was referred to as Principal Component Pursuit (PCP) in \cite{rpca}:
\[
\min_{\tL, \tS}  \|\tL\|_* + \lambda \|\tS\|_1  \text{ subject to } \M = \tL + \tS
\]
Here $\|\A\|_1$ denotes the vector $l_1$ norm of the matrix $\A$ (sum of absolute values of all its entries) and $\|\A\|_*$ denotes the nuclear norm (sum of its singular values). The nuclear norm can be interpreted as the $l_1$ norm of the vector of singular values of the matrix. In other literature, it is also called the Schatten-1 norm.
PCP is the first known polynomial time solution to RPCA that is also provably correct. The paper \cite{rpca} both gave a  simple correctness result, {\em and} showed how PCP outperformed existing work at the time for the video analytics application.

{\em Why it works. }
It is well known from compressive sensing literature (and earlier) that the vector $l_1$ norm serves as a convex surrogate for the support size (number of nonzero entries) of a vector (or vectorized matrix). In a similar fashion, the nuclear norm serves as a convex surrogate for the rank of a matrix. Thus, while the program that tries to minimize the rank of $\tL$ and sparsity of $\tS$ involves an impractical combinatorial search, the above program is convex and solvable in polynomial time.

{\em Time and memory complexity. } The complexity of solving a convex program depends on the iterative algorithm (solver) used to solve it. Most solvers have time complexity that is more than linear in the matrix dimension: a typical complexity is $O(n \tmax^2)$ per iteration \cite{robpca_nonconvex}. Also they typically need $O(1/\epsilon)$ iterations to return a solution that is  within $\epsilon$ error of the true solution of the convex program \cite{robpca_nonconvex}. Thus the typical complexity is $O(n \tmax^2/\epsilon)$.
PCP operates on the entire matrix so memory complexity is $O(n \tmax)$.

\subsection{Non-convex solutions: Alternating Minimization}


%
To obtain faster algorithms, in more recent works, authors have tried to develop provably correct algorithms that rely on either alt-min or projected gradient descent (GD). Both alt-min and GD have been used for a long time as practical heuristics for trying to solve various non-convex programs. The initialization either came from other prior information, or multiple random initializations were used to run the algorithm and the ``best" final output was picked.
%
The new ingredient in these provably correct solutions is a carefully designed initialization scheme that already outputs an estimate  that is ``close enough'' to the true one.

For RPCA, the first provably correct  non-convex solution was Alternating-Projection (AltProj) \cite{robpca_nonconvex}. This borrows some ideas from an earlier algorithm called GoDec \cite{godec}. AltProj works by projecting the residual at each step onto the space of either sparse matrices or onto the space of low-rank matrices. The approach proceeds in stages with the first stage projecting onto the space of rank $\hat{r}=1$ matrices, while increasing the support size of the sparse matrix estimate at each iteration in the stage. In the second stage, $\hat{r}=2$ is used and so on for a total of $r$ stages.
Consider stage one. AltProj is initialized by thresholding out the very large entries of the data matrix $\M$ to return the first estimate of the sparse matrix. Thus $\Shat_0= HT(\M;{\zeta_0})$ where $HT$ denotes the hard thresholding operator and $\zeta_0$ denotes the threshold used for it. After this, it computes the residual $\M - \Shat_0$ and projects it onto the space of rank one matrices. Thus $\Lhat_1 = \pP_1(\M - \Shat_0)$ where $\pP_1$ denotes a projection onto the space of rank one matrices. It then computes the residual $\M - \Lhat_1$ and projects it again onto the space of sparse matrices but with using a  carefully selected threshold $\zeta_1$ that is smaller than $\zeta_0$. Thus $\Shat_1 = HT(\M - \Lhat_1;{\zeta_1})$. This process is repeated until a halting criterion is reached. The algorithm then moves on to stage two. This stage proceeds similarly but each low rank projection is now onto the space of rank $\hat{r}=2$ matrices. This process is repeated for $r$ stages.

{\em Why this works. }
Once the largest outliers are removed, it is expected that projecting onto the space of rank one matrices returns a reasonable rank one approximation of $\L$, $\Lhat_1$.
This means that the residual $\M - \Lhat_1$ is a better estimate of $\S$ than $\M$ is. Because of this, it can be shown that $\Shat_1$ is a better estimate of $\S$ than $\Shat_0$  and so the residual $\M - \Shat_1$ is a better estimate of $\L$ than $\M - \Shat_0$. This, in turn, means $\Lhat_2$ will be a better estimate of $\L$ than $\Lhat_1$ is.
The proof that the initial estimate of $\L$ is good enough  relies on incoherence of left and right singular vectors of $\L$ and the fact that no row or column has too many outliers.
These two facts are also needed to show that each new estimate is better than the previous.

{\em Time and memory complexity. } The time complexity of AltProj is $O(n \tmax r^2 \log(1/ \epsilon))$. It operates on the entire matrix so memory complexity is $O(n \tmax)$.

\subsection{Non-convex Solutions: Projected Gradient Descent (RPCA-GD and NO-RMC)}

While the time complexity of AltProj was lower than that of PCP, there was still scope for improvement in speed. In particular, the outer loop of AltProj that runs $r$ times seems unnecessary (or can be made to run fewer times). Two more recent works \cite{rpca_gd,rmc_gd} try to address this issue.
The question asked in \cite{rpca_gd} was can one solve RPCA with computational complexity that is of the same order as a single $r$-SVD? This has complexity $O(n \tmax r (-\log \epsilon))$.
The authors of \cite{rpca_gd} show that this is indeed possible with an extra factor of $\kappa^2$ in the complexity and with a tighter bound on outlier fractions. Here $\kappa$ is the condition number of $\L$. To achieve this, they developed an algorithm that relies on projected gradient descent (GD). We will refer to this algorithm as RPCA-GD.
The authors of \cite{rmc_gd} use a different approach. They develop a projected GD solution for robust matrix completion (RMC), and  argue that, even in the absence of missing entries, the same algorithm provides a very fast solution to RPCA as long as the {\em data matrix is nearly square.} For solving RPCA, it deliberately under-samples the available full matrix $\M$ in order to speed up the algorithm. We will refer to this algorithm as NO-RMC (which is short for nearly optimal RMC).

Projected GD is natural heuristic for using GD to solve constrained optimization problems. To solve $\min_{x \in \cal{C}} f(x)$, after each GD step, it projects the  output onto the set $\cal{C}$ before moving on to the next iteration.


{\em RPCA-GD \cite{rpca_gd}. } Notice that the matrix $\L$ can be decomposed as $\L = \tU \tV'$ where $\tU$ is an $n \times r$ matrix and $\tV$ is a $\tmax \times r$ matrix. The algorithm alternately solves for $\S, \tU, \tV$. Like AltProj, it also begins by first estimating the sparse component $\Shat_0$. Instead of hard thresholding, it uses a more complicated approach called ``max-sorting-thresholding'' for doing this. It then initializes $\tUhat_0$ and $\tVhat_0$ via $r$-SVD on $\M - \Shat_0$ followed by ``projecting onto the set of incoherent matrices'' (we explain this in the next para).
After this, it repeats the following three steps at each iteration: (1) use ``max-sorting-thresholding'' applied to $\M - \Lhat_{i-1}$ to obtain $\Shat_i$; (2a)  implement one gradient descent step for minimizing the cost function $\mathcal{L}(\tU,\tV,\S):=\|\tU \tV' + \S - \M\|_F^2 + 0.25 \|\tU'\tU - \tV'\tV\|_F$ over $\tU$ while keeping $\tV,\S$ fixed at their previous values, and (2b) obtain $\tUhat_i$ by ``projecting the output of step 2a onto the set of incoherent matrices''; and (3) obtaining $\tVhat_i$ in an analogous fashion. 

The step ``projecting onto the set of incoherent matrices'' involves the following. Recall from earlier that incoherence (denseness) of left and right singular vectors is needed for static RPCA solutions to work. 
To ensure that the estimate of $\tU$ after one step of gradient descent satisfies this, RPCA-GD projects the matrix onto the ``space of incoherent matrices''. This is achieved by clipping: if a certain row has norm larger than $\sqrt{2 \mu r / n} \|\tUhat_0\|_2$, then each entry of that row is re-scaled so that the row norm equals this value.


{\em NO-RMC \cite{rmc_gd}. }
The NO-RMC algorithm is designed to solve the more general RMC problem. Let $\Omega$ denote the set of observed entries and let $\Pi_\Omega$ denote projection onto a subspace formed by matrices supported on $\Omega$ \footnote{Thus $\Pi_\Omega(\A)$ zeroes out entries of $\A$ whose indices are not in $\Omega$ and keeps other entries the same.}. The set $\Omega$ is generated uniformly at random.
NO-RMC modifies AltProj as follows. First, instead of running the algorithm in $r$ stages, it reduces the number of stages needed. In the $q$-th outer loop it projects onto the space of rank $k_q$ matrices (instead of rank $q$ matrices in case of AltProj). The second change is in the update step of $\Lhat$ which now also includes a gradient descent step before the projection.
Thus, the $i$-th iteration of the $q$-th outer loop now computes $\Lhat_i =\pP_{k_q} (\Lhat_{i-1} + (1/p) \pP_\Omega(\M - \Lhat_{i-1} - \Shat_{i-1}))$. On first glance, the projection onto the space of rank $k_q$ matrices should need $O(n \tmax k_q)$ time. However, as the authors explain, because the matrix that is being projected is a sum of a low rank matrix and a matrix with many zeroes (sparse matrix as used in fast matrix algorithms' terminology), the computational cost for this step is actually only $O(|\Omega|k_q + (n+\tmax)k_q + k_q^3)$ (see page 4 of the paper). The required number of observed entries is roughly $|\Omega|\approx n r^3 \log n$ whp. This is what results in a significant speed-up.
However NO-RMC requires $\tmax$ to be of the same order as $n$. This is a stringent requirement for high-dimensional datasets for which $n$ is large, e.g., it is hard to find videos that have as many frames as the image size $n$.%



\newcommand{\tPhat}{\hat{\bm{P}}}
\usetikzlibrary{decorations.markings}
\usetikzlibrary{decorations.pathreplacing}
\def\MarkLt{4pt}
\def\MarkSep{2pt}

\tikzset{
  TwoMarks/.style={
    postaction={decorate,
      decoration={
        markings,
        mark=at position #1 with
          {
              \begin{scope}[xslant=0.2]
              \draw[line width=\MarkSep,white,-] (0pt,-\MarkLt) -- (0pt,\MarkLt) ;
              \draw[-] (-0.5*\MarkSep,-\MarkLt) -- (-0.5*\MarkSep,\MarkLt) ;
              \draw[-] (0.5*\MarkSep,-\MarkLt) -- (0.5*\MarkSep,\MarkLt) ;
              \end{scope}
          }
       }
    }
  },
  TwoMarks/.default={0.5},
}

\usetikzlibrary{positioning}
\usetikzlibrary{calc}
%
\tikzstyle{block}  = [rectangle, draw, rounded corners, text width=4cm, text centered, minimum height=1em]
\tikzstyle{block}  = [rectangle, draw, rounded corners, text width=5cm, text centered, minimum height=1em]
\tikzstyle{smallblock}  = [rectangle, draw, rounded corners,text width=1.8cm, text centered, minimum height=1em]
\tikzstyle{input}  = [rectangle, draw, text width=1.2cm, text centered, minimum height=1em]
\tikzstyle{output}  = [rectangle, draw, text width=1.2cm, text centered, minimum height=1em]
\tikzstyle{block1}  = [rectangle, draw, rounded corners,text width=5cm, text centered, minimum height=1em]
\tikzstyle{blockl1}  = [rectangle, draw, rounded corners,text width=4cm, text centered, minimum height=1em]
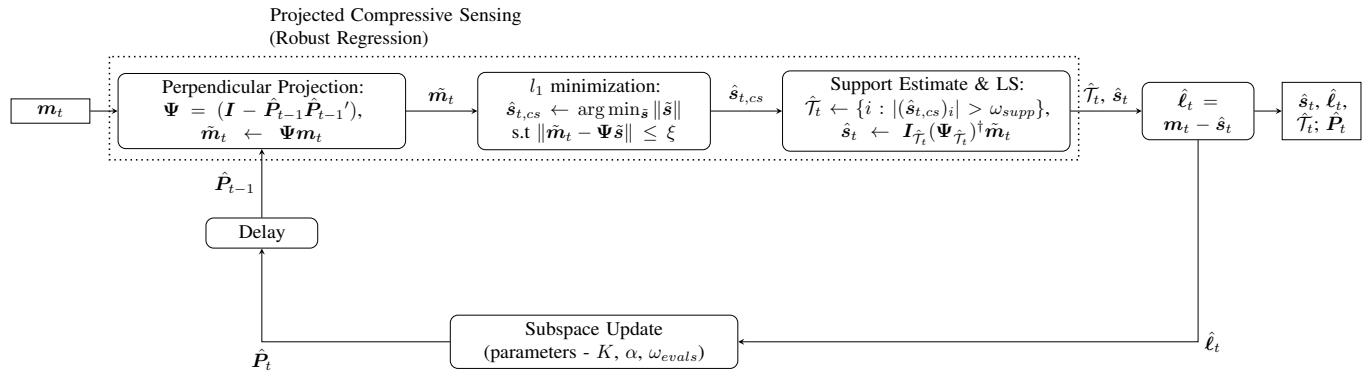
\begin{figure*}[t!]
\begin{center}
\resizebox{\linewidth}{!}{
\begin{tikzpicture}
	\node[input] (0) {$\yt$};
    \node[block] (1) [right = .5cm of 0] {Perpendicular Projection: \\ $\bpsi = (\I - \tPhat_{t-1} \tPhat_{t-1}{}')$, $\tilde{\y}_t \leftarrow \bpsi \yt$};
    \node[blockl1] (2) [right = 1.3cm of 1]  {$l_1$ minimization: \\ $\xhat_{t,cs} \leftarrow \arg\min_{\tilde{\x}} \norm{\tilde{\x}}$ \\ s.t $\norm{\tty_t - \bpsi \tilde{\x}} \leq \xi$};
    \node[block] (3) [right = 1.3cm of 2] {Support Estimate \& LS: \\ $\That_t \leftarrow$ $\{i : |(\xhat_{t,cs})_i| > \omega_{supp} \}$,
    \\ $\xhat_t \leftarrow \I_{\That_t}(\bpsi_{\That_t})^{\dagger} \tty_t$};
	\node[smallblock] (4) [right = 1.3cm of 3] {$\lhat_t= \yt-\xhat_{t}$};
	\node[output](100) [right = 0.5cm of 4] {$\xhat_t$, $\lhat_t$, $\That_t$; $\tPhat_t$};
	\node[smallblock](5) [below = 1.25cm of 1]{Delay};
	\node[block1](6) [below = 3cm of 2]{Subspace Update \\ (parameters - $K$, $\alpha$, $\omega_{evals}$)};
    \draw[->, >=stealth, scale=5] (0) -- (1);
    \draw[->, >=stealth] (1) --  node[above] {$\tilde{\yt}$} (2);
	\draw[->, >=stealth] (2) --  node[above] {$\xhat_{t,cs}$} (3);
    \draw[->, >=stealth] (3) --  node[above] {$\That_t$, $\xhat_t$} (4);
    \draw[->, >=stealth] (4) -- (100);
	\draw[->, >=stealth]  (4.south) |- node[right]{$\lhat_t$} (6.east);
	\draw[->, >=stealth]  (6.west) -| node[below]{$\tPhat_t$} (5.south);
	\draw[->, >=stealth]  (5) -- node[left]{$\tPhat_{t-1}$} (1);
	 \draw[thick,dotted] ($(1.north west)+(-0.15,0.3)$)  rectangle ($(3.south east)+(0.15,-0.15)$);
	 \node[text width=5cm, anchor=north] at (6.5,2.0)
    {Projected Compressive Sensing (Robust Regression)};
\end{tikzpicture}
}
\end{center}
\vspace{-0.15in}
\caption{\small{
The Recursive Projected Compressive Sensing (ReProCS) solution framework for robust subspace tracking or dynamic RPCA.
Subspace Update: is either in ``detect'' phase or in ``update'' phase. Suppose that the $j$-th change is detected at time $\that_j$. The update phase proceeds as follows.
At time $t=\that_j+k \alpha$, for $k=1,2,\dots,K$, it computes the $j$-th estimate of $\P_j$ (denoted $\Phat_{j,k}$) as the top $r$ singular vectors of $\Lhat_{t;\alpha}:=[\lhat_{t-\alpha+1},\lhat_{t-\alpha+2}, \dots, \lhat_t]$ and it sets $\Phat_t$ equal to this. At other times it just sets $\Phat_t \leftarrow \Phat_{t-1}$.
The detect phase uses the following idea: to detect the $j$-th change, it checks if the maximum singular value of $(\I - \Phat_{j-1}\Phat_{j-1}{}')\Lhat_{t;\alpha}$ is larger than $\sqrt{\alpha \omega_{evals}}$.
}}
\vspace{-0.2in}
\label{reprocs_block}
\end{figure*}


\subsection{Non-convex online solution for RST and RPCA: Recursive Projected Compressive Sensing (ReProCS)} \label{reprocs_details}
The RST problem has received much lesser attention than RPCA \cite{rrpcp_perf,rrpcp_aistats,rrpcp_medrop}. In fact, even the easier subspace tracking or subspace tracking with missing data problems are not fully solved. Many algorithms exist, e.g., see \cite{past,petrels,grouse}, however all existing guarantees are asymptotic results for the statistically stationary setting of data being generated from a {\em single unknown} subspace; {\em and} most are partial guarantees. Of course, as explained earlier, the most general nonstationary model that allows the subspace to change at each time is not even identifiable. The ReProCS approach \cite{rrpcp_allerton,rrpcp_perf} described here resolves this issue by using the piecewise constant subspace change model described earlier. 

%
The simplest reprocs algorithm called ReProCS-NORST \cite{rrpcp_medrop,rrpcp_merop} starts with a ``good'' estimate of the initial subspace, which is obtained by $C (\log r)$ iterations of AltProj applied to $\Y_{[1,t_\train]}$ with $t_\train=C r$.
After this at each time, it first solves {\em Projected CS (Robust Regression)}\footnote{(To understand the equivalence, see Section \ref{projCS_RR})} in order to estimate the sparse outliers, $\xt$, and then $\lt$ as $\lhat_t = \mt- \xhat_t$; and then (b) {\em Subspace Update} to update the subspace estimate $\Phat_{(t)}$.
 Projected CS (Robust Regression) proceeds as follows. At time $t$, let $\Phat_{(t-1)}$ denote the subspace estimate from the previous time instant, $(t-1)$. If this estimate is accurate enough, because of slow subspace change, projecting $\mt := \xt + \lt + \vt$ onto its orthogonal complement nullifies most of $\lt$. To be precise, we compute $\tty_t:= \bm\Psi  \mt$ where $\bm{\Psi} := \I - \Phat_{(t-1)}\Phat_{(t-1)}{}'$. Thus, $\tty_t = \bm{\Psi} \xt + \b_t$ where $\b_t:= \bm\Psi (\lt + \vt)$ and $\| \bm{b}_t \|$ is small. Recovering $\xt$ from $\tty_t$ is thus a regular compressive sensing (CS) / sparse recovery problem in small noise \cite{candes_rip}. Notice that, even though $\bm\Psi$ is square, it is rank deficient: it has rank $n-r$. This is why we cannot just invert it to estimate $\xt$, but instead need to solve the CS problem. For solving the CS problem, any approach can be used. As an example, one can use $l_1$ minimization: compute $\xhat_{t,cs}$ using $l_1$ minimization, followed by thresholding based support estimation to get $\That_t$, and then a Least Squares (LS) based debiasing step on $\That_t$ to get $\xhat_t$. The reason we can accurately recover $\xt$ by solving the CS problem is the following: denseness (incoherence) of the $\P_{(t)}$'s along with slow subspace change can be used to show that the matrix $\bm\Psi$ satisfies restricted isometry property (RIP) \cite{candes_rip} with a small enough RIP constant \cite{rrpcp_perf}. Once $\xhat_t$ is available, one can then recover  $\lt$ by subtraction: thus $\lhat_t = \yt - \xhatt$.

The $\lhatt$'s are used in the subspace update step which runs every $\alpha$ frames. In its simplest form, this involves (i) detecting subspace change, and (ii) obtaining progressively improved estimates of the changed subspace via $K$ steps of $r$-SVD, each done with a new set of $\alpha$ frames of $\lhatt$. Here $r$-SVD means compute the top $r$ left singular vectors of $\Lhat_{t;\alpha}:=[\lhat_{t-\alpha+1},\lhat_{t-\alpha+2}, \dots, \lhat_t]$.
This step was designed assuming a piecewise constant subspace change model; however, the algorithm itself works even without this assumption (it works for real videos as well).
A block diagram of the ReProCS framework is shown in Fig. \ref{reprocs_block}. 


{\em Why it works. }
It is not hard to see that the ``noise'' $\b_t:=\bm\Psi (\lt+ \vt)$ seen by the projected CS step is proportional the error between the subspace estimate from $(t-1)$ and the current subspace. Using the RIP arguments given above, and the outlier magnitudes' lower bound assumption, one can ensure that the CS step output is accurate enough and the outlier support $\T_t$ is correctly recovered. With this, it is not hard to see that $\lhat_t = \lt + \vt - \et$ where $\et:=\x_t -\xhat_t$ satisfies
\[
\et= \I_{\T_t} (\bm\Psi_{\T_t}{}'\bm\Psi_{\T_t})^{-1} \I_{\T_t}{}' \bm\Psi' \lt.
\]
and $\|\et\| \le C \|\b_t\|$. 
Consider subspace update. Every time the subspace changes, one can show that the change can be detected within a short delay. After that, the $K$ SVD steps help get progressively improved estimates of the changed subspace. To understand this, observe that, after a subspace change, but before the first update step, $\b_t$ is the largest and hence, $\et$, is also the largest for this interval. However, because of good initialization or because of slow subspace change and previous subspace correctly recovered (to error $\zz$), neither is too large. Both are proportional to $(\zz + \dif)$, or to the initialization error. Recall that $\dif$ quantifies the amount of subspace change. For simplicity suppose that $\SE(\Phat_0,\P_0) = \dif$.
Using the idea below, we can show that we get a ``good'' first estimate of the changed subspace.

The input to the PCA step is $\lhat_t$ and the noise seen by it is $\et$. Notice that $\et$ depends on the true data $\lt$ and hence this is a setting of PCA in data-dependent noise \cite{corpca_nips,pca_dd}. From \cite{pca_dd}, it is known that the subspace recovery error of the PCA step is proportional to the ratio between the time averaged noise power plus time-averaged signal-noise correlation, $(\|\sum_t \E[\et \et{}']\| + \|\sum_t \E[\lt \et{}'\|)/\alpha$, and the minimum signal space eigenvalue, $\lambda^-$.
The instantaneous values of both noise power and signal-noise correlation are of order $(\dif+\zz)$ times $\lambda^+$. However, using the fact that $\et$ is sparse with support $\T_t$ that changes enough over time so that $\outfracrow^\alpha$ is bounded, their time averaged values are at least $\sqrt{\outfracrow^\alpha}$ times smaller. This follows using Cauchy-Schwartz. As a result, after the first subspace update, the subspace recovery error is below $3 \sqrt{\outfracrow} (\lambda^+/\lambda^-)$ times $(\dif+\zz)$. Since $9 \outfracrow (\lambda^+/\lambda^-)^2$ is bounded by a constant $c_2 < 1$, this means that, after the first subspace update, the subspace error is below $\sqrt{c_2}$ times $(\dif+\zz)$.

This, in turn, implies that $\|\b_t\|$, and hence $\|\et\|$, is also $\sqrt{c_2}$ times smaller in the second subspace update interval compared to the first. This, along with repeating the above argument, helps show that the second estimate of the changed subspace is $\sqrt{c_2}$ times better than the first, i.e., its error is $(\sqrt{c_2})^2$ times $(\dif+\zz)$. Repeating the argument $K$ times, the $K$-th estimate has error $(\sqrt{c_2})^K$ times $(\dif+\zz)$. Since $K = C \log(1/\zz)$, this is an $\zz$ accurate estimate of the changed subspace.%

{\em Time and memory complexity. } ReProCS memory complexity is $n \alpha$ in online mode and $K n \alpha$ in offline mode. With the choices of $K$ and $\alpha$ that suffice for its guarantee (see the ReProCS-NORST Theorem given earlier), its memory complexity is just $O(n r \log n)$ in online mode and $O(n r \log n \log (1/\epsilon) )$ in offline mode. Both are within logarithmic factors of $nr$, and thus nearly optimal: $nr$ is the memory needed to output the estimate of an $r$-dimensional subspace in $\Re^n$.
%
The computational complexity is $O(n \tmax r \log (1/\epsilon) )$: this is the time needed to compute an $r$-SVD for an $n \times \tmax$ matrix with good eigen-gap. Thus, the speed is as fast as it can get without deliberately under-sampling the input data for algorithm speed-up.

{\em Summary of ReProCS-based solutions. }
%
In \cite{rrpcp_allerton,rrpcp_perf}, a novel solution framework called Recursive Projected Compressive Sensing (ReProCS) was introduced to solve the RST or the dynamic RPCA problem. In later works, this was shown to be provably correct under progressively weaker assumptions \cite{rrpcp_isit15,rrpcp_aistats, rrpcp_dynrpca,rrpcp_medrop}.
Under two extra assumptions, the ReProCS framework provides an online (after initialization), fast, memory-efficient, and {\em highly-robust} solution (see the ReProCS-NORST Theorem given earlier). By highly-robust we mean that, after initialization, it can tolerate an order-wise larger maximum fractions of outliers in any row than other RPCA approaches. For the video application, this means that it tolerates slow-moving or occasionally static foreground objects better than other RPCA approaches. This is also evident from the experiments; see the column for Intermittent Object Motion (IOM) category of videos in Table \ref{table1}. ReProCS can also tolerate a larger fraction of outliers per column especially when $\rmat \gg r$.
The extra assumptions needed are simple and practically valid: (i) slow subspace change (or fixed subspace in case of static RPCA); and (ii) outlier magnitudes are either all large enough, or most are large enough and the rest are very small.
(i) is a natural assumption for static camera videos (with no sudden scene changes) and (ii) is also easy because, by definition, an ``outlier'' is a large magnitude corruption. The small magnitude ones get classified as $\vt$. In practice, intermediate magnitude outliers are the most problematic of course (for all approaches).
We explain the improved outlier tolerance and these extra requirements in Sec. \ref{outrow_explain}. Also, as noted there, the outlier magnitude lower bound can be relaxed.

We have explained the latest and simplest version of ReProCS here \cite{rrpcp_medrop}. Under slightly stronger assumptions on subspace change, for example, if it is known that, at each subspace change time, only one subspace direction change, the SVD step can be replaced by a slightly faster approach called projection-SVD \cite{rrpcp_perf,rrpcp_tsp,rrpcp_dynrpca}. The simplest version of the resulting algorithm is simple-ReProCS studied in \cite{rrpcp_dynrpca}.

\subsubsection{Looser bound on $\outfracrow$ and outlier magnitudes' lower bound}\label{outrow_explain}
As explained earlier, if the allowed maximum outlier fraction in any row is more than $1/\rmat$, the original RPCA problem becomes unidentifiable. ReProCS tolerates a constant maximum fraction per row because it uses extra assumptions. We explain these here. It recovers $\xt$ first and then $\lt$ and does this at each time $t$. To recover $\xt$ accurately, it does require that $\lt$ not be sparse (incoherence or denseness of the $\P_j$'s ensures this). Moreover, it also exploits  ``good'' knowledge of the subspace of $\lt$ (either from initialization or from the previous subspace's estimate and slow subspace change). But it has no way to deal with the residual error, $\b_t:= (\I - \Phat_{(t-1)} \Phat_{(t-1)}{}') \lt$, in the subspace knowledge. Since the individual vector $\b_t$ does not have any structure that can exploited\footnote{However the $\b_t$'s arranged into a matrix do form a matrix that is approximately rank $r$ or even lower (if not all directions change). If we try to exploit this structure we end up with the modified-PCP approach described next.}, the error in recovering $\xt$ cannot be lower than $C \|\b_t\|$.
Thus, to enable correct recovery of the support of $\xt$, $\xmint$ needs to be larger than $C\|\b_t\|$. This is where the $\xmint$ lower bound comes from.
Correct support recovery is needed to ensure that the subspace estimate can be improved with each update. 
This step also requires element-wise boundedness of the $\at$'s along with mutual independence and identical covariances (these together are similar to right incoherence of $\L$, see \cite{rrpcp_medrop}).%
%
\begin{remark}[Relax outlier magnitudes lower bond assumption]
The outlier magnitudes lower bound assumption of  the ReProCS-NORST Theorem given in the Overview section can be relaxed to a lower bound on most outlier magnitudes. In particular, the following suffices: assume that
$\xt$ can be split into $\xt = (\xt)_{small} + (\xt)_{large}$ that are such that, in the $k$-th subspace update interval,
$\|(\xt)_{small}\|  \le 0.3^{k-1} (\zz+\dif)\sqrt{r \lambda^+} $ and
the smallest nonzero entry of $(\xt)_{large}$ is larger than $C 0.3^{k-1} (\zz+\dif)\sqrt{r \lambda^+} $.

If there were a way to bound the element-wise error of the CS step (instead of the $l_2$ norm error), the above requirement could be relaxed further.
\label{relax_lowerbound}
\end{remark}

\subsubsection{Relating Projected CS and Robust Regression} \label{projCS_RR}
Projected CS is equivalent to solving the robust regression problem with a sparsity model on the outliers. To understand this simply, let $\Phat= \Phat_{(t-1)}$. The exact version of robust regression assumes that the data vector $\yt$ equals $\Phat \at + \x_t$, while its approximate version assumes that this is only approximately true.  Since $\Phat$ is only an approximation to $\P_{(t)}$, even in the absence of unstructured noise $\vt$, approximate robust regression is the correct problem to solve for our setting.
Its unconstrained version involves $\min_{\a,\x} \lambda \|\x\|_1+ \|\yt - \Phat \a - \x\|^2$. Observe that one can solve for $\a$ in closed form to get $\hat\a = \Phat'(\yt-\x)$. Substituting this, the minimization simplifies to $\min_{\x} \|\x\|_1+ \|(\I - \Phat \Phat') (\yt - \x)\|^2$. This is exactly the same as the Lagrangian version of projected CS. The version for which we obtain guarantees solves $\min_{\x} \|\x\|_1 \text{ s.t. } \|(\I - \Phat \Phat') (\yt - \x)\|^2 \le \xi^2$, but we could also have used other CS results and obtained guarantees for the Lagrangian version with minimal changes.


\subsection{Online solution 2: Modified-PCP for RPCA with Partial Subspace Knowledge and for Dynamic RPCA}
A simple extension of PCP, called modified-PCP, provides a nice solution to the problem of RPCA with partial subspace knowledge \cite{zhan_pcp_jp}. This problem often occurs in face recognition type applications: the training dataset may be outlier free, but, due to limited training data, it may not contain all possible directions of variation of the face subspace.
%
To understand the solution idea, let $\G$ denote the basis matrix for the partial subspace knowledge.  If $\G$ is such that the matrix $(\I-\G\G')\L$ has rank smaller than $\rmat$, then the following is a better idea.
\[
 \text{min}_{\tL,\tS}    \|(\I - \G \G') \tL\|_* + \lambda \|\tS\|_1      \text{ subject to }  \tL + \tS =  \M
\]
This solution was called {\em modified-PCP} because it was inspired by a similar idea, called modified-CS \cite{modcsjournal}, that solves the problem of compressive sensing (or sparse recovery) when using partial support knowledge is available.
More generally, even if the approximate rank of $(\I-\G\G')\L$ is much smaller, i.e., suppose that $\L = \G \A + \W + \L_\new $ where $\L_\new$ has rank $r_\new \ll \rmat$ and $\|\W\|_F \le \zz$ is small, the following simple change to the same idea works:
\[
 \text{min}_{\tL_\new,\tS, \tilde{\A}}   \|\tL_\new\|_* + \lambda \|\tS\|_1    \text{ s.t. }  \| \M - \tS - \G \tilde\A  \|_F \le \zz
\]
and output $\Lhat = \G \hat\A + \Lhat_\new$. 
To solve the tracking problem using modified-PCP, the subspace estimate from the previous set of $\alpha$ frames serves as the partial subspace knowledge $\G$ for the current set. It is initialized using PCP.

The Modified-PCP guarantee was proved using ideas borrowed from \cite{rpca} for PCP (PCP(C)). It shows that, when $r_\new \ll r$, mod-PCP needs a much weaker version of the strong incoherence condition needed by PCP. However, like PCP (and unlike ReProCS), it also needs uniformly randomly generated support sets which is a strong requirement.
But, also like PCP (and unlike ReProCS) it does not need a lower bound on outlier magnitudes. However, its slow subspace change assumption is unrealistic. Also, it does not detect the subspace change automatically, i.e., it assumes $t_j$ is known. This last limitation can be removed by using the ReProCS approach to detecting subspace change \cite{rrpcp_aistats,rrpcp_dynrpca}.%

\subsection{Summary discussion of all simple (and provably correct) approaches}
A summary is provided in Table \ref{compare_assu}. The results for PCP(C) and modified-PCP both require uniform random support (impractical requirement) but, because of that, both allow a constant outlier fraction in any row or column of the data matrix. The PCP(H) result as well as all follow up works (AltProj, ReProCS, GD, NO-RMC) do not require any outlier support generation model but, because of that, they require a tighter bound on the outlier fractions in any row or column. AltProj and NO-RMC allow this to be $O(1/\rmat)$ while GD requires this to be $O(1/\rmat^{1.5})$.
ReProCS has the best outlier fraction tolerance: for the first $C r$ frames it needs the fractions to be $O(1/r)$; after that it needs the fraction in any row of a $\alpha$ sample mini-batch to only be $O(1)$, while it needs the fraction in any column to be $O(1/r)$. The row bound is better than the information theoretic upper bound on what can be tolerated without extra assumptions. ReProCS uses fixed or slow subspace change and most outlier magnitudes lower bounded to obtain this advantage.
These two extra assumptions (and a different way to ensure right incoherence) also help get an online algorithm that has near-optimal memory complexity and near-optimal tracking delay. 

In terms of time complexity, PCP is the slowest because it needs to solve a convex program, AltProj is much faster. GD and ReProCS are faster than AltProj, and NO-RMC is the fastest. Up to log factors, its time complexity is order $n\rmat^2$. For small $\rmat$, this is almost linear in the number of unknowns. However, to achieve this, it deliberately undersamples the full data matrix and then solves an RMC problem; and for this to work correctly it requires the data matrix to be nearly square. This is a strong requirement. Also it should be clarified that this discussion does not incorporate the undersampling complexity. At least the simple way to generate a Bernoulli support (for the undersampling) will incur a cost that is linear in the matrix size, i.e., $O(n \tmax)$.  

In terms of practical performance, for a video analytics (foreground-background separation) application, ReProCS has the {\em best} performance among all the provable methods, and it is also among the fastest methods. In fact, for the Dynamic Background (DB) and Intermittent Object Motion (IOM) category of videos, it has the {\em best} performance compared to all methods that do not use more assumptions. The IOM videos are examples of large outlier fractions per row, while the DB videos are examples of large $\rmat$ (but $r$ may be smaller).
%
In terms of MATLAB time taken, mod-PCP and ReProCS are the fastest. The others are at least three times slower. Of these AltProj is the third fastest while both GD and NO-RMC are even slower than it experimentally. One reason for RMC to be slow would be that undersampling is expensive. 

All approaches need a form of left and right incoherence as discussed earlier.

\subsection{Other online heuristics for RPCA and dynamic RPCA (robust subspace tracking)}
After the ReProCS algorithm was introduced in 2010 \cite{rrpcp_allerton,rrpcp_allerton11}, many other online algorithms for RPCA were proposed. None of these come with complete guarantees. 

{\em Bilinear-Dec) \cite{mateos_anomaly2}. }
Mateos and Giannakis proposed a robust PCA approach using bilinear decomposition with sparsity control. Their approach uses a Least-Trimmed Squares (LTS) PCA estimator that is closely related to an estimator obtained from an $l_0$-norm regularized criterion, adopted to fit a low-rank bilinear factor analysis model that explicitly incorporates an unknown sparse vector of outliers per datum. Efficient approximate solvers are employed by surrogating the $l_0$-norm of the outlier matrix with its closest convex approximation. This leads to an M-type PCA estimator.

{\em GoDec \cite{godec}. } GoDec can be understood most easily as a precursor to the AltProj algorithm described in detail earlier. Its initialization is a little different, the first step involves low rank projection of the data matrix instead of first removing large outliers. Also, it does not proceed in stages.


{\em Stochastic Optimization   \cite{xu_nips2013_1}. }
In \cite{xu_nips2013_1}, an online algorithm was developed to solve the PCP convex program using stochastic optimization. This relies the fact that the nuclear norm of a rank $r$ matrix $\L$ can be computed as the minimizer of the squared sum of the Frobenius norms of matrices $\tU$ and $\tV$ under the constraint $\L = \tU \tV'$. We list ORPCA as a heuristic because it only comes with a partial guarantee: the guarantee assumes that the subspace estimate outputted at each time $t$ is full rank.
%
%

{\em  Adaptive Projected Subgradient Method (APSM) \cite{rob_ss_track,chouvardas2015robust}. } For the APSM algorithm, at each time instant, a cost function is defined based on the incoming data. This cost function scores a zero loss for a non-empty set of points/possible solutions. The aim is to find a point which belongs to the intersection of all the sets associated with the incoming data. The time instant at which the data contain outlier noise are identified. CoSAMP is used to estimate the sparse outlier vector. 

{\em Grassmannian Robust Adaptive Subspace Tracking Algorithm (GRASTA) \cite{grass_undersampled}. } This borrows the main idea of ReProCS -- projected  CS (a.k.a. approximate Robust Regression) followed by subspace update using the estimated true data vectors. But it replaces both the steps by different and approximate versions.
GRASTA solves the exact version of robust regression which involves recovering $\at$ as $\arg\min_{\a} \|\yt - \Phat_{(t-1)} \a\|_1$. This formulation ignores the fact that $\Phat_{(t-1)}$ is only an approximation to the current subspace $\P_{(t)}$. This is why, in experiments, GRASTA fails when there are significant subspace changes: it ends up interpreting the subspace tracking error as an outlier. In its subspace update step, the SVD or projected-SVD used in different variants of ReProCS \cite{rrpcp_allerton11,rrpcp_tsp,rrpcp_medrop} are replaced by a faster but approximate subspace tracking algorithm called GROUSE \cite{grouse}. 


{\em pROST \cite{seidel2013prost} and ROSETA. } Both these algorithms modify the GRASTA approach, and hence, indirectly rely on the basic ReProCS framework of alternating projected sparse recovery (robust regression) and subspace update.  pROST replaces $l_1$ minimization in the robust regression step by non-convex $l_0$-surrogates ($l_p$ norm minimization for $p<1$). 
Therefore, the authors call their algorithm $l_p$ norm Robust Online Subspace Tracking (pROST).
ROSETA uses the same framework but projection coefficients and sparse outliers  are computed using ADMM and the subspace is updated using a proximal point iteration with adaptive parameter selection.

\subsection{Heuristics that use extra application-specific spatial and/or temporal constraints}
The algorithms described above only solve the robust PCA or dynamic robust PCA problems, without using extra application specific constraints.  Using such constraints can significantly improve algorithm performance when they are valid. There has been a lot of work in this area and it is hard to discuss or even mention all of it in this review article.
We mention a few approaches as examples, all of these are motivated by the video analytics application. The incPCP algorithm is a robust PCA based solution for video analytics  that is online and near real-time and that uses extra heuristics to deal with camera jitter and to handle panning and camera motion \cite{incPCP}.
GOSUS (Grassmannian Online Subspace Updates with Structured-sparsity) \cite{gosus} is another incremental algorithm that uses structured sparsity of the outlier terms in conjunction with a GRASTA-like (or ReProCS-like) algorithm. 

Finally two useful modifications of the ReProCS idea exploit the fact that, for many video applications, the foreground (sparse outlier) support changes in a correlated fashion over time. The first, called modified-ReProCS \cite{rrpcp_allerton11,rrpcp_tsp}, simply assumes slow support change of the sparse outliers. Thus, instead of recovering the outlier vector and its support via $l_1$ minimization (or CoSaMP etc), it uses the modified-CS idea \cite{modcsjournal}. Modified-CS is designed to exploit partial support knowledge in sparse signal recovery.  The estimate of the outlier support from the previous time instant serves as the partial support knowledge. Replacing Modified-CS by weighted-$l_1$ further helps in certain settings. 
The support-predicted modified ReProCS approach \cite{rrpcp_isit} generalizes this basic idea further to include a support prediction step based on a simple object motion model and a Kalman filter to track the object's motion and velocity.
Modified-ReProCS \cite{rrpcp_tsp} is a truly practically useful algorithm since it can be used without any assumptions on how many moving objects, or regions, are there in a video; it only assumes that the foreground support does not change drastically. Its practical version uses a simple heuristic to decide when to use modified-CS and when to stick with simple $l_1$ minimization. We compare it in Table \ref{table1}.
\begin{table*}[ht!]
\caption{\small{Comparing assumptions, time and memory complexity for RPCA solutions. For simplicity, we ignore all dependence on condition numbers. 
All the algorithms require \emph{left and right} incoherence, and thus we do not list this in the third column.
}}
\vspace{-0.12in}
\begin{center}
\renewcommand*{\arraystretch}{1.15}
\resizebox{\linewidth}{!}{
\begin{tabular}{lllll}
\toprule
Algorithm & Outlier tolerance, rank of $(\L)$ & Assumptions  & Memory, Time, &  \# params.\\
\midrule
PCP(C)\cite{rpca}  & $\outfracrow = \mathcal{O}(1)$& strong incoh, & Mem: $\mathcal{O}(n \tmax)$   & zero \\
(offline)        & $\outfraccol = \mathcal{O}(1) $ &       unif. rand. support,  &             Time: $\mathcal{O}(n \tmax^2 \frac{1}{\epsilon})$   & \ \\
         &  $\rmat \leq \frac{c\min(n,\tmax)}{\log^2 n}$ &     & \   & \ \\
\midrule
AltProj\cite{robpca_nonconvex},  & $\outfracrow = O\left(1/\rmat\right)$ & & Mem: $\mathcal{O}(n \tmax)$   & 2 \\
(offline)   & $\outfraccol = O\left( 1/\rmat\right)$      &    & Time: $\mathcal{O}(n \tmax \rmat^2 \log \frac{1}{\epsilon})$   &   \\
\midrule
RPCA-GD & $\outfracrow = \mathcal{O}(1/\rmat^{1.5})$       &  & Mem: $\mathcal{O}(n \tmax)$   & 5 \\
 \cite{rpca_gd} (offline)       & $\outfraccol = \mathcal{O}(1/\rmat^{1.5})$ & & Time: $\mathcal{O}(n \tmax \rmat \log \frac{1}{\epsilon})$   &  \\
\midrule
PG-RMC& $\outfracrow = O\left(1/\rmat\right)$ & $\tmax  \approx n$  & Mem: $\mathcal{O}(n \tmax)$   & 4 \\
  \cite{rmc_gd} (offline)      & $\outfraccol  = \mathcal{O}(1/\rmat)$            &  & Time: $\mathcal{O}(n \rmat^3 \log^2 n \log^2 \frac{1}{\epsilon})$   &  \\
\midrule
{ReProCS-NORST} & {$\outfracrow  = \mathcal{O}(1) $} & outlier mag. lower bounded       & { Mem: $\mathcal{O}(n \rmat \log n \log \frac{1}{\epsilon} )$ }  & 4 \\
 \cite{rrpcp_merop,rrpcp_medrop} ({\em online})    & $\outfraccol = \mathcal{O}(1/\rmat) $ & slow subspace change or fixed subspace &  Time: $\mathcal{O}(n \tmax \rmat \log \frac{1}{\epsilon})$   & \ \\
 {\cred detects \& tracks}      &   & first $Cr$ samples: AltProj assumptions    & Track delay: $O(r\log n \log(1/\epsilon))$  & \ \\
{\cred subspace change}   & & \ & & \\
%
\bottomrule
\end{tabular}
}
\label{compare_assu}
\end{center}
\vspace{-0.15in}
\end{table*}

\renewcommand{\rmat}{r_L}
\begin{table*}[t!]
\caption{\small{Comparing all RPCA and RST solutions - both with and without guarantees. The methods are chronologically ordered.
}}
\vspace{-0.1in}
\begin{center}
\renewcommand*{\arraystretch}{1.1}
\resizebox{\linewidth}{!}{
\begin{tabular}{|l|l|l|l|l|}
\hline
algorithm & outlier tolerance \& rank of $\L$ & assumptions  & memory, time complexity & mode (\# par.) \\
&  or how well it works in practice &  if guarantee exists & or practical speed  &      \\ \hline
RSL \cite{Torre03aframework} & works well & no guarantee & slower & batch ($0$) \\ \hline
i-RSL \cite{Li03anintegrated,ipca_weightedand} & never works & no guarantee & slower & online ($0$) \\
\hline

PCP(C) \cite{rpca}  & $\outfracrow \in O(1)$   & unif. random support, & memory: $O(n \tmax)$   & batch ($0$) \\
 & $\outfraccol \in O(1) $ &       strong incoherence  &             time: $O(n \tmax^2 /{\epsilon})$   &   \\
         &  $\rmat \leq \frac{c\min(n,\tmax)}{\log^2 n}$ &     & \   & \ \\
\hline
PCP(H) \cite{rpca2,rpca_zhang},  & $\outfracrow \in O (1/\rmat)$ &  no extra assump.  & memory: $O(n \tmax)$   & batch ($2$)  \\
 & $\outfraccol \in O(1/\rmat)$      &    & time: $O(n \tmax^2 /{\epsilon})$  & \\
\hline
{ReProCS \cite{rrpcp_allerton,rrpcp_tsp}} &  works well & partial guarantee & fast & online ($4$) \\
\hline

GRASTA \cite{grass_undersampled} & works & no guarantee & faster  & online ($0$) \\ \hline
Bilinear-dec\cite{mateos_anomaly2} & never works & partial guarantee& & online ($1$) \\
 \hline
ORPCA \cite{xu_nips2013_1} & works well & partial guarantee & faster & online ($3$) \\
\hline
GoDec \cite{godec}  & works & partial guarantee & fast & batch ($3$) \\    
\hline
pRoST \cite{seidel2013prost} & works & no guarantee & slower & online ($2$) \\
\hline
AltProj \cite{robpca_nonconvex},  & $\outfracrow \in O\left(1/\rmat\right)$ &  no extra assump.  & memory: $O(n \tmax)$   & batch ($2$)  \\
 & $\outfraccol \in O\left(1/\rmat\right)$      &    & time: $O(n \tmax \rmat^2 \log \frac{1}{\epsilon})$   & \\
\hline
Original-ReProCS  &         {$\outfracrow  \in O(1) $} & ReProCS-NORST assump's,  and   & {memory: $O(nr \log n /{\epsilon^2} )$}  & online ($4$) \\
 \cite{rrpcp_isit15,rrpcp_aistats}    & {$\outfraccol \in O(1/r)$} &  outlier supp. change  model, & {time: $O(n \tmax r \log \frac{1}{\epsilon})$}  & \\ 
 & \ & slow subspace change (strong) & \ & \\
\hline
Modified-PCP \cite{zhan_pcp_jp} & $\outfracrow \in O(1)$           & unif. random support,           & memory: $O(n r \log^2 n )$   & online ($1$) \\
 & $\outfraccol \in O(1)$ & strong incoherence (weaker),      & time: $O(n \tmax r \log^2 n   \frac{1}{\epsilon} )$   &  \\
        &  $\rmat \le \frac{c\min(n,\tmax)}{\log^2 n}$    & slow subspace change (strong), & \   &  \\
\                           &   & init data: AltProj assump.    &   &  \ \\
\hline
APSM \cite{chouvardas2015robust} & \ & partial guarantee &   & online ($3$)  \\
\hline
RPCA-GD \cite{rpca_gd} & $\outfracrow \in O(1/\rmat^{1.5})$       & no extra assump. & memory: $O(n \tmax)$   & batch ($5$)  \\
        & $\outfraccol \in O(1/\rmat^{1.5})$ & \ & time: $O(n \tmax \rmat \log \frac{1}{\epsilon})$   & \\
\hline
NO-RMC \cite{rmc_gd} & $\outfracrow \in O\left(1/\rmat\right)$ & $\tmax  \approx n$  & memory: $O(n \tmax)$   & batch ($5$) \\
   & $\outfraccol  \in O(1/\rmat)$            &  & time: $O(n \rmat^3 \log n \log \frac{1}{\epsilon})$   &  \\
\hline
%
%
ReProCS-NORST &         {$\outfracrow  \in O(1) $} & outlier mag. lower bounded,      & {memory: $O(nr \log n \log \frac{1}{\epsilon} )$}  & online ($4$) \\
\& simple-ReProCS & {$\outfraccol \in O(1/r)$} & slow subspace change (mild), & {time: $O(n \tmax r \log \frac{1}{\epsilon})$}  & \\ 
\cite{rrpcp_medrop,rrpcp_dynrpca}  &   &  first $Cr$ samples: AltProj assumptions    &   & \\ 
\hline

\end{tabular}
}
\label{compare_assu2}
\end{center}
\raggedright \small{\emph{Note}: 
For algorithms that come without guarantees or with partial guarantees, we specify this in the ``assumptions" column. For all these algorithms in the ``outlier tolerance" column, we list ``works", ``works well" or ``never works". This information is based on the comparisons shown in  \cite{rrpcp_tsp} (for iRSL and bilinear-dec) or in the next section. In the memory, time complexity column, we list ``fast", ``faster", or ``slower". This information is again based on experiments described in the next section.
The dynamic RPCA solutions assume subspace at any time has dimension $r$ and there are a total of $J$ subspace changes. Thus $\rmat=rJ$.
%
%
}
\vspace{-0.15in}
\end{table*}

\section{Robust Subspace Recovery}

\subsection{History}
The robustness of PCA methods was first addressed in the fields of statistics and neural networks. In the statistics literature in 1980s, the predominant approach consists in replacing the standard estimation of the covariance matrix with a robust estimator of the covariance matrix \cite{RP_Campell,RS_Huber}. This formulation weights the mean and the outer products which form the covariance matrix. Calculating the eigenvalues and eigenvectors of this robust covariance matrix gives eigenvalues that are robust to sample outliers. The result is more robust, but unfortunately is limited to relatively low-dimensional data. The second approach to robust PCA uses this idea along with projection pursuit techniques \cite{RPCAPP_Croux}. 
In the old neural network literature in 1990s, Xu and Yuille \cite{rpca_neu} first addressed the problem of robust PCA by designing a neural network that relied on self-organizing rules based on statistical physics. The PCA energy function proposed by Oja was generalized by adding a binary decision field with a given prior distribution in order to take into account the outliers. The binary variables are zero when a data sample is considered an outlier.

\subsection{Low-rank plus column-sparse interpretation: outlier pursuit}\label{out_pursuit}
Recall that Robust Subspace Recovery assumes that an entire data vector is either an inlier or an outlier. Borrowing ideas from the S+LR literature, one way to reformulate this problem is as follows. The data matrix $\M$ can be split as $\M = \L + \S_C$ where $\S_C$ is a column-sparse matrix (many columns are zero and some are nonzero) \cite{outlier_pursuit}.
With this, an easy extension of the PCP idea can solve this problem. This was called ``outlier pursuit'' in \cite{outlier_pursuit} and solves
\[
\min_{\tL, \tS}  \|\tL\|_* + \lambda \|\tS_C\|_{2,1}  \text{ subject to } \M = \tL + \tS_C
\]
Here $\|\tS\|_{2,1}$ is the $l_1$ norm of the vector of column-wise $l_2$ norms. It is a commonly used convex surrogate for promoting group sparsity in the compressive sensing literature. The same program was also proposed in almost parallel work by McCoy and Tropp \cite{mccoy_tropp11} where it was called Low-Leverage Decomposition (LLD).
The guarantee given by Xu et. al \cite{outlier_pursuit} for outlier pursuit is given earlier in Theorem \ref{out_pursuit_thm}.

\subsection{Maximizing Mean absolute Deviation via MDR}
 In \cite{mccoy_tropp11}, McCoy and Tropp developed a randomized algorithm that they call Mean absolute Deviation Rounding (MDR) to approximate the robust first principal component (PC). This is defined as $v_{MDR}:=\max_{v:\|v\|=1} \|\Y'v\|_1 :=\max_{v:\|v\|=1} \sum_i |v'\y_i|$. This cost is called ``mean absolute deviation'' (MAD) and is  a classical metric for robust estimation known since the work of Huber going back to the 1980s. The later principal components can be computed by projecting the data vectors orthogonal to the subspace spanned by the previous ones (the projection pursuit PCA idea of Huber).
 The MDR solution is a computationally efficient algorithm to approximate $v_{MDR}$. It first solves a Semi-Definite Programming (SDP) relaxation of the original maximizing MAD problem. The SDP can be solved in polynomial time where as the original problem is non-convex. Next, it uses the output of the SDP in a randomization procedure to generate $K$ candidate guesses of $v_{MDR}$. The algorithm then picks the ``best'' one (the one that maximizes the MAD) and outputs that as an estimate of $v_{MDR}$.
 The authors show that the MDR algorithm is guaranteed to provide an estimate of the first robust PC whose mean absolute deviation is at least $(1-\epsilon)$ times that of $v_{MDR}$ w.h.p. The probability is high enough when $K$ is large enough. 


\subsection{Outlier-Robust PCA \cite{xu2013outlier}}
Given a mix of authentic and corrupted points, Xu et al. \cite{xu2013outlier} proposed the High-dimensional Robust PCA (HR-PCA) approach. This attempts to find a low-dimensional subspace that captures as much variance of the authentic points as possible.
As explained in \cite{xu2013outlier}, HR-PCA alternates between a PCA and a ``random removal" step in each iteration. For the PCA step, a candidate subspace is found via PCA on the ``clean points'' identified by the previous iteration. Its quality is measured by computing its trimmed variance (TV) which is defined as $TV(\Phat):= \sum_{j=1}^r \sum_{k=1}^{r} |(\Phat)_j{}' \y|_{(k)}^2 / r$. Here $|(\Phat)_j' \y|_{(k)}$ is the $k$-th order statistic ($k$-th largest entry) among the ``clean'' data points in the current iteration $|(\Phat)_j{}' \y_i|,i=1,2,\dots, \tmax$ and $(\Phat)_j$ denotes the $j$-th column of $\Phat$.
In the first iteration, all points are treated as ``clean points''. In consecutive iterations, this number is reduced by one by using ``random removal'' done as follows: the probability of a point $\y_i$ being removed is proportional to $\sum_{j=1}^r ((\Phat)_j{}'\y_i)^2$. In the end, the algorithm compares the subspace estimates from all iterations and picks the one with largest TV. The guarantee given for HR-PCA provides a lower bound on the expressed variance of the computed principal components. 
The obtained bound is hard to parse. As explained by the authors, it shows that the expressed variance is greater than zero (the solution is not meaningless) whenever the fraction of outliers is less than 50\%. On the other extreme, if the fraction of outliers goes to zero with $n$, then asymptotically the expressed variance goes to one (this corresponds to perfect recovery).

\subsection{Solutions for ``Robust Subspace Recovery''}
Robust subspace recovery as addressed by Lerman et al. \cite{novel_m_estimator} aims to recover the  underlying subspace, $\P$, from a dataset consisting of inlier and outlier points by using the following robust M-estimator: 
\begin{equation}
\min_{\P: \P'\P=\I} \sum_{i=1}^\tmax \left\| \y_i-\pP_{\P} \y_i \right\|=\sum_{i=1}^\tmax\left\|\pP_{\P_\perp} \y_i \right\|. \nn
\label{rsr_eq}
\end{equation}
Here
$\P_\perp$ is the orthogonal complement of $\P$, and $\pP_{\P}$ and $\pP_{\P_\perp}$ denote orthogonal projections onto $\P$ and $\P_\perp$ respectively. Since one needs to minimize over all matrices $\P$ with orthonormal columns, the above is a non-convex problem. However, notice that $\sum_{i=1}^\tmax\left\|\pP_{\P_\perp} \y_i \right\| = \sum_{i=1}^\tmax\left\| \P_\perp \P_\perp{}' \y_i \right\|$. The matrix $Q:=\P_\perp \P_\perp{}'$ is a projection matrix with properties $Q=Q'$, $Q^2 = Q$ and $tr(Q)=n-r$. This intuition leads to the following convex relaxation based approach.
The Geometric Median Subspace (GMS) algorithm \cite{novel_m_estimator} proceeds as follows: define the set $\mathbf{H}=\left\{Q \in \Re^{n \times n} : Q=Q', tr(Q)=1 \right\}$ and then solve
\[
\hat{Q}= \arg\min_{Q \in \mathbf{H}}   F(Q):=\sum_{i=1}{\tmax} \left\|Q \y_i \right\|
\]
For the noiseless case, the subspace $\P$ is estimated  as the eigenvectors corresponding to the smallest $r$ eigenvalues of $\hat{Q}$.  
%
In later work, Lerman et al. \cite{lerman_FCM} suggested a tighter convex relaxation of the original problem. Their approach, called REAPER, suggests solving the following:
\[
\hat{B}= \arg\min_B \sum_{i=1}^\tmax \left\| \y_i- B \y_i \right\| \ \text{ s.t. } 0 \preceq B \preceq \I, \ tr(B)=r
\]
This is the tightest convex relaxation because its constraint set is the convex hull of the set of basis matrices.
An estimate of $\P$ is then obtained as the top $r$ eigenvectors of $\hat{B}$.

Both GMS and REAPER work in the batch setting and therefore do not scale to big data. To address these limitations, Lerman et al. \cite{RSPCA_lerman} provided three stochastic approximation algorithms that combined GMS  and REAPER with the Stochastic Gradient Descent (SGD) idea.
Each improved algorithm presents less running time and space complexity requirements than the batch GMS and REAPER.

\subsection{Online PCA with contaminated data \cite{xu_nips2013_2}}
Feng et al. \cite{xu_nips2013_2} addressed the case of online PCA where outliers have no structural assumptions on them and data vectors come in sequentially. They develop an online algorithm which employed a probabilistic admiting/rejection procedure when a new sample comes in. Its guarantee makes assumptions on the initialization step (which is not specified), and the guarantee itself is asymptotic. The assumption on outlier fraction is very mild: anything less than 50\% corruption works.

\section{Parameter Setting and Experimental Comparisons}
We first explain how to set the parameters for the key approaches. After this, we provide a quantitative performance evaluation of these and many other RPCA methods for foreground-background separation (background subtraction) using \textit{Change Detection} (CDnet) 2012 dataset \cite{CDNET}. For all the videos, the authors' code with their final parameter settings was used. No extra tuning was done for any approach.

\newcommand{\lthres}{\omega_{evals}}
\subsection{Parameter Setting}
{\em Parameter setting for PCP \cite{rpca}. } There is only one parameter $\lambda$ in the PCP convex program. A smaller value of $\lambda$ allows the solution $\Shat$ to be less sparse. Thus its value should be chosen based on the expected number of outliers in the entire data matrix. However the PCP convex program itself requires a solver (an iterative algorithm that solves it to within $\epsilon$ error in a finite number of iterations). The solver has various other parameters. The selection of which solver to use can itself be another parameter.

{\em Parameter setting for AltProj \cite{robpca_nonconvex}. } The AltProj algorithm requires $2$ parameters, and a third to indicate the desired tolerance $\epsilon$. The first is the rank. No guidance is provided on how to set this except cross-validation. The second, is referred to as a thresholding parameter, $\beta$ which controls the rate at which the sparse outliers are detected. A larger value of $\beta$ indicates that after each ``inner loop'', the smallest non-zero value of the outlier is larger. A larger value of $\beta$ indicates that the algorithm converges quickly if the outliers are very large/small.
To detect outliers that have intermediate range requires a smaller value for $\beta$. The authors show that setting $\beta = 1/n$ works well in video experiments. 

{\em Parameter setting for RPCA-GD \cite{rpca_gd}:} There are $5$ parameters to be tuned for RPCA-GD. It needs an estimate of the rank, $\rmat$, the corruption fraction (the maximum fraction of outliers per each row and column), $\alpha_{GD}$, a ``safe-bound'' on the estimate of corruption fraction, $\gamma$, the step size for the projected gradient descent sub-routine, $\eta$, and the number of iterations, $T$. In the paper \cite{rpca_gd} the authors show that the step size can be set as $O(1/\sigma_1(\bm{U}_0\bm{V}_0{}'))$ where $\bm{U}_0$ and $\bm{V}_0$ denote the first estimate of the factorized low-rank estimate, i.e., $\bm{M} - \hat{\bm{S}}_0 = \bm{U}_0 \bm{V}_0{}'$. No intuition is given on how to set the other parameters; please see their code.

{\em Parameter setting for NO-RMC \cite{rmc_gd}:} The parameter setting here is very similar to that of AltProj. However, in addition to these parameters, NO-RMC also requires an estimate of the incoherence parameter $\mu$, and an estimate of the largest singular value of $\bm{L}$, $\sigma$. Both these estimates are required to set the number of iterations for the inner loop and ensures that the rank $k_q$ estimate is ``good enough''. For default parameters, please see their code.

{\em Parameter setting for ReProCS \cite{rrpcp_dynrpca,rrpcp_medrop}}: The first step for setting algorithm parameters for ReProCS based approaches is to obtain an initialization. There are two approaches that can be used for this. If there is outlier free training data (e.g., for long surveillance videos obtained from CCTV cameras), one can use simple SVD to obtain the initial subspace estimate: compute the top singular vectors that contain at least a certain percentage, e.g., 95\%, of the total energy (sum of squares of all singular values).
Another approach is to use few iterations of a batch RPCA method such as PCP or AltProj applied to the first $t_\train = C r $ frames of data. PCP requires no tuning parameters but is slower. AltProj is faster but needs to know $r$. By cross-validation, we conclude that $r=40$ and $t_\train = 400 = 10r$ suffices in all our video experiments. We set $K = C \log(1/\zz) = 3$, and $\alpha= C r \log n = 60$ as required by the guarantee. Since the dependence of $\alpha$ on $n$ is only logarithmic, for all practically used values of $n$ ($n=1000$ to $n=320*240$) $\alpha=60$ suffices.
The parameter $\lambda^+$ is computed as the square of the largest singular value of $\Lhat_{\init}/\sqrt{t_\train}$; and $\lambda^-$ is the square of its $r$-th largest singular value. The two parameters for the projected CS step are $\xi$ and $\omega_{supp}$. The required theoretical values of these depend on $\xmint$, the theorem requires $\xi=\xmint/15$ and $\omega_{supp}=\xmint/2$. One way to estimate $\xmint$ is to let it be time-varying and set it as $\min_{i \in \That_{t-1}}|(\xhat_{t-1})_i|$ at $t$. An approach that works better for the video application is to set $\xi_t =  \|\bpsi\lhat_{t-1}\|$, $\omega_{supp,t} = \sqrt{\|\mt\|^2/n}$. The former relies on the idea that $\xi_t$ needs to be a bound $\|\bpsi \lt\|$. The latter is the  Root Mean Square value of the measurement vector at time $t$. If there are not too many outliers, most outlier magnitudes will be larger than this threshold (the problematic outliers being large magnitude corruptions) and hence this is a useful support estimation threshold. Finally the subspace change detection threshold $\omega_{evals}$ can be set equal to a very small fraction of $\lambda^+$, $\omega_{evals}=0.0007 \lambda^+$ is used in all video experiments.

\subsection{Experimental Comparisons for Video Layering (foreground-background separation)}
We evaluated the performance of the current state-of-the-art RPCA-based methods for foreground-background separation (background subtraction) using \textit{Change Detection} (CDnet) 2012 dataset \cite{CDNET}.
We compared a total of $26$ existing methods comprising $16$ batch methods and eight online methods.
These methods can be classified into three main categories.
\ben
\item Provable methods comprise \textit{Principal Component Pursuit} (PCP) \cite{rpca,rpca2}, \textit{non-convex Alternating Projections based RPCA} (AltProj) \cite{robpca_nonconvex},  \textit{Near Optimal RMC} (NO-RMC) \cite{rmc_gd}, \textit{RPCA via Gradient Descent} (RPCA-GD) \cite{rpca_gd}, \textit{Recursive Projected Compressive Sensing} (ReProCS) \cite{rrpcp_tsp,rrpcp_perf}, \textit{simple Recursive Projected Compressive Sensing} (simple-ReProCS) \cite{rrpcp_dynrpca}, and Modified-PCP \cite{zhan_pcp_jp}.

\item  Heuristics methods include
\textit{Grassmannian Robust Adaptive Subspace Tracking Algorithm} (GRASTA) \cite{grass_undersampled}, \textit{Online RPCA} (OR-PCA) \cite{xu_nips2013_1}, \textit{Three Term Decomposition} (3TD) \cite{oreifej2013simultaneous}, \textit{Two-Pass RPCA} (2PRPCA) \cite{2PRPCA}, \textit{Go Decomposition} (GoDec) \cite{godec},  \textit{$l_{p}$ Robust Online Subspace Tracking} (pROST) \cite{seidel2013prost}, and \textit{Probabilistic Robust Matrix Factorization} (PRMF) \cite{wang2012probabilistic}. %

 \item Heuristics methods with application specific constraints consist of \textit{modified Recursive Projected Compressive Sensing} (modified-ReProCS) \cite{rrpcp_tsp},
\textit{incremental Principal Component Pursuit} (incPCP) \cite{incPCP},  \textit{Motion-assisted Spatiotemporal Clustering of Low-rank} (MSCL) \cite{MSCL}, \textit{Detecting Contiguous Outliers in the LOw-rank Representation} (DECOLOR) \cite{DECOLOR}, \textit{Low-rank Structured-Sparse Decomposition} (LSD) \cite{liu2015background}, \textit{Total Variation RPCA} (TVRPCA) \cite{TVRPCA}, \textit{Spatiotemporal RPCA} (SRPCA) \cite{SRPCA}, \textit{Robust Motion Assisted Matrix Restoration} (RMAMR) \cite{ye2015foreground}, \textit{Generalized Fussed Lasso} (GFL) \cite{xin2015background}, \textit{Grassmannian Online Subspace Updates with Structured-sparsity} (GOSUS) \cite{gosus}, \textit{Contiguous Outliers Representation via Online Low-rank Approximation} (COROLA) \cite{COROLLA}, and \textit{Online Mixture of Gaussians for Matrix Factorization with Total Variation} (OMoGMF$+$TV) \cite{MOGMF}, \textit{Online RPCA with illumination constraints} (OR-PCA-illum) \cite{javed2015robust}.
 \een
We also used the original author's implementations of TVRPCA, 2PRPCA, GRASTA, GoDec, DECOLOR, 3TD, LSD, SRPCA, GFL, GOSUS, RMAMR, OMoGMF$+$TV, simple-ReProCS, modified-ReProCS, AltProj, NO-RMC, and RPCA-GD whereas we reported the results which were obtained in their respective studies for the remaining methods.
The implementation of all the methods is also available in the low-rank and sparse library \cite{lrslibrary2015}.
The execution times required by all of the algorithms were compared on a machine with a 3:0 GHz Intel core i5 processor and 4GB of RAM.

\begin{figure*}[t!]
\centering
\includegraphics[width=\textwidth]{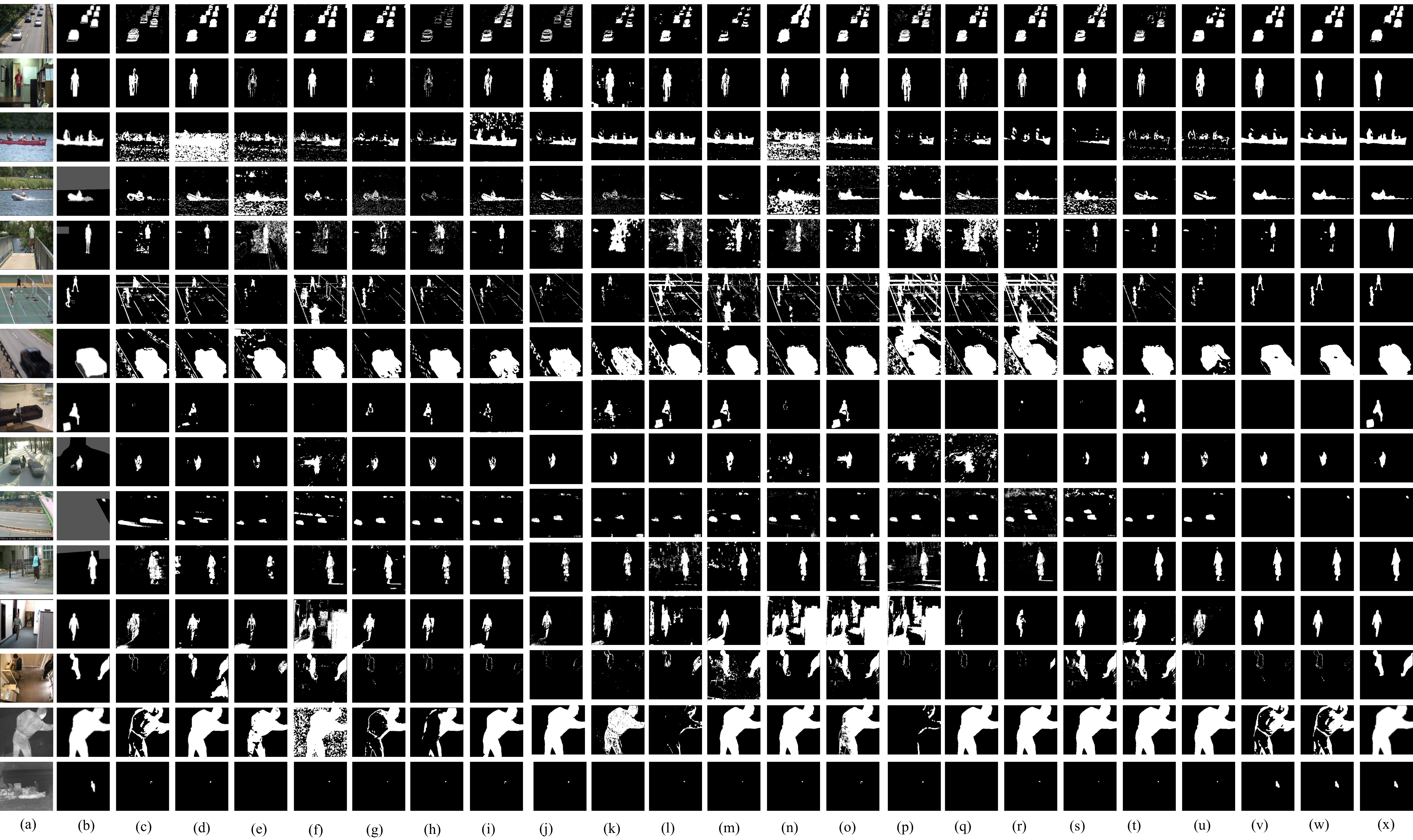}
\vspace{-1.0\baselineskip}
\caption{Comparison of the visual results obtained by the current state-of-the-art RPCA-based methods for background subtraction.
From left to right: comparison of the qualitative results of the $15$ input images from change detection dataset are shown.
(a) set of $15$ input images.
(b) the ground truth of foreground objects.
(c) background subtraction estimated by RPCA via PCP method.
(d) GoDec.
(e) RPMF.
(f) RPCA-GD.
(g) 3TD.
(h) pROST.
(i) incPCP.
(j) RMAMR.
(k) GRASTA.
(l) ReProCS.
(m) TVRPCA.
(n) SRPCA.
(o) NO-RMC.
(p) LSD.
(q) GOSUS.
(r) OMoGMF$+$TV.
(s) COROLLA.
(t) OR-PCA-illum.
(u) 2PRPCA.
(v) DECOLOR.
(w) GFL.
(x) MSCL.
From to bottom: Rows (1)-(2): Sequences `Highway' and `office' from category `Baseline'.
Rows (3)-(5): Sequences `canoe', `boat', and `overpass' from category DB.
Rows (6)-(7): Sequences `badminton' and `traffic' from category `Camera Jitter'.
Rows (8)-(10): Sequences `sofa', `winter Driveway', and `streetLight' from category IOM.
Rows (11)-(13): Sequences `BackDoor', `cubicle', and `copyMachine' from category `Shadow'.
Rows (14)-(15): Sequences `library' and `lakeside' from category `Thermal'.
}
\label{fig1}
\end{figure*}

Change detection 2012 dataset (CDnet) \cite{CDNET} is the real-world region-level benchmark obtained by human experts.
This dataset contains almost $31$ video sequences which are divided into six different video categories comprising `Baseline', `Dynamic Backgrounds' (DB), `Intermittent Object Motion' (IOM), `Thermal', `Camera Jitter', and `Shadows'.
The resolution of the videos also varies from $320\times240$  to $480 \times 720$ with hundred to thousand number of frames.

Visual results were reported using $15$ challenging sequences from CDnet dataset for comparison purpose.
This contained two sequences namely `highway' and `office' from `Baseline' category, three sequences `canoe', `boats', and `overpass' from DB category, two sequences `traffic' and `badminton' from `Camera Jitter' category, three sequences `winterDriveway', `sofa', and `streetLight' from IOM category, three sequences `backdoor', `copyMachine' and `cubicle' from `Shadows' category, and two sequences `library' and `lakeside' from `Thermal' category.
Fig.~\ref{fig1} provides qualitative results and comparisons of $22$ current state-of-the-art RPCA-based methods on $15$ sequences.

We used ground truth based metrics computed from the \textit{True Positives} (TP), \textit{True Negatives} (TN), \textit{False Positives} (FP) and \textit{False Negatives} (FN) for quantitative validation.
FP and FN refer to pixels misclassified as foreground (FP) or background (FN) while TP and TN account for accurately classified pixels respectively as foreground and background.
Then, we computed the metrics used in the CD.net \cite{CDNET} dataset evaluation such as the recall, the precision, and the $F_{1}$-measure score.
Recall gives the percentage of correct pixels classified as background when compared with the total number of background pixels in the ground truth as $Recall=\frac{TP}{TP+FN}$.
Precision gives the percentage of correct pixels classified as background as compared at the total pixels classified as background by the method as $Precision=\frac{TP}{TP+FP}$.
A good performance is obtained when the detection rate also known as recall is high without altering the precision.
Based on these metrics, we  computed the $F_{1}$-measure (or effectiveness measure) as $F_{1}=\frac{2 \times Recall \times Precision}{Recall+Precision}.$
The F-measure characterizes the performance of classification in precision-recall space.
The aim is to maximize $F_{1}$-measure closed to one.
Table \ref{table1} shows the quantitative results in terms of average $F_{1}$ measure score of all of the compared categories reviewed in this study.

On average, among all methods that do not use extra constraints, PRMF, 2PRPCA, simple-ReProCS, had the best performance with $F_1$ scores of 74-78\%. On average for all datasets, only two of the methods that use extra constraints -  MSCL and GOSUS - were better and only by a little - achieved 83 and 81\% scores.

We also reported the computational time in seconds for all of the compared methods. We selected a very large video sequence known as \textit{boats} from DB category for this purpose.
The \textit{boats} video contains about 8,000 image sequences of $320 \times 240$ image resolution.
We divided this sequence into temporal segments for most of the batch methods because of memory issue.
Table \ref{table1} also presents the computational time for all of the compared methods.
Simple-ReProCS and Modified-PCP are the fastest methods in provable methods category. Of these simple-ReProCS also has the best performance. From the heuristics, OR-PCA-illum and GRASTA are even faster, but have  bad performance.

The category \textit{Baseline} contained four simple videos where the background was always static in all of the video frames.
Fig.~\ref{fig1} (rows 1 and 2), showed that most of the compared methods in each category produced a good quality of foreground objects for these sequences.
%
The \textit{Dynamic Backgrounds} (DB) category comprised six challenging videos (three of them were shown in rows 3, 4, and 5 in Fig.\ref{fig1}) depicting outdoor scenes.
This was the most difficult among all categories for mounted camera object detection, which contained sequences exhibiting dynamic background motions because of rippling of water surfaces and swaying of bushes.
ONe scheme in provable methods category (ReProCS), three in the heuristics methods category (3TD, 2PRPCA, PRMF), and many in the heuristics methods with specific constraints category estimated a better quality of foreground objects than all of the compared methods in these categories.
Table \ref{table1} shows an average $F_{1}$ score of close to $80 \%$ and more than $80 \%$ for these methods because of the over-smoothing and spatio-temporal constraints enforced on these methods. In contrast, all other methods in these categories generated a noisy foreground segments because of highly dynamic background regions.
This experiment demonstrates that ReProCS can tolerate more background changes (larger rank $r$) for a given maximum outlier fraction per column (maximum support size of any foreground image).

The \textit{Intermittent Object Motion} (IOM) category included six videos (three were shown in rows 8, 9, and 10 in Fig.\ref{fig1}), which contained ghosting artifacts in the detected motion.
In these sequences, the moving foreground objects were motionless most of the time. This is a setting of large outlier fractions per row. As explained earlier, all static robust PCA methods that do not exploit slow subspace change will fail for this setting. This is indeed observed in our experiment. Most compared methods categories were not able to handle this challenge and obtained a low $F_{1}$ score (shown in Table \ref{table1}). ReProCS and ReProCS-provable achieved the best performance among all methods (provable or not) that do not exploit extra problem-specific assumptions. It had an $F_1$ score of 70\% since ReProCS {\em does} exploit the subspace dynamics.
Only two methods in heuristics with additional constraints category (SRPCA and MSCL) were better than ReProCS because they include specific heuristics to detect and remove motionless frames and they use spatiotemporal regularization in the low-rank background model.

The \textit{Shadows} category comprised six videos (some of them were shown in rows 11 to 13 in Fig.\ref{fig1}) exhibiting both strong and faint shadows.
For most of the compared methods in provable and heuristics methods category these videos posed a great challenge (see Table \ref{table1}).
Provable methods such as RPCA-GD and ReProCS, heuristics methods such as PRMF, and many heuristic methods with additional constraints achieved promising performance as compared to other methods. We observed that some hard shadows on the ground were still a major limitation of the top performing algorithms.
The \textit{Thermal} category comprised five sequences captured by the far-IR camera as shown in rows 14 and 15 in Fig. \ref{fig1}. Color saturation was the main challenge in this category, which degraded the performance of majority of the  compared methods in each category ($F_{1}$ score less than $80 \%$ in Table\ref{table1}). Only one provable method (ReProCS), and eight heuristics methods (ReProCS, 3TD, 2PRPCA, PRMF, and OR-PCA-illum, MSCL, GOSUS, and COROLLA) were able to discriminate the background-foreground pixels effectively in the presence of color saturation.
The \textit{Camera Jitter} category comprised one indoor and three outdoor videos (rows 6 and 7 in Fig. \ref{fig1}) where the background scene undergoes jitter induced motion throughout the video frames. 

\begin{table*}[t!]
\caption{The average $\textrm{F}_{1}$ score using CDnet dataset for the comparison of provable, heuristics, and heuristics with specific constraints methods for background subtraction.
Time is shown for a video having $320 \times 240$  resolution of $8,000$ frames.
The best and second best performing methods are shown in red and blue, respectively.}
\vspace{-0.1in}
\begin{center}
\makebox[\linewidth]{
\scalebox{0.90}{
\begin{tabu}{|c|ccccccc|c|}
\tabucline[1pt]{-}
Provable Methods&Baseline&DB&Camera Jitter&Shadow&Thermal&IOM&Average&Time (secs/frame)\\
\tabucline[1pt]{-}
PCP (batch) Fig. \ref{fig1} (c) \cite{rpca,rpca2}&0.75&0.69&0.62&0.73&0.65&0.48&0.65&4.19  \\
AltProj (batch) \cite{robpca_nonconvex}  &\textcolor{red}{\textbf{0.78}}&\textcolor{blue}{\textbf{0.71}}&0.60&\textcolor{red}{\textbf{0.76}}&0.69&0.58&\textcolor{blue}{\textbf{0.68}}&2.38\\
NO-RMC  (batch) Fig. \ref{fig1} (o) \cite{rmc_gd}  &0.71&0.64&0.64&0.66&\textcolor{blue}{\textbf{0.71}}&0.50&0.64&2.85\\
RPCA-GD  (batch) Fig. \ref{fig1} (f) \cite{rpca_gd} &0.74&0.62&0.68&\textcolor{blue}{\textbf{0.75}}&0.66&0.49&0.65&2.46\\
simple-ReProCS (online) \cite{rrpcp_aistats,rrpcp_dynrpca} & \textcolor{blue}{\textbf{0.77}} &\textcolor{red}{\textbf{0.77}}&\textcolor{blue}{\textbf{0.69}}&0.71&\textcolor{red}{\textbf{0.74}}&\textcolor{red}{\textbf{0.70}}&\textcolor{red}{\textbf{0.73}}& \textcolor{blue}{\textbf{0.74}}\\
Mod-PCP (online) \cite{zhan_pcp_jp} &{{0.75}}&0.64&\textcolor{red}{\textbf{0.70}}&0.65&0.69&\textcolor{red}{\textbf{0.70}}&\textcolor{blue}{\textbf{0.68}}&\textcolor{red}{\textbf{0.44}}\\
\tabucline[1pt]{-}
Heuristics Methods&Baseline&DB&Camera Jitter&Shadow&Thermal&IOM&Average&Time\\
\tabucline[1pt]{-}
GRASTA (online) Fig. \ref{fig1} (k) \cite{grass_undersampled} &0.66&0.35&0.43&0.52&0.42&0.35&0.45&1.16\\
3TD (batch) Fig. \ref{fig1} (g) \cite{oreifej2013simultaneous} &\textcolor{blue}{\textbf{0.88}}&0.75&0.72&0.68&\textcolor{blue}{\textbf{0.78}}&0.55&0.72&2.17\\
2PRPCA (batch) Fig. \ref{fig1} (u) \cite{2PRPCA}&\textcolor{red}{\textbf{0.92}}&\textcolor{red}{\textbf{0.79}}&\textcolor{blue}{\textbf{0.81}}&\textcolor{blue}{\textbf{0.80}}&0.76&\textcolor{blue}{\textbf{0.65}}&\textcolor{red}{\textbf{0.78}}&1.63\\
GoDec (batch) Fig. \ref{fig1} (d) \cite{godec} &0.77&0.58&0.48&0.51&0.62&0.38&0.55&1.56\\
OR-PCA (online)  \cite{xu_nips2013_1}  & 0.62 &	0.45 &	0.36	& 0.52 & 0.66	& 0.59 &	0.53 & 0.17  \\ 
pROST (online) Fig. \ref{fig1} (h) \cite{seidel2013prost}&0.79&0.59&0.79&0.70&0.58&0.48&0.65&2.03\\
PRMF  Fig. \ref{fig1} (e) \cite{wang2012probabilistic} &\textcolor{red}{\textbf{0.92}}&\textcolor{blue}{\textbf{0.77}}&\textcolor{red}{\textbf{0.85}}&\textcolor{red}{\textbf{0.88}}&\textcolor{red}{\textbf{0.83}}&0.48&\textcolor{red}{\textbf{0.78}}&2.40\\
\tabucline[1pt]{-}
Heuristics Methods with Specific Constraints&Baseline&DB&Camera Jitter&Shadow&Thermal&IOM&Average&Time\\
\tabucline[1pt]{-}
modified-ReProCS (online) Fig. \ref{fig1} (l) \cite{rrpcp_tsp} &0.80&0.76&0.72&0.75&0.77&\textcolor{red}{\textbf{0.69}}&\textcolor{blue}{\textbf{0.74}}&\textcolor{blue}{\textbf{0.61}}\\
incPCP (online) Fig. \ref{fig1} (i) \cite{incPCP}&0.81&0.71&0.78&0.74&0.70&\textcolor{blue}{\textbf{0.75}}&0.74& \textcolor{blue}{\textbf{0.41}}\\
OR-PCA-illum (online) Fig. \ref{fig1} (t) \cite{javed2015robust}&0.86&0.75&0.70&0.74&0.76&0.56&0.72&\textcolor{red}{\textbf{0.22}}\\
MSCL (batch) Fig. \ref{fig1} (x)  \cite{MSCL} &0.87&\textcolor{blue}{\textbf{0.85}}&\textcolor{red}{\textbf{0.83}}&0.82&\textcolor{red}{\textbf{0.82}}&\textcolor{red}{\textbf{0.80}}&\textcolor{red}{\textbf{0.83}}&1.68\\
DECOLOR (batch) Fig. \ref{fig1} (v) \cite{DECOLOR} &\textcolor{red}{\textbf{0.92}}&0.70&0.68&\textcolor{blue}{\textbf{0.83}}&0.70&0.59&0.73&1.88\\
LSD (batch) Fig. \ref{fig1} (p) \cite{liu2015background}&\textcolor{red}{\textbf{0.92}}&0.71&0.78&0.81&0.75&0.67&0.77&1.43\\
TVRPCA (batch) Fig. \ref{fig1} (m) \cite{TVRPCA}&0.84&0.55&0.63&0.71&0.69&0.57&0.66&1.48\\
SRPCA (batch) Fig. \ref{fig1} (n) \cite{SRPCA}&0.82&0.84&0.78&0.77&0.79&\textcolor{red}{\textbf{0.80}}&0.80& \textcolor{blue}{\textbf{0.59}}\\
RMAMR (batch) Fig. \ref{fig1} (j) \cite{ye2015foreground} &0.89&0.82&0.75&0.73&0.75&0.66&0.76&1.32\\
GFL (batch) Fig. \ref{fig1} (w) \cite{xin2015background}&0.83&0.74&0.78&0.82&0.76&0.59&0.75&2.40\\
GOSUS (online) Fig. \ref{fig1} (q) \cite{gosus} &\textcolor{blue}{\textbf{0.90}}&0.79&\textcolor{blue}{\textbf{0.82}}&\textcolor{red}{\textbf{0.84}}&\textcolor{blue}{\textbf{0.80}}&0.74&\textcolor{blue}{\textbf{0.81}}&0.89\\
COROLA (online) Fig. \ref{fig1} (s) \cite{COROLLA} &0.85&\textcolor{red}{\textbf{0.86}}&\textcolor{blue}{\textbf{0.82}}&0.78&\textcolor{blue}{\textbf{0.80}}&0.71&0.80&\textcolor{blue}{\textbf{0.39}}\\
OMoG$+$TV (online) Fig. \ref{fig1} (r) \cite{MOGMF} &0.85&0.76&0.78&0.68&0.70&0.71&0.74&\textcolor{red}{\textbf{0.19}}\\
\tabucline[1pt]{-}
Non-RPCA Methods&Baseline&DB&Camera Jitter&Shadow&Thermal&IOM&Average&Time\\
\tabucline[1pt]{-}
PAWCS \cite{st2015self}&{{0.88}}&{{0.85}}&{{0.78}}&{{0.86}}&{{0.80}}&{{0.74}}&{{0.81}}&{{2.82}}\\
SuBSENSE  \cite{st2015subsense} &{{0.91}}&{{0.78}}&0.78&{{0.88}}&0.78&0.63&{{0.79}}&{{1.89}}\\
LOBSTER \cite{st2014improving} &0.87&0.74&{{0.80}}&0.72&{{0.79}}&{{0.72}}&0.75&3.7\\
\tabucline[1pt]{-}
\end{tabu}
}}
\end{center}
\label{table1}
\end{table*}

\subsection{Comparison with non-RPCA methods}
We evaluated the RPCA methods against the top methods in the CDnet 2012 dataset. We excluded the supervised methods based on deep learning because RPCA methods are unsupervised ones. We selected three top performing unsupervised methods called SuBSENSE \cite{st2015subsense}, PAWCS \cite{st2015self}, and LOBSTER \cite{st2014improving}. For a fair comparison, we re-tested these methods with their source code publicly available either on the author's respective web pages or BGSLibrary, and we also reported the time taken by these methods on the same machine as the one used for the RPCA methods. From our comparisons, the F scores of SUBSENSE, PAWCS, and LOBSTER are little bit lower than the ones reported in the CDnet 2012 dataset. This is likely because we used the same set of parameters for all of the categories and the videos  of the dataset whilst the authors of SuBSENSE, PAWCS, and LOBSTER have optimized the set of parameters for each case. 
In addition, the computational time is reported for these methods by including the training time.  This is again done to keep comparisons with RPCA based methods fair.
 On average, SuBSENSE, PAWCS, and LOBSTER reached an F-score equal to $0.79$, $0.81$ and $0.75$, respectively which is better than many RPCA methods. But they are also slower - the time taken is 2.8, 1.9, 3.7 seconds per frame. This is much slower many of the RPCA methods that also work well (are among the top 5 methods), e.g., ReProCS takes only 0.7 seconds per frame.

 The reasons for both the accuracy and slow speed are that (i) SuBSENSE, PAWCS, and LOBSTER used many more color and texture features than RPCA methods; and (ii) they use multiple cues (use additional process to deal with shadows and/or sudden illumination changes). SuBSENSE uses an advanced texture features called spatio-temporal binary features in addition to the color feature. For PAWCS, a background dictionary is iteratively updated using a local descriptor based on different low-level features. For LOBSTER, it is also based on a texture feature called Local Binary Patterns (LBP) to cope with sudden lighting variations in the background scenes. 

\section{Future Directions}
The RPCA problem has been extensively studied in the last seven to ten years.  Dynamic RPCA or robust subspace tracking has received significant attention much more recently and there are many unsolved important questions.  The first question is can we obtain a guarantee for it under even weaker assumptions - can the lower bound on outlier magnitudes be relaxed? 
Two important extensions of RPCA are robust matrix completion and undersampled robust PCA. While these have been studied, their dynamic extensions have received almost no attention. These remain important questions for very long datasets where a changing subspace assumption is a more appropriate one.
 There is an algorithm and a partial guarantee for the undersampled case in \cite{rrpcp_allerton11,rrpcp_globalsip}, but a complete correctness result still does not exist; and careful experimental evaluations of the proposed techniques on real dynamic MRI datasets are missing too.

A question motivated by video analytics is how to develop an RPCA and an RST solution for data from moving cameras? To understand the problem, suppose that the camera motion corresponds to just x-y translation. Then, this means that  the video data matrix $\M$ itself is not sparse + low-rank, but one needs to motion-compensate the video images in such way that the resulting new matrix $\tilde\M$ is accurately modeled as sparse + low-rank. While many heuristics exist to address this issue, a simple provably correct and useful approach does not exist.
Another related issue of practical importance is how to reliably detect sudden subspace changes? For the video application, this would occur due to sudden scene changes (camera being turned around for example). Slow subspace change is easy to detect and reliably estimate (as explained earlier), but sudden change results in a complete loss of track and hence the low-rank component also gets misclassified as outlier.  One heuristic that works in practice is to use a sudden significant increase in outlier support size as an indicator of sudden scene change. It would be interesting to study if this is indeed a reliable indicator of sudden change.

Another interesting and practically relevant question is how to develop a simple and provable RST solution that is streaming, i.e., it requires only one pass through the data and  needs just enough memory to only store the output subspace estimate (an $n \times r$ matrix). A streaming RPCA solution has been developed in a 2016 arXiv preprint, but it works only for one-dimensional RPCA. On the other hand, ReProCS is a nearly memory optimal solution to dynamic RPCA \cite{rrpcp_dynrpca,rrpcp_medrop}, but it requires more than one pass through the data (enough passes to solve SVD up to the error level at a given iteration).


An open question is how can robust and dynamic robust PCA ideas be successfully adapted to solve other more general related problems. One such problem is subspace clustering which involves clustering a given dataset into one of $K$ different low-dimensional subspaces.  This can be interpreted as a generalization of PCA which tries to represent a given dataset using a single low-dimensional subspace. 
Another completely open question is whether one can solve the phaseless robust PCA or S+LR problem. In many  applications, such as ptychography, sub-diffraction imaging or astronomy, one can only acquire magnitude-only measurements. If the unknown signal or image sequence is well modeled as S+LR, can this modeling be exploited to recover it from under-sampled phaseless measurements? Two precursors to this problem, low rank phase retrieval \cite{lrpr_tsp} and phase retrieval for a single outlier-corrupted signal \cite{chi_pr_outliers}, have been recently studied.
Finally, this article does not review the literature on deep learning based approaches to RPCA, e.g., \cite{rpca_nn}, nor does it overview the recent work on robust or dynamic robust PCA for tensor data, e.g. \cite{selin_reprocs}. 
All of these are active research areas with many open questions, but will hopefully be the foci of other review papers.

\bibliographystyle{IEEEbib} 
\bibliography{tipnewpfmt_kfcsfullpap,addTB,bare_jrnl}

\section*{Author Biographies}

\begin{IEEEbiography}[{\includegraphics[width=1in,height=1.25in,clip,keepaspectratio]{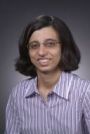}}]{Namrata Vaswani}
(Email: namrata@iastate.edu) is a Professor of Electrical and Computer Engineering, and (by courtesy) of Mathematics, at Iowa State University. She received a Ph.D. in 2004 from the University of Maryland, College Park and a B.Tech. from Indian Institute of Technology (IIT-Delhi) in India in 1999. Her research interests lie at the intersection of statistical machine learning / data science, computer vision, and signal processing. Her recent research has focused on provably correct and practically useful online (recursive) algorithms for the following two structured (big) data recovery problems: (a) dynamic compressive sensing (CS) and (b) dynamic robust principal component analysis (RPCA).

Vaswani is an Area Editor for IEEE Signal Processing Magazine, has served twice as an Associate Editor for IEEE Transactions on Signal Processing and is the Lead Guest Editor for a Proceedings IEEE Special Issue on Rethinking PCA for Modern Datasets that will appear in 2018. She is also the Chair of the Women in Signal Processing (WiSP) Committee and a steering committee member of SPS's Data Science Initiative.  In 2014, Prof Vaswani received the IEEE Signal Processing Society (SPS) Best Paper Award for her Modified-CS work that was co-authored with her graduate student Lu in the IEEE Transactions on Signal Processing in 2010.
\end{IEEEbiography}

\begin{IEEEbiography}[{\includegraphics[width=1in,height=1.25in,clip,keepaspectratio]{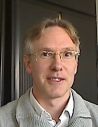}}]{Thierry Bouwmans} (Email: tbouwman@univ-lr.fr) is currently an Associate Professor at the University of La Rochelle (France) since 2000. He obtained the diploma of HDR for full professor position in 2014. His research interests consist mainly in the detection of moving objects in challenging environments. He has coauthored two books in CRC Press (background/foreground separation for video surveillance, robust PCA via decomposition in low rank and sparse matrices). His research investigated particularly the use of robust PCA in video surveillance. He is also the main organizers of the RSL-CV workshops hosted at ICCV in 2015 and 2017. He is a reviewer for international journals including IEEE (Trans. on Image Processing, Trans. on Multimedia, Trans. on CSVT, etc.), SPRINGER (IJCV, MVA, etc.) and ELSEVIER (CVIU, PR, PRL, etc.), and top-level conferences such as CVPR, ICPR, ICIP, AVSS, etc.
\end{IEEEbiography}

\begin{IEEEbiography}[{\includegraphics[width=1in,height=1.25in,clip,keepaspectratio]{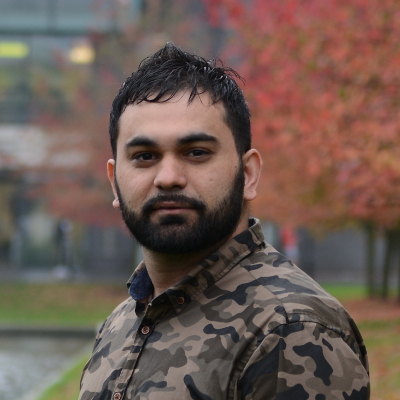}}]{Sajid Javed} (Email: S.Javed.1@warwick.ac.uk)
  is currently a Post-doc research fellow in the Department of Computer Science, University of Warwick, United Kingdom, from October, 2017. Dr. Sajid is working actively on computer vision and image analytics for cancer under the supervision of world leader Prof. Nasir Rajpoot at TIA Lab. Before joining the Tissue Image Analytics  Lab, he obtained his Bs.c (hons) degree in Computer Science from University of Hertfordshire, UK, in 2010. He then joined the Virtual Reality Laboratory of Kyungpook National University, Republic of Korea, in 2012, where he completed his combined Master's and Doctoral degrees in Computer Science under the supervision of Prof. Soon Ki Jung and co-supervision of Prof. Thierry Bouwmans from MIA Lab, France in 2017. Dr. Sajid has co-authored about 30 publications, this includes several journals and international conferences publications in the area of Robust Principal Component Analysis for Background-Foregorund Modeling. His other research interests are salient object detection, visual object tracking, semantic segmentation, subspace clustering, and social analysis of cancer cells.
\end{IEEEbiography}

\begin{IEEEbiography}[{\includegraphics[width=1in,height=1.25in,clip,keepaspectratio]{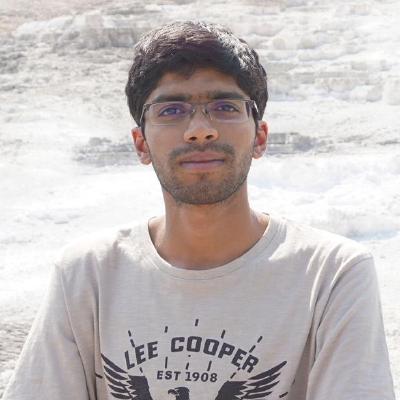}}]{Praneeth Narayanamurthy} (Email: pkurpadn@iastate.edu) (S' 18) is a Ph.D. student in the Department of Electrical and Computer Engineering at Iowa State University (Ames, IA). He previously obtained his B.Tech degree in Electrical and Electronics Engineering from National Institute of Technology Karnataka (Surathkal, India) in 2014, and subsequently worked as a Project Assistant at the Department of Electrical Engineering of Indian Institute of Science. His research interests include the algorithmic and theoretical aspects of High-Dimensional Statistical Signal Processing, and Machine Learning.
\end{IEEEbiography}

\end{document}